\documentclass[12pt,a4paper]{article}
\pdfoutput=1

\usepackage{graphicx,array}
\usepackage{hyperref}
\usepackage{comment}
\usepackage[normalem]{ulem} 
\usepackage{cite}
\usepackage{color}
\usepackage{appendix}
\usepackage{epsfig}
\usepackage{latexsym}
\usepackage{amsmath}
\usepackage{amssymb}
\usepackage{bm}
\usepackage{hyperref}
\usepackage{slashed}
\usepackage{bbold}
\usepackage{subfig}
\usepackage{multirow}

\numberwithin{equation}{section}

\definecolor{darkblue}{cmyk}{1,0.3,0,0.2}
\definecolor{violet}{cmyk}{0,1,0,0.2}
\hypersetup{colorlinks, bookmarksnumbered, citecolor=darkblue, linkcolor=darkblue, pdfstartview=FitH, urlcolor=darkblue, linktocpage}

\newcommand{\be}{\begin{equation}}
\newcommand{\ee}{\end{equation}}
\newcommand{\bea}{\begin{eqnarray}}
\newcommand{\eea}{\end{eqnarray}}

\newcommand{\MeV}{\textrm{ MeV}}
\newcommand{\GeV}{\textrm{ GeV}}
\newcommand{\TeV}{\textrm{ TeV}}
\newcommand{\SU}{\textrm{SU}}
\newcommand{\SO}{\textrm{SO}}
\newcommand{\Sp}{\textrm{Sp}}
\newcommand{\U}{\textrm{U}}
\newcommand{\Tr}{\textrm{Tr}}
\newcommand{\SM}{\textrm{SM}}
\newcommand{\gsim}{\lower.7ex\hbox{$\;\stackrel{\textstyle>}{\sim}\;$}}
\newcommand{\lsim}{\lower.7ex\hbox{$\;\stackrel{\textstyle<}{\sim}\;$}}

\newcommand{\LL}{\mathcal{L}}
\newcommand{\cL}{\mathcal{L}}

\newcommand{\OO}{\mathcal{O}}
\newcommand{\GG}{\mathcal{G}}

\newcommand{\MM}{\mathcal{M}}

\newcommand{\BR}{\mathcal{B}}

\newcommand{\ba} {\begin{eqnarray}}
\newcommand{\ea} {\end{eqnarray}}

\begin{document}
 
 \hfill

\vspace{1.0cm}

\begin{center}
{\LARGE\bf Addressing the $B$-physics anomalies\\[0.3cm] in a fundamental Composite Higgs Model}
\\ 

\bigskip\vspace{1cm}{
{\large \mbox{David Marzocca}  }
} \\[7mm]
{\em INFN, Sezione di Trieste, SISSA, Via Bonomea 265, 34136, Trieste, Italy}  \\ 

\vspace*{0.5cm}
   
\end{center}
\vspace*{1.5cm}

\centerline{\large\bf Abstract}
\medskip\noindent 
I present a model addressing coherently the naturalness problem of the electroweak scale and the observed pattern of deviations from the Standard Model in semi-leptonic decays of $B$ mesons. The Higgs  and the two scalar leptoquarks responsible for the $B$-physics anomalies, $S_1 = ({\bf \bar 3}, {\bf 1}, 1/3)$ and $S_3 = ({\bf \bar 3}, {\bf 3}, 1/3)$, arise as pseudo Nambu-Goldstone bosons of a new strongly coupled sector at the multi-TeV scale. I focus on an explicit realization of such a dynamics in terms of a new strongly coupled gauge interaction and extra vectorlike fermions charged under it.
The model presents a very rich phenomenology, ranging from flavour observables, Higgs and electroweak precision measurements, and direct searches of new states at the LHC.

\vspace{0.3cm}

\newpage
\tableofcontents


\section{Introduction}

The search for phenomena beyond those described by the Standard Model (SM) at the Large Hadron Collider (LHC) has been motivated mainly by the naturalness problem of the electroweak (EW) scale. Indeed, its solutions predict new physics (NP) not too far from the electroweak scale. However, none of the expected spectacular signatures have been observed thus far, pushing the mass scale of new physics particles to uncomfortably high values and implying that at least some amount of tuning most likely has to be accepted.

While the experimental situation in high-$p_T$ searches might look somewhat depressing, a set of interesting deviations from the SM predictions started to appear a few years ago in semileptonic decays of $B$ mesons, particularly in observables testing lepton-flavour universality (LFU). The first deviations were observed by the BaBar collaboration in the charged-current transition $b \to c \tau \bar\nu$ via the observables $R(D^{(*)}) = \BR(B \to D^{(*)} \tau \nu) / \BR(B \to D^{(*)} l \nu)$ \cite{Lees:2012xj,Lees:2013uzd}. All subsequent measurements of the same observables by the Belle and LHCb experiments provided results consistently above the robust SM prediction \cite{Aaij:2015yra,Huschle:2015rga,Sato:2016svk,Hirose:2016wfn}. Global fits \cite{Amhis:2016xyh} put the combined statistical significance just above the $4 \sigma$ level.
A second set of deviations has been observed by the LHCb experiment in rare neutral-current $b \to s \mu^+\mu^-$ transitions. First hints appeared when studying angular distributions in the $B \to K^* \mu^+ \mu^-$ decay \cite{Aaij:2013qta,Aaij:2015oid} as well as in decay rates of other processes with the same partonic transition. These observables, however, face difficulties in the SM prediction since non-perturbative QCD effects can be sizeable and challenging to control \cite{Ciuchini:2015qxb}. Theoretically cleaner observables probing the same partonic transition are the LFU ratios $R(K^{(*)}) = \BR(B \to K^{(*)} \mu^+ \mu^-) / \BR(B \to K^{(*)} e^+ e^-)$, which also show consistent deviations from the SM \cite{Aaij:2014ora,Aaij:2017vbb}. The overall significance of the deviations in neutral-current processes is also above the $4\sigma$ level (the precise number depending on the theory error estimate in the angular observables) \cite{Altmannshofer:2015sma,Descotes-Genon:2015uva,Altmannshofer:2017yso,DAmico:2017mtc,Capdevila:2017bsm,Ciuchini:2017mik,Hiller:2017bzc}.

Further data to be gathered by the LHCb and Belle II experiments will provide a conclusive answer as to the nature of these anomalies within the next few years. Therefore it is now timely to attempt an explanation of the observed pattern of deviations in terms of some new physics, even more so since such an exercise provides correlations with other low- and high-energy observables. First attempts towards combined explanations of the two sets of anomalies have been studied in Refs.~\cite{Bhattacharya:2014wla, Alonso:2015sja,Greljo:2015mma,Calibbi:2015kma,Bauer:2015knc,Fajfer:2015ycq, Barbieri:2015yvd,Buttazzo:2016kid,Das:2016vkr, Boucenna:2016qad,Becirevic:2016yqi,Hiller:2016kry, Bhattacharya:2016mcc,Barbieri:2016las,Becirevic:2016oho,Bordone:2017anc,Megias:2017ove,Crivellin:2017zlb,Cai:2017wry,Altmannshofer:2017poe,Sannino:2017utc}. Most of these scenarios, however, face very challenging constraints from $\tau \tau$ searches at the LHC \cite{Greljo:2015mma,Faroughy:2016osc} and from electroweak precision data \cite{Feruglio:2016gvd,Feruglio:2017rjo,Cornella:2018tfd}, mainly due to the low scale of new physics required by the charged-current anomaly. A first simple solution to these issues was found in Ref.~\cite{Buttazzo:2017ixm}, where a sizeable $b-s$ mixing allows to raise the new physics scale enough to pass both the collider and the electroweak precision data bounds. A large flavour-violating coupling, however, carries potential problems in $B \to K^* \nu \nu$ and $B$-meson mixing, which must be addressed in a realistic scenario.

At the level of simplified models, a classification of the new particles, and their properties, which can generate the required operators when integrated out at the tree-level while avoiding other constraints, has also been presented in Ref.~\cite{Buttazzo:2017ixm}. These are:
\begin{itemize}
	\item a vector leptoquark, $U_1^\mu = ({\bf 3}, {\bf 1}, 2/3)$,
	\item a pair of scalar leptoquarks, $S_1 = ({\bf \bar 3}, {\bf 1}, 1/3)$ and $S_3 = ({\bf \bar 3}, {\bf 3}, 1/3)$,
\end{itemize}
where I show the representation under the SM gauge group $\GG_{\SM} = \SU(3)_c \times \SU(2)_w \times \U(1)_Y$.

Going beyond simplified models, embedding these leptoquarks (LQ) in a more complete theory can offer further insight and new correlations with different observables, such as direct searches of other particles predicted by the UV theory.
A first observation to be made when thinking about possible UV realisations is that the mass scale of the leptoquarks required to fit the $B$-physics anomalies is close to $\sim 1 \TeV$, which corresponds also to the scale where new physics related to the electroweak hierarchy problem is supposed to be. This coincidence of scales is a strong motivation to look for UV theories which address both issues in a coherent manner.

Some examples of embedding the vector LQ $U_1^\mu$ in a more complete theory have been presented in the literature. For example, it can be recognised as one of the heavy gauge bosons in Pati-Salam unification, or variations thereof \cite{Assad:2017iib,Calibbi:2017qbu,DiLuzio:2017vat,Bordone:2017bld,Greljo:2018tuh,Blanke:2018sro}. In these scenarios, however, the naturalness problem remains unaddressed. Alternatively, $U_1^\mu$ could arise as a composite vector resonance of a new strongly coupled sector lying at the TeV scale \cite{Barbieri:2016las,Barbieri:2017tuq,Cline:2017aed}. In some of these setups the same sector could also generate the Higgs boson as a pseudo-Nambu-Goldstone boson (pNGB), as in composite Higgs models.
In all these scenarios other states, such as neutral or color-octet vectors, are necessarily present with a mass close to the LQ one. They usually generate undesired too large effects in $\Delta F = 2$ processes and direct searches, inducing some tension in the models. 
The problem can be summarised as the fact that the mass scale of the other resonances contributing significantly to flavour is naturally at the same scale as the vector LQ: $m_{VLQ} \sim \Lambda$.

The scalar leptoquarks $S_1$ and $S_3$, on the other hand, can be naturally lighter than the other states in the theory if they arise as pNGB of some spontaneously broken global symmetry of a new strongly coupled sector:
\be
	m_{SLQ} \ll \Lambda~.
\ee
This splitting naturally explains why the effects of the scalar leptoquarks in flavour observables are the leading ones.
This idea was explored in Refs.~\cite{Gripaios:2009dq,Gripaios:2014tna} in an effective field theory (EFT) approach, where however only the neutral-current anomalies were considered.
In such a setup it is natural to consider also the Higgs boson as a pNGB of the same dynamics, thereby realising a composite Higgs model \cite{Georgi:1984af,Kaplan:1983fs} and addressing the naturalness problem of the electroweak scale.
The $S_1$ and $S_3$ LQs have already been considered, also separately, as possible mediators for either the neutral- or charged-current anomalies (or both) in Refs.~\cite{Gripaios:2009dq,Sakaki:2013bfa,Hiller:2014yaa,Gripaios:2014tna,Bauer:2015knc,Das:2016vkr,Becirevic:2016oho,Hiller:2016kry,Crivellin:2017zlb,Cai:2017wry,Dorsner:2017ufx,Buttazzo:2017ixm,Fajfer:2018bfj}.

Following this route, in this work I present a natural model able to address at the same time both the charged- and neutral-current $B$-physics anomalies via the exchange of the $S_1$ and $S_3$ scalar leptoquarks. They arise as pNGB, together with the Higgs boson, from a new strongly coupled sector at the $\sim 10\TeV$ scale.
Rather than employing an EFT-like approach, in order to be more predictive and to provide a more realistic and UV-complete setup I also specify the strong dynamics as a four-dimensional fermionic confining gauge theory \cite{Galloway:2010bp,Schmaltz:2010ac,Barnard:2013zea,Ferretti:2013kya,Cacciapaglia:2014uja,Ferretti:2014qta,Vecchi:2015fma,Ma:2015gra,Ferretti:2016upr}. This puts strong constraints on the viable global symmetry-breaking patterns, therefore on the low-energy chiral Lagrangian.

The structure of the paper is as follows.
In Section~\ref{sec:CompHiggs} I introduce the specific fundamental Composite Higgs model, its global symmetries and the low-energy pNGB field content, which includes two Higgs doublets and the two scalar LQ among other fields.
In Section~\ref{sec:4FermOp} I discuss the way by which elementary fermions couple to the composite sector, thereby generating the Higgs Yukawa and leptoquark couplings. These couplings, together with SM gauge interactions and fermion masses break explicitly the global symmetry of the strong sector. This generates a scalar potential for the pNGB, which is studied in Section~\ref{sec:potential}. This potential is responsible for the Higgs non-vanishing vacuum expectation value (vev) and for electroweak symmetry breaking (EWSB), Section~\ref{sec:EWSB}.
The flavour phenomenology arising from the LQ couplings to fermions, including the fit to the $B$-physics anomalies, is studied in Section~\ref{sec:flavour}. The most interesting collider signatures, as well as the present limits from direct searches, are presented in Section~\ref{sec:collider}. Finally, I conclude in Section~\ref{sec:conclusions}.

\section{A fundamental Composite Higgs Model}
\label{sec:CompHiggs}

The naturalness problem of the electroweak scale can be solved by assuming that the Higgs boson is a composite state of a new strong dynamics at a scale $\Lambda \sim \TeV$. Furthermore, the splitting $m_h \ll \Lambda$, required by phenomenological constraints, can be naturally realised if the Higgs arises as a pseudo Nambu-Goldstone boson from the spontaneous breaking of an (approximate) global symmetry of the strong dynamics \cite{Kaplan:1983fs,Georgi:1984af}, in close analogy to the pions in QCD.

Extending this idea to include the scalar leptoquarks $S_1$ and $S_3$, I construct a fermionic fundamental description of a composite model, from which both the scalar LQ and the Higgs arise as pNGBs. See App.~\ref{app:Requirements} for a general discussion on the requirements such a UV setup should satisfy.

\subsection{The explicit model}

As sketched already in Ref.~\cite{Buttazzo:2017ixm}, and in analogy with Refs.~\cite{Buttazzo:2016kid,Ma:2015gra,Vecchi:2015fma}, I add a new non-abelian gauge group $\GG_{HC} = \SU(N_{HC})$, assumed to confine at a scale $\Lambda_{HC} \sim 10 \TeV$, and a vectorlike set of fermions in the fundamental (and anti-fundamental) representation of this new gauge group and charged under the SM group as well. The extra matter content considered in this work, classified in representations of $\SU(N_{HC}) \times \SU(3)_c \times \SU(2)_w \times \U(1)_Y$, is shown in Table~\ref{tab:fields}.
\begin{table}
\centering
\begin{tabular}{c | c | c | c | c }
			& $\SU(N_{HC})$ & $\SU(3)_c$ & $\SU(2)_w$ & $\U(1)_Y$ \\\hline
	$\Psi_L$ & ${\bf N_{HC}}$ & ${\bf 1}$ & ${\bf 2}$ &  $Y_L$ \\
	$\Psi_N$ & ${\bf N_{HC}}$ & ${\bf 1}$ & ${\bf 1}$ &  $Y_L + 1/2$ \\
	$\Psi_E$ & ${\bf N_{HC}}$ & ${\bf 1}$ & ${\bf 1}$ &  $Y_L - 1/2$ \\
	$\Psi_Q$ & ${\bf N_{HC}}$ & ${\bf 3}$ & ${\bf 2}$ &  $Y_L - 1/3$
\end{tabular}
\caption{\small\label{tab:fields} Extra Dirac fermions charged under the hypercolor $\SU(N_{HC})$ gauge group. $Y_L$ is a free parameter.}
\end{table}
The kinetic term of the Lagrangian for the theory above $\Lambda_{HC}$ reads
\be
	\LL_{HC} = - \frac{1}{4} \sum_{X = HC, c, w, Y} F^X_{\mu\nu} F^{X\mu\nu} + \sum_{j = Q,L,N,E} \bar{\Psi}_j i \gamma^\mu D_\mu \Psi_j~,
	\label{eq:LagrKinHC}
\ee
where $D_\mu = \partial_\mu - i g_{HC} t^a A_\mu^a - i \sum_{x \in c,w,Y} g_{x}^{\SM} t^x_{\SM} A_\mu^{\SM,x}$ and $t^a$ are the generators of $\SU(N_{HC})$ in the fundamental representation while $t^x_{\SM}$ are the generators of the SM gauge groups.\footnote{To this Lagrangian one should also add the $\theta$ terms for QCD and for the HC group. The former experimentally has to be very small while the latter might induce new sources of CP violation and might also address the strong CP problem \cite{Dimopoulos:1979qi}.}
Since the total number of HC flavours is 10, in the absence of SM gauging and other explicit symmetry-breaking terms, the global symmetry group of the theory is
\be
	G = \SU(10)_L \times \SU(10)_R \times \U(1)_{HB}~,
	\label{eq:Ggroup}
\ee
where $\U(1)_{HB}$ is the hyper-baryon number, which is conserved at this stage.
The HC-fermion masses explicitly break the global symmetry $G$:
\be
	\LL_{m_\Psi} = - m_L \bar{\Psi}_L \Psi_L - m_E \bar{\Psi}_E \Psi_E - m_N \bar{\Psi}_N \Psi_N - m_Q \bar{\Psi}_Q \Psi_Q = - \bar{\Psi}^a \MM_{ab} P_L \Psi^b + h.c. ~,
	\label{eq:pNGBmassesLagr}
\ee
where $P_L = (1-\gamma^5)/2$. The mass matrix $\MM$ can also be seen as a spurion transforming under $G$ as $\MM \to g_R \MM g_L^\dagger$.
The phenomenological requirement of custodial symmetry imposes $m_E = m_N$, since otherwise a large breaking of custodial symmetry would appear in the Higgs potential.
As shown in Section~\ref{sec:EWSB}, these arbitrary masses should be slightly below the electroweak scale $m_\Psi \lesssim v$. A possible way to address this apparent coincidence of scales is mentioned at the end of this subsection.

The three fields $\Psi_L$, $\Psi_N$, and $\Psi_E$ reproduce the minimal viable composite Higgs scenario with complex representations, with a $G \to H$ pattern $\SU(4)_L \times \SU(4)_R \to \SU(4)_V$~\cite{Schmaltz:2010ac,Vecchi:2015fma,Ma:2015gra}.
The field $\Psi_Q$, containing six flavours from the HC point of view, is required to have also the two scalar LQ $S_1$ and $S_3$ as pNGBs. The only difference in the field content with respect to Ref.~\cite{Vecchi:2015fma} is in the fact that here $\Psi_Q$ is a $\SU(2)_w$ doublet rather than a singlet.

Since the HC gauge interaction must confine at the scale $\Lambda_{HC}$, it has to be asymptotically free in the ultraviolet. In App.~\ref{app:RGevolution} I show that, with the field content in Table.~\ref{tab:fields}, this is true for any $N_{HC} \geq 2$. Furthermore, depending on $Y_L$ and $N_{HC}$, the SM gauge couplings can remain perturbative up to the Planck scale.
Nevertheless, the extra dynamics which must be introduced slightly above the scale $\Lambda_{HC}$ in order to generate the SM Yukawas and leptoquark couplings, is expected to alter the RG evolution of all the gauge couplings.

Another interesting possibility is for $\GG_{HC}$ to be approximately conformal above $\Lambda_{HC}$, up to a scale $\Lambda_{FP}$ \cite{Luty:2004ye,Luty:2008vs,Galloway:2010bp}. This would allow a larger separation between the flavour and compositeness scales.
Also, in this case $\Lambda_{HC}$ could be generated by the soft breaking of the conformal symmetry due to the HC-fermion masses, thus potentially explaining dynamically the approximate coincidence between $\Lambda_{HC}$ and $m_{\Psi}$.
Perturbative computations suggest that for $\GG_{HC} = \SU(3)$ the strong dynamics has a strongly coupled IR fixed point in the window $9 \leq N_F \leq 16$ \cite{Vecchi:2015fma}, which includes this setup. See also Refs.~\cite{Hasenfratz:2009ea,Fodor:2009wk,Aoki:2012eq,Aoki:2013xza,Hasenfratz:2014rna,Appelquist:2014zsa,Aoki:2014oha,Fodor:2015baa} for lattice studies for different values of the number of flavours. 
%

\subsection{Condensate and pNGBs}

This theory is expected to form a condensate \cite{Raby:1979my,Preskill:1981sr,Vafa:1983tf}
\be
	\langle \bar{\Psi}_i \Psi_j \rangle = - B_0 f^2 \delta_{ij}~,
	\label{eq:condensate}
\ee
where $B_0$ is a non-perturbative constant (see e.g. Refs.~\cite{Gasser:1983yg,Pich:1995bw} for the QCD case), which in the QCD case is approximately given by $B_0 \approx 20 f$.
For $N_{HC} = 3$ and $N_F = 10$ also the condition quoted in Ref.~\cite{Sannino:2016sfx} for the condensate to form is satisfied.

This condensate spontaneously breaks the global symmetry $G$, Eq.~\eqref{eq:Ggroup}, to the diagonal subgroup 
\be
	G = \SU(10)_L \times \SU(10)_R \times \U(1)_{HB} \quad \to\quad
	H = \SU(10)_D \times \U(1)_{HB}~,
\ee
generating a set of $99$ real pNGBs transforming in the adjoint of $\SU(10)_D$. They can be described in terms of the matrix $U(\phi) \equiv u(\phi)^2$,
\be
	U[\phi(x)] = \exp \left( 2 i \frac{\phi^\alpha(x)}{f} T^\alpha \right)~,
	\label{eq:Udef}
\ee
transforming under $(g_L, g_R) \in G$ as $U \to g_R U g_L^\dagger$ \cite{Coleman:1969sm,Callan:1969sn}. In the expression above, $f$ is the NGB decay constant and $T^\alpha$ are the $SU(10)$ generators normalised as $\text{Tr}[ T^\alpha T^\beta ] = \frac{1}{2} \delta^{\alpha\beta}$.
The complete list of generators and the SM embedding is detailed in App.~\ref{app:SU10generators}.
The pNGBs are arranged into representations of $G_{\SM} = \SU(3)_c \times \SU(2)_w \times \U(1)_Y$ as (see App.~\ref{app:pNGBdef} for details):
\be\begin{array}{l l | l l | c}
	\text{valence} & \text{irrep.} & \text{valence} & \text{irrep.} & \text{d.o.f.} \\\hline
	H_1 \sim i \sigma^2 (\bar\Psi_L \Psi_N) &  ({\bf 1}, {\bf 2})_{1/2} &
	H_2 \sim (\bar\Psi_E \Psi_L) &  ({\bf 1}, {\bf 2})_{1/2} & 4 + 4 \\
	S_1 \sim (\bar\Psi_Q \Psi_L) &  ({\bf \bar{3}}, {\bf 1})_{1/3} &
	S_3 \sim (\bar\Psi_Q \sigma^a \Psi_L) &  ({\bf \bar{3}}, {\bf 3})_{1/3} & 6 + 18 \\
	\omega^\pm \sim (\bar\Psi_N \Psi_E) &  ({\bf 1}, {\bf 1})_{-1} &
	\Pi_L \sim (\bar\Psi_L \sigma^a \Psi_L) &  ({\bf 1}, {\bf 3})_{0} & 2 + 3 \\
	\tilde R_2 \sim (\bar\Psi_E \Psi_Q) &  ({\bf 3}, {\bf 2})_{1/6} &
	T_2 \sim (\bar\Psi_Q \Psi_N) &  ({\bf \bar{3}}, {\bf 2})_{5/6} & 12 + 12 \\
	\tilde \pi_1 \sim (\bar\Psi_Q T^A \Psi_Q) &  ({\bf 8}, {\bf 1})_{0} &
	\tilde \pi_3 \sim (\bar\Psi_Q T^A \sigma^a \Psi_Q) &  ({\bf 8}, {\bf 3})_{0} & 8 + 24 \\
	\Pi_Q \sim (\bar\Psi_Q \sigma^a \Psi_Q) &  ({\bf 1}, {\bf 3})_{0} &
	\eta_i \sim 3 \times c^a_i (\bar{\Psi}_a \Psi_a) &  ({\bf 1}, {\bf 1})_{0} & 3 + 3 \\
\end{array}~.
\label{eq:pNGB}\ee
These include two Higgs doublets $H_{1,2}$ as well as the two leptoquarks $S_{1,3}$. A general bottom-up study of composite Higgs models with two Higgs doublets can be found in Ref.~\cite{Mrazek:2011iu}.

In order to estimate the size of various operators in the low energy chiral Lagrangian, I assume na\"ive dymensional analysis (NDA) as the power counting scheme \cite{Manohar:1983md}, opportunely extended to the fermion sector (see e.g. Ref.~\cite{Panico:2011pw}):
\be
	\cL^{\rm eff} \sim \Lambda^2 f^2 \left( \frac{\Lambda}{4\pi f} \right)^{2L} \left( \frac{\phi^a}{f} \right)^{E_\phi} \left( \frac{g V_\mu}{\Lambda} \right)^{E_V} \left( \frac{\psi}{\sqrt{\Lambda} f} \right)^{E_\psi} \left( \frac{\partial_\mu}{\Lambda} \right)^{d} \left( \frac{m_\Psi}{\Lambda} \right)^{\chi} \left( \frac{g f}{\Lambda} \right)^{2 \mu} \left( \frac{g_\psi f}{\Lambda} \right)^{E_{4f}}  ~,
	\label{eq:NDA}
\ee
where $\Lambda \sim g_* f \sim 4\pi f$, $L$ counts the loop level at which the operator is generated, $E_{\phi, V, \psi}$ count the insertions of pions, elementary SM gauge bosons and fermions, $d$ counts the derivatives and $\chi$ the mass insertions. Finally, $\mu \geq 0$ takes into account if some operator is further suppressed due to symmetry arguments \cite{Panico:2011pw} while $E_{4f} \geq 0$ counts insertions of $G$-breaking effective four-fermion operators such as those responsible for the SM Yukawas.

The leading-order chiral Lagrangian contains the pNGB kinetic term, a mass term, and their gauge interactions:
\be
	\LL^{\rm eff}_{kin} = \frac{f^2}{4} \left( \text{Tr}\left[ (D_\mu U)^\dagger D^\mu U \right] + \text{Tr}\left[ U^\dagger \chi + \chi^\dagger U \right]\right) + \OO(f^2\frac{D^4}{\Lambda^2}) ~,
	\label{eq:LOchiralpNGB}
\ee
where $\chi = 2 B_0 \MM$ and the covariant derivative is given by $D_\mu U = \partial_\mu U - i [A^{\SM}_\mu, U]$, with $A^{\SM}_\mu \equiv g_s G_\mu^A T^A_{\SU(3)_c} + g_w W_\mu^i T^i_{\SU(2)_w} + g_Y B_\mu T_Y$.
The HC fermion mass term $\MM$ and the SM gauge interactions are two of the sources of explicit breaking of the global symmetry $G$, alongside the coupling of the pNGB to SM fermions. These terms are reponsible for generating a potential for the pNGBs and giving them all a mass (more details in Section~\ref{sec:potential}).

I require the pNGB potential to generate a minimum for non-zero Higgs fields, which thus take a vacuum expectation value, breaking spontaneously the electroweak gauge symmetry to the electromagnetic subgroup.
It can be shown that, up to an unphysical phase, the most general vev that preserves custodial symmetry at the tree level and leaves the SM color unbroken is \cite{Ma:2015gra}
\be
	\Omega(\theta) \equiv \langle U \rangle = {\bf 1}_{6\times 6} \otimes
	\left( \begin{array}{cccc}
	\cos \theta & 0 & \sin \theta & 0 \\
	0 & \cos \theta & 0 & \sin \theta \\
	-\sin \theta & 0 & \cos \theta & 0 \\
	0 & -\sin \theta & 0 & \cos \theta 
	\end{array}\right)~,
\label{eq:Uvev}\ee
where I factorized the $10\times 10$ matrix $U$ in two diagonal $6\times 6$ and $4\times 4$ blocks. The angle $\theta$ describes the misalignment between the EW-preserving vacuum and the true one \cite{Contino:2011np}.\footnote{The embedding of the Higgs fields in $U$ can be found in Eq.~\eqref{eq:HiggsPNGBmatr} and App.~\ref{app:SU10gen}.} Inserting this in Eq.~\eqref{eq:LOchiralpNGB} one gets mass terms for the $W$ and $Z$ bosons, from which one can recognise \footnote{The relation between the scale $f$ defined here and the one of Ref.~\cite{Ma:2015gra}, $f_{MC}$, is $f = 2 f_{MC}$.}
\be
	2 \sin^2 \theta \equiv \xi \equiv \frac{v^2}{f^2}~,
\ee
where $v \approx 246 \GeV$ is the SM Higgs vev and I introduced the traditional $\xi$ parameter of composite Higgs models. More details on EWSB are described in Section~\ref{sec:EWSB}.

\section{SM fermion masses and LQ couplings}
\label{sec:4FermOp}

In order to generate Yukawa couplings between the composite Higgs and the elementary SM fermions at low energy, the two sectors must be coupled.
In this case, also the couplings of the scalar $S_{1,3}$ leptoquarks to quarks and leptons must have to be generated in a similar way.

In modern Composite Higgs models, this is usually achieved by coupling each elementary SM fermion to a fermionic operator of the composite sector, with the same quantum numbers: $\LL \sim \sum_\psi \lambda_\psi \bar\psi_{\SM} \OO_\psi$ . After diagonalising the mass matrix before EWSB, the resulting massless eigenvalues (i.e. the SM fermions) are \emph{partially composite}, and a coupling with the Higgs is obtained \cite{Kaplan:1991dc}.
On the one hand, this setup usually requires light composite fermionic top partners \cite{Matsedonskyi:2012ym,Redi:2012ha,Marzocca:2012zn} as well as partners for each SM fermion.
On the other hand, in models with a fundamental fermionic description of the HC sector these composite fermions are baryonic resonances, which are expected to have a mass near $\Lambda_{HC}$, far too heavy to be viable top partners in a partial compositeness setup. Furthermore, devising a UV completion of this mechanism has proven to be challenging.\footnote{Possible 4d UV completion of the partial compositeness scenario have been obtained by introducing extra elementary HC-colored scalars \cite{Kaplan:1991dc,Sannino:2016sfx,Sannino:2017utc} or in a supersymmetric setup \cite{Caracciolo:2012je,Marzocca:2013fza}. Partial compositeness also arises naturally in extra-dimensional holographic Higgs models \cite{Agashe:2004rs}.}

For all these reasons, I assume instead that the bilinears of SM fermions couple to scalar operators of the strong sector, which at low energy are interpolated by pNGB fields such as the Higgses or the leptoquarks, as in original Technicolor models \cite{Dimopoulos:1979es,Eichten:1979ah}: $\LL \sim \sum_\psi y_\psi \bar\psi_{\SM} \psi_{\SM} \OO$.
These couplings can arise from four-fermion operators with two SM and two HC-charged fermions:
\be
	\LL_{4-Fermi} \sim \frac{c_{\psi\Psi}}{\Lambda_t^{d-1}} \bar{\psi}_{SM} \psi_{SM} \bar{\Psi}\Psi
		\quad \stackrel{E\lesssim \Lambda_{HC}}{\longrightarrow} \quad
	\sim c_{\psi\Psi} f \left(\frac{\Lambda_{HC}}{\Lambda_t} \right)^{d-1} \bar{\psi}_{SM} \psi_{SM} \frac{\phi}{f}~,
	\label{eq:EFTtoYukCoupl}
\ee
where the scaling dimension of the scalar operator $(\bar\Psi \Psi)$ is given by $d = 3 - \delta$, where $\delta > 0$ is the anomalous dimension of the operator.
At the scale $\Lambda_t$ some dynamics should be responsible for generating these operators. A sizeable part of the Technicolor (TC) literature focussed on the study of such a dynamics: Extended TC, Walking TC, etc.. See e.g. Refs.~\cite{Hill:2002ap,Lane:2002wv} for reviews of this topic and a list of references. For this first exploration of the model I take a bottom-up approach and do not discuss UV completions of these operators, leaving it for a future dedicated analysis. Using simply the NDA estimate of Eq.~\eqref{eq:NDA} with $E_{4f} = 1$ one obtains that the final Yukawa coupling is $y_{\psi\phi} \sim \mathcal{O}(1)$.

One of the main problems of such a setup is due to the fact that the dynamics responsible for generating these operators is also likely to produce four-fermion operators of the form
\be
	\LL_{4-Fermi} \supset 
	\frac{c_{\psi\psi}}{\Lambda_t^2} \bar{\psi}_{SM} \psi_{SM} \bar{\psi}_{SM} \psi_{SM} 
	+ \frac{c_{\Psi\Psi}}{\Lambda_t^2} \bar{\Psi}\Psi \bar{\Psi}\Psi~.
\ee
The effect of $(\Psi)^4$ operators is to generate further effective contributions to the pNGB masses in Eq.~\eqref{eq:pNGBmasses}. Since these pNGB should be heavy enough to pass the phenomenological constraints, this is not an unwanted feature. On the contrary, if they generate large enough masses for the singlets pNGBs, it could be possible to eliminate the need of fundamental HC fermion masses.
The $(\psi_{\SM})^4$ operators, instead, could generate dangerous effects in flavour physics (particularly in meson-antimeson mixing and lepton flavour violating processes).

If the strong sector is close to an interactive IR conformal fixed point above the scale $\Lambda_{HC}$, a sizeable value of the anomalous dimension $\delta$ could allow to increase the gap between $\Lambda_{HC}$ and $\Lambda_t$, thus suppressing the flavour-violating operators. See e.g. Refs.~\cite{Luty:2004ye,Luty:2008vs,Galloway:2010bp} for modern realisations of this idea and for a discussion of the problems one may encounter in this approach. 

If, instead, the anomalous dimension $\delta$ is small, the scale $\Lambda_t$ should be not much above the compositeness scale $\Lambda_{HC}$ in order to generate the required top Yukawa coupling. In this case an approximate flavour symmetry is required in order to protect the theory from unwanted flavour violation effects.
In the following I take this approach and assume that the sector responsible for generating these four-fermion operators enjoys a global approximate, possibly accidental, $\SU(2)^5$ flavour symmetry \cite{Barbieri:2011ci,Barbieri:2012uh,Barbieri:2012tu}:
\be
	G_F = \SU(2)_q \times \SU(2)_u \times \SU(2)_d \times \SU(2)_l \times \SU(2)_e~.
\ee
I also assume that the UV dynamics is such that in the symmetric limit only the third generation fermions are coupled to the strong sector. All other terms are generated via small symmetry-breaking effects. These are encoded in a small set of spurions. The mass of the first two SM families can be generated by a set of bi-doublets:
\be
	\Delta Y_u = ({\bf 2}, {\bf \bar{2}}, {\bf 1}, {\bf 1}, {\bf 1})~, \quad
	\Delta Y_d = ({\bf 2}, {\bf 1}, {\bf \bar{2}}, {\bf 1}, {\bf 1})~,\quad
	\Delta Y_e = ({\bf 1}, {\bf 1}, {\bf 1}, {\bf 2}, {\bf \bar{2}})~.
	\label{eq:GFspurions1}
\ee
The mixing between these and the third generation, instead, can be successfully described by only two doublets:
\be
	V_q = ({\bf 2}, {\bf 1}, {\bf 1}, {\bf 1}, {\bf 1})~, \qquad
	 V_l = ({\bf 1}, {\bf 1}, {\bf 1}, {\bf 2}, {\bf 1})~.
	 \label{eq:GFspurions2}
\ee
While $V_q$ is related to the CKM matrix elements, the leptonic spurion $V_l$ is unconstrained. Due to the smallness of the first two generation fermion masses, these two doublets provide the leading effects in most flavour observables. 
The smallness of the bottom and $\tau$ Yukawa couplings could be explained by introducing two approximate $\U(1)_d \times \U(1)_e$ symmetries, under which all the right-handed down quarks and leptons are charged \cite{Barbieri:2012uh}.
The flavour symmetry and this set of spurions also provide a good structure to fit the $B$-physics anomalies \cite{Greljo:2015mma,Barbieri:2015yvd,Bordone:2017anc,Buttazzo:2017ixm} while at the same protecting the model from other flavour and high-$p_T$ constraints.
Indeed, possible dangerous effects of the $\frac{1}{\Lambda_t^2}(\psi_{\SM})^4$ operators are suppressed by the $G_F$ symmetry and the large $\Lambda_t$ scale.

Another class of possible bilinear operators are those built in terms of vector currents. At low energies these are interpolated by vector resonances of the strong sector as well as pNGB vector currents:
\be
	\LL \supset \frac{c}{\Lambda_t^2}(\bar \psi_{\SM} \gamma^\mu \psi_{\SM}) (\bar \Psi_a \gamma_\mu \Psi_b) \to
	g_{\rho\psi} (\bar \psi_{\SM} \gamma^\mu \psi_{\SM}) \text{Tr}( c_{ab} i U^\dagger D_\mu U + c_{ab}  \rho_\mu)~,
	\label{eq:rhoffCoupl}
\ee
where by NDA, Eq.~\eqref{eq:NDA} with $E_{4f} = 1$, one has $g_{\rho\psi} \sim \mathcal{O}(f/\Lambda) \sim \mathcal{O}(1 / 4\pi)$. Their effect is discussed in Section~\ref{sec:HiggsPheno}.

\subsection{HC-fermion bilinears}

I construct the coupling of the SM fermions to the two Higgses and the $S_{1,3}$ scalar leptoquarks via operators like $\bar \psi_{\SM} \psi_\SM \bar{\Psi}_i \Psi_j$, where $\bar \Psi \Psi_j$ interpolates the pNGBs below $\Lambda_{HC}$.

In general, both baryon ($B$) and lepton ($L$) numbers are broken by adding non-renormalizable operators (as happens in the SM EFT).
In order to avoid proton decay and other unwanted effects, one could impose $B$ and $L$ conservation in the operators at the scale $\Lambda_t$ while assigning suitable quantum numbers to the HC fermions.\footnote{For the purpose of this paper I neglect the non-perturbative breaking of $B+L$.}
Focussing in particular on the $\bar\psi_\SM \psi_\SM \bar\Psi \Psi$ effective operators, an equally successful but more minimal requirement is to impose conservation of a combination of $B$ and $L$, such as for example $F_+ = 3B + L$ or $F_- = 3B - L$.
Requiring only that the operators generating the Higgs Yukawa couplings and the $S_{1,3}$ leptoquark couplings to SM fermions are allowed provides the following charge assignment for the HC fermions:
\be
	F_+( \Psi_L ) = F_+( \Psi_N ) = F_+( \Psi_E ) = F_L~, \qquad
	F_+( \Psi_Q ) = F_L + 2~,
	\label{eq:BLassignment}
\ee
where $F_L$ is an arbitrary charge. Assuming $F_-$ conservation, instead, all HC fermions should have the same (arbitrary) $F_-$ charge.

The complete list of possible $\bar\psi_\SM \psi_\SM \bar\Psi \Psi$ operators compatible with gauge symmetries and $F_\pm$ conservation, given the assignment of Eq.~\eqref{eq:BLassignment},
is (schematically):
\be\begin{split}
	(\bar{q}_L u_R + \bar{d}_R q_L + \bar{e}_R l_L) (\bar\Psi_N \Psi_L) ~, \qquad
	&(\bar{q}_L u_R + \bar{d}_R q_L + \bar{e}_R l_L) (\bar\Psi_L \Psi_E) ~, \\
	(\bar q_L^c l_L + \bar e^c_R u_R) (\bar\Psi_Q \Psi_L )~,  \qquad
	&(\bar q_L^c \sigma^a l_L) (\bar\Psi_Q \sigma^a \Psi_L ) ~,
\label{eq:2f2Fop}
\end{split}\ee
where all indices have been suppressed.
Comparing the HC bilinears with Eq.~\eqref{eq:pNGB}, one recognises the Yukawa couplings for the two Higgs doublets in the first line, while the second line corresponds to the desired couplings of the $S_{1,3}$ leptoquarks to SM fermions. Note that, given the assumptions above, also a coupling of $S_1$ with right-handed fermions $\bar e^c_R u_R$ is allowed.

The remaining scalar operators, allowed by gauge symmetries but forbidden by $F_\pm$ conservation, are
\be
	(\bar q_L^c q_L + \bar u^c_R d_R) (\bar\Psi_L \Psi_Q ) ~, \qquad
	(\bar d_R l_L ) (\bar\Psi_E \Psi_Q ) ~, \qquad
	(\bar l_L^c l_L) (\bar\Psi_E \Psi_N ) ~,
\ee
corresponding to couplings of the $S_{1,3}$ to diquark, of $\tilde R_2$ to quarks and leptons, and of $\omega$ to di-leptons.
It is remarkable that, once the $F_\pm$ quantum numbers are assigned to the HC fermions to allow the desired Higgs and LQ couplings, automatically the $B$ and $L$-violating operators are forbidden and none of the other pNGBs is allowed to have a linear coupling to SM fermions.\footnote{On the contrary, requiring only $B-L$ conservation would allow also the coupling of $S_{1,3}$ to diquark, which would mediate proton decay.}

For each of the interactions in Eq.~\eqref{eq:2f2Fop} it is clearly possible to write two independent terms, one for each chiral structure of the HC bilinears: $\bar \Psi_{i, L} \Psi_{j, R}$ or $\bar \Psi_{i, R} \Psi_{j, L}$. By comparing Green functions in the high- and low-energy theory it is easily shown that the HC fermions bilinears correspond to the following expressions below the scale $\Lambda_{HC}$ (see e.g. the QCD case in Ref.~\cite{Pich:1995bw}):
\be\begin{split}
	& \bar\Psi_{i, L} \Psi_{j, R}  \to - B_0 f^2 U(\phi)_{ji}~, \qquad
	 \bar\Psi_{i, R} \Psi_{j, L}  \to - B_0 f^2 U^\dagger(\phi)_{ji}~,\\
	& \bar\Psi_i \Psi_j  \to - B_0 f^2 \left(U(\phi) + U^\dagger(\phi) \right)_{ji}~, \qquad
	\bar\Psi_i \gamma_5 \Psi_j  \to - B_0 f^2  \left(U(\phi) - U^\dagger(\phi) \right)_{ji}~,
\end{split}\ee
where $B_0$ is defined in Eq.~\eqref{eq:condensate}.
Upon expanding $U(\phi)$ in powers of the pNGB, Eq.~\eqref{eq:Udef}, it is clear that only the pseudoscalar combination is linear in the pNGB and thus can generate the desired couplings.
The scalar combination can give some effects in the pNGB potential \cite{Ma:2015gra} but, in order to keep the discussion simple, I will set it to zero in the following.

\subsection{SM Yukawas}

The four-fermion operators generated at the scale $\Lambda_t$ responsible for the SM Yukawas are
\be\begin{split}
	\LL_{F} \supset& \frac{1}{\Lambda_t^2} \left(\bar u_R c_{1,u}^{\dagger} q_L + \bar q_L c_{1,d} d_R \, \epsilon + \bar l_L c_{1,e} e_R \, \epsilon \right) (\bar\Psi_L \gamma_5 \Psi_N) \; + \\
	+& \frac{1}{\Lambda_t^2} \left(\bar u_R c_{2,u}^{\dagger} q_L \epsilon + \bar q_L c_{2,d} d_R  + \bar l_L c_{2,e} e_R \right) (\bar\Psi_E \gamma_5 \Psi_L) + h.c.~,
	\label{eq:4FYuk}
\end{split}\ee
where flavour and gauge indices have been suppressed and $\epsilon \equiv i \sigma^2$ acts on $\SU(2)_w$.
In order to track the explicit breaking of the global symmetry $G$ due to these operators one can introduce a set of spurions $\Delta_{H_{1,2}}^\alpha$ defined from (the explicit expression is in App.~\ref{app:spurions})
\be
	\bar\Psi_L^\alpha \gamma_5 \Psi_N = \epsilon^{\alpha\beta} \bar\Psi_i (\Delta_{H_1}^\beta)_{ij} \gamma_5 \Psi_j~, \qquad
	\bar\Psi_E \gamma_5 \Psi_L^\alpha = \bar\Psi_i (\Delta_{H_2}^\alpha)_{ij} \gamma_5 \Psi_j ~,
	\label{eq:4FermiH12Spurions}
\ee
where $\alpha,\beta = 1,2$ are $\SU(2)_w$ indices. They transform under $G$ as $\Delta_{H_{1,2}}^{\alpha, LR} \to g_L \Delta_{H_{1,2}}^{\alpha, LR} g_R^\dagger$, $\Delta_{H_{1,2}}^{\alpha, RL} \to g_R \Delta_{H_{1,2}}^{\alpha, RL} g_L^\dagger$, with the identification $\Delta_{H_{1,2}}^{\alpha, LR} = \Delta_{H_{1,2}}^{\alpha, RL} = \Delta_{H_{1,2}}^{\alpha}$.
Below the HC-confinement scale the corresponding chiral operators can be written as
\be\begin{split}
	\LL^{\rm eff}_{\rm Yuk} =&  \frac{f}{2} \left(\bar u_R \tilde y_{1,u}^{\dagger} q^\beta_L \epsilon^{\beta\alpha} + \bar q^\alpha_L \tilde y_{1,d} d_R + \bar l^\alpha_L \tilde y_{1,e} e_R  \right) \Tr[ \Delta_{H_1}^\alpha (U - U^\dagger) ] + \\
	& + \frac{f}{2} \left(\bar u_R \tilde y_{2,u}^{\dagger} q^\beta_L \epsilon^{\beta\alpha} + \bar q^\alpha_L \tilde y_{2,d} d_R + \bar l^\alpha_L \tilde y_{2,e} e_R  \right) \Tr[ \Delta_{H_2}^\alpha (U - U^\dagger) ] + h.c.~,
	\label{eq:pNGBYuk}
\end{split}\ee
where the NDA estimate of the Yukawa couplings in terms of the high-energy EFT coefficients is $\tilde y_f \sim \frac{B_0 f}{\Lambda_{t}^2} c_f$. By expanding the pNGB matrix one gets 
\be
	\Tr[ \Delta_{H_{1,2}}^\alpha (U - U^\dagger) ] = i \frac{2 \sqrt{2}}{f} H_{1,2}^\alpha + \OO(\phi^2/f^2)~,
\ee
Substituting $U$ with its EW symmetry-breaking vev, Eq.~\eqref{eq:Uvev}, one has $\Tr[ \Delta_{H_{1(2)}} (\langle U\rangle - \langle U^\dagger \rangle) ] = (-1)^{1(2)} (0, 2 \sin \theta)^T$. The SM fermion mass matrices are given by (in a $\bar f_L m_f f_R$ notation)
\be
	m_{f} = f \sin \theta (\tilde y_{1,f} - \tilde y_{2,f}) = \frac{v}{\sqrt{2}} (\tilde y_{1,f} - \tilde y_{2,f}) \equiv \frac{v}{\sqrt{2}}  y_{f}~,
\ee
where $f = u,d,e$. As shown in Ref.~\cite{Ma:2015gra}, in order to avoid any undesired misalignment of the pNGB vev in a custodial-breaking direction also the condition 
\be
	\tilde y_{1,f} = - \tilde y_{2,f} = \frac{y_f}{2}
	\label{eq:YukCond}
\ee
should be imposed. This condition can be obtained by imposing a symmetry under the exchange $P_H: H_1 \leftrightarrow - H_2$, which is automatically satisfied by the kinetic and gauge terms, as well as by the HC-masses under the condition $m_E = m_N$. This symmetry is instead broken by higher-order terms proportional to the LQ couplings to fermions which, however, do not affect the Higgs potential at this order in the chiral expansion.

Furthermore, to suppress dangerous tree-level flavour-changing neutral currents mediated by the Higgses, the two proto-Yukawa matrices should be aligned, see e.g. the discussion in Ref.~\cite{Mrazek:2011iu}, so Eq.~\eqref{eq:YukCond} is imposed at the matrix level.
If also the scalar HC currents were kept, a slightly more general condition can be derived, see Ref.~\cite{Ma:2015gra} for a detailed discussion of this point.

The $G_F$ flavour symmetry and its spurions (\ref{eq:GFspurions1},\ref{eq:GFspurions2}) dictate the structure of the Yukawa matrices. At leading order in the spurions and up to possible $O(1)$ factors multiplying each term one has \cite{Barbieri:2011ci} (in $\bar{L}R$ notation):
\be
	y_u \sim y_t \left( \begin{array}{cc}
	{\Delta Y_u} & V_q \\
	0 & 1
	\end{array}\right)~,\quad
	y_d \sim y_b \left( \begin{array}{cc}
	{\Delta Y_d} & V_q \\
	0 & 1
	\end{array}\right)~,\quad
	y_e \sim y_\tau \left( \begin{array}{cc}
	{\Delta Y_e} & V_l \\
	0 & 1
	\end{array}\right)~.\quad
\ee
In the left-handed quark sector this can be put in correspondence with the CKM matrix elements:
\be
	V_q = a_q  \left( \begin{array}{c}
	V_{td}^* \\
	V_{ts}^*
	\end{array}\right)~,
\ee
where $a_q$ is an $O(1)$ parameter. As shown in Section~\ref{sec:flavour}, in order to fit the flavour anomalies while avoiding dangerous effects involving electrons, the left-handed lepton spurion can be taken approximately as
\be
	V_l \approx \left( \begin{array}{c}
	0 \\
	\lambda_{\tau\mu}
	\end{array}\right)~,
	\label{eq:SU2ellspurions}
\ee
where $\lambda_{\tau\mu} \ll 1$.

\subsection{$S_{1,3}$ LQ couplings}
\label{sec:S13LQyuk}

The operators responsible for generating the leptoquark couplings to fermions are
\be\begin{split}
	\LL_{F} \supset
	\frac{1}{\Lambda_t^2} \left[ \left(\bar q_L^c c_{1,ql} \epsilon l_L + \bar e^c_R c_{1,eu} u_R \right) (\bar\Psi_Q \gamma_5  \Psi_L) + \left(\bar q_L^c c_{3,ql} \epsilon \sigma^A l_L \right) (\bar\Psi_Q \gamma_5 \sigma^A \Psi_L) \right] + h.c.~.
	\label{eq:4FLQ}
\end{split}\ee
Also in this case one can introduce a set of spurions of $G$ to keep track of the explicit breaking of the global symmetry (see App.~\ref{app:spurions}):
\be\begin{split}
	\bar\Psi_Q^{a} \gamma_5 \Psi_L &= \bar\Psi \Delta_{S_1}^a \gamma_5 \Psi ~, \\
	\bar\Psi_Q^a \sigma^A \gamma_5 \Psi_L &= \bar\Psi \Delta_{S_3}^{A,a} \gamma_5 \Psi~,
	\label{eq:4FermiS13Spurions}
\end{split}\ee
where the index $a$ runs in the fundamental of $\SU(3)_c$ while $A$ is in the adjoint of $\SU(2)_w$.
Below $\Lambda_{HC}$ one can write the couplings of both scalar LQ to SM fermions as\footnote{In presence of EWSB, a factor of $\cos \frac{\theta}{2}$ should muliply all terms in the last line of Eq.~\eqref{eq:S13lagr}. Since this is $\approx 1$ up to a small $\mathcal{O}(\xi)$ correction, I neglect it in the following.}
\bea
	\!\!\!\!\!\!\LL^{\rm eff}_{\rm LQ} \!\!\!\!&=&\!\!\!\! \; i \frac{f}{4} \left(g_1 \bar q_L^{c, a} \beta_{1} \epsilon l_L + g_1^u \bar e^c_R \beta_{1}^u u_R^a \right) \Tr[ \Delta_{S_1}^a (U - U^\dagger) ] + h.c. \nonumber\\
	\!\!\!\!&&\!\! +i \frac{f}{4} \left( g_3 \bar q_L^{c, a} \beta_{3} \epsilon  \sigma^A l_L  \right) \Tr[ \Delta_{S_3}^{A,a} (U - U^\dagger) ] + h.c. = \label{eq:S13lagr}\\
	\!\!\!\!&=&\!\! - g_1 \beta_{1,i\alpha} (\bar q_L^{c \, i} \epsilon l_L^\alpha) S_1 
	- g_1^u (\beta^u_1)^T_{\alpha i} (\bar e_R^{c \, \alpha} u_R^i) S_1
	- g_3 \beta_{3,i\alpha} (\bar q_L^{c \, i} \epsilon \sigma^A l_L^\alpha) S_3^A 
	+ h.c. + \mathcal{O}(\phi^2)~, \nonumber
\eea
where $i$ and $\alpha$ are quark and lepton flavour indices, respectively. As for the Higgs Yukawa couplings, also these ones are related to the high-energy coefficients via relations as in Eq.~\eqref{eq:EFTtoYukCoupl}.
The flavour structure of the couplings is given by the $G_F$ symmetry and its breaking spurions. Up to $\mathcal{O}(1)$ coefficients one has
\be
	\beta_{1,3} \sim \left( \begin{array}{cc}
	V_q^* V_l^\dagger & V_q^* \\
	V_l^\dagger & 1
	\end{array}\right)~,\quad
	\beta_{1}^u \sim \left( \begin{array}{cc}
	0 & (V_q^\dagger \Delta Y_u)^T \\
	V_l^\dagger \Delta Y_e & 1
	\end{array}\right)~,\quad
	\label{eq:LQflavStructure}
\ee
where, without loss of generality, the (33) element has been reabsorbed in the definition of the overall couplings $g_{1,3}^{(u)}$ and I also show the terms quadratic in the spurions, since they are relevant to the $b\to s \mu\mu$ anomalies. One can immediately notice that, with this choice of flavour spurions, the off-diagonal entries in $\beta_1^u$ are suppressed by the small Yukawa couplings of the light fermions. By adding spurions transforming as doublets of the right-handed fields, these terms might also be larger. For this reason I leave them arbitrary in the flavour analysis.

Integrating out the two scalar leptoquarks at tree-level one generates a set of dimension-6 operators, $\LL^{\rm eff} =  - \frac{1}{v^2} \sum_x C_x O_x$, with \cite{deBlas:2017xtg}
\be\begin{split}
	(C_{l q}^{(1)})_{\alpha \beta i j} &= - |\epsilon_1|^2 \; \beta_{1,i\alpha}^* \beta_{1,j\beta} - 3 |\epsilon_3|^2 \; \beta_{3,i\alpha}^* \beta_{3,j\beta} ~, \\
	(C_{l q}^{(3)})_{\alpha \beta i j} &= |\epsilon_1|^2 \; \beta_{1,i\alpha}^* \beta_{1,j\beta} - |\epsilon_3|^2  \; \beta_{3,i\alpha}^* \beta_{3,j\beta} ~, \\
	(C_{l equ}^{(1)})_{\alpha \beta i j} &= - 2 \epsilon_1^u \epsilon_1^* \; \beta_{1,j\beta}^{u} \beta_{1,i\alpha}^{*} ~, \\
	(C_{l equ}^{(3)})_{\alpha \beta i j} &= \frac{1}{2} \epsilon_1^u \epsilon_1^* \; \beta_{1,j\beta}^{u} \beta_{1,i\alpha}^{*} ~, \\
	(C_{eu})_{\alpha \beta i j} &= - 2 |\epsilon^u_1|^2 \; \beta_{1,i\alpha}^{u\,*} \beta^u_{1,j\beta} ~,
	\label{eq:EFTS13match}
\end{split}\ee
where the corresponding operators are
\be\begin{array}{r l  r l}
	(O_{l q}^{(1)})_{\alpha \beta i j} = &  (\bar{l}_L^\alpha \gamma_\mu l_L^\beta) (\bar{q}_L^i \gamma^\mu q_L^j)~, &
	(O_{l q}^{(3)})_{\alpha \beta i j} = &  (\bar{l}_L^\alpha \gamma_\mu \sigma^a l_L^\beta) (\bar{q}_L^i \gamma^\mu \sigma^a q_L^j)~,  \vspace{0.3em}\\
	(O_{l e q u}^{(1)})_{\alpha \beta i j} = &  (\bar{l}_L^\alpha e_R^\beta) \epsilon (\bar{q}_L^i u_R^j)~,  &
	\qquad (O_{l e q u}^{(3)})_{\alpha \beta i j} = &  (\bar{l}_L^\alpha \sigma_{\mu\nu} e_R^\beta) \epsilon (\bar{q}_L^i \sigma^{\mu\nu} u_R^j)~,   \vspace{0.3em} \\
	(O_{e u})_{\alpha \beta i j} = &  (\bar{e}_R^\alpha \gamma_\mu e_R^\beta) (\bar{u}_R^i \gamma^\mu u_R^j)~,   &
\end{array}\label{eq:semilepOperators}\ee
and the $\epsilon_i$ contain the relevant combinations of masses and couplings:
\be
	\epsilon_1 = \frac{g_1 v}{2 m_{S_1}}~, \quad
	\epsilon_3 = \frac{g_3 v}{2 m_{S_3}}~, \quad
	\epsilon^u_1 = \frac{g^u_1 v}{2 m_{S_1}}~.
	\label{eq:definitionepsilon}
\ee

\section{The pNGB potential}
\label{sec:potential}

The compositeness scale $\Lambda_{HC} \sim 4 \pi f$ sets the mass of most of the resonances of the strong sector. The exception are the pNGB, whose mass is proportional to the various explicit symmetry-breaking terms: HC-fermion masses, SM gauging, and four-fermion operators.
In this section I present the leading operators in the chiral expansion which constitute the pNGB potential and generate their masses, and discuss the conditions required to achieve a successful EWSB.

\subsection{Potential from the HC fermion masses}

The contribution to the pNGB potential from the explicit breaking due to the HC fermion masses is controlled by the spurion $\MM$ and the leading chiral operator describing this is given in Eq.~\eqref{eq:LOchiralpNGB}.
Upon expanding $U$ in powers of pNGBs one gets the mass terms which, for the non-singlets pNGB is
\be
	m^2_{(\bar{\Psi}_i \Psi_j)} = B_0 (m_i + m_j)~,
	\label{eq:pNGBmasses}
\ee
where $i,j = Q,L,N,E$ represent the \emph{valence} fundamental HC fermion constituting the pNGB, according to Eq.~\eqref{eq:pNGB}. I recall that $m_N = m_E$ to avoid custodial symmetry breaking.
In particular, the contribution to the two Higgs doublets mass is
\be
	V_{m_\Psi} = - \frac{f^2}{4} \Tr[U^\dagger \chi + \chi^\dagger U] \supset B_0 (m_E + m_L) ( |H_1|^2 + |H_2|^2 ) ~.
	\label{eq:LHCmassHiggs}
\ee
In order to obtain the singlets masses one needs the expression of the 3 Cartan generators of $\SU(10)_D$ transforming as singlets of $\GG_{SM}$. They are given in Appendix~\ref{app:SU10gen}, Eq.~\eqref{eq:singlGen}.
In the unbroken EW symmetry limit one gets:
\be
	m^2_{\eta_1} = 2 B_0 m_E~, \quad
	M^2_{\eta_2,\eta_3} = \left(
	\begin{array}{cc}
	B_0 (m_E + m_L) & - \sqrt{\frac{3}{5}} B_0 (m_E - m_L) \\
	- \sqrt{\frac{3}{5}} B_0 (m_E - m_L)  & \frac{1}{5} B_0 (3 m_E + 3 m_L + 4 m_Q)
	\end{array}\right),
	\label{eq:singlmassMatr}
\ee
where in general $\eta_2$ and $\eta_3$ mix with each other. For $m_E = m_L$ the mixing vanishes and:
\be
	m^2_{\eta_1} = m^2_{\eta_2} = 2 B_0 m_L~, \qquad
	m^2_{\eta_3} = \frac{2}{5} B_0 (3 m_L + 2 m_Q)~.
	\label{eq:singletmasses}
\ee
Since this is the only contribution to the three singlets masses, the fundamental HC-fermion masses are required in order to make them heavy enough to pass phenomenological bounds (discussed in Section~\ref{sec:searches}). A possible alternative could be if a sufficiently large contribution is generated via the $\frac{1}{\Lambda_t^2}\Psi^4$ operators as mentioned in Section~\ref{sec:4FermOp}. The effect of these operators in the potential has been briefly considered in Ref.~\cite{Galloway:2010bp}, where it is argued to be suppressed.

\subsection{Potential from the SM gauging}
\label{sec:mass_gauge}

The explicit breaking of the global symmetry $G$ due to the gauging of the SM subgroup is analogous to the one due to the QED gauging in the QCD chiral Lagrangian, responsible for the $\pi^\pm$ - $\pi^0$ mass splitting. It can be described in terms of spurions, defined from the SM gauge interactions of the HC fermion currents, Eq.~\eqref{eq:LagrKinHC}:
\be
	\LL_{HC} \supset g_s G_\mu^A J_{s,\mu}^A + g_w W_\mu^i J_{w,\mu}^i + g_Y B_\mu J^Y_\mu =
	\left( G_\mu^A \GG_{s,A}^\alpha + W_\mu^i \GG_{w,i}^\alpha  + B_\mu \GG_{Y}^\alpha \right) J_\mu^\alpha~,
\ee
where $J_\mu^\alpha = \bar\Psi_L \gamma_\mu T^\alpha_L \Psi_L + \bar{\Psi}_R \gamma_\mu T^\alpha_R \Psi_R$, $T^\alpha_{L,R}$ are the generators of $G$, and the various $\GG_X^\alpha$ are the spurions. They represent the embedding of the SM gauging within $G$ (see App.~\ref{app:SMgauging} for the explicit expression). One can define the generators associated with a given SM gauge field as the combinations:
\be
	\GG_{s,A}^{L,R} \equiv \GG_{s,A}^\alpha T^\alpha_{L,R}~, \qquad
	\GG_{w,i}^{L,R} \equiv \GG_{w,i}^\alpha T^\alpha_{L,R}~, \qquad
	\GG_{Y}^{L,R} \equiv \GG_{Y}^\alpha T^\alpha_{L,R}~. \qquad
\ee
Their transformation properties under $SU(10)_L \times SU(10)_R$ are
\be
	\GG_{X}^{L,R} \to g_{L,R} ~ \GG_{X}^{L,R} ~ g_{L,R}^\dagger~.
\ee
Since the HC theory is vectorlike, the left and right spurions are identical.
The leading operator in the chiral Lagrangian built with these spurions is
\be
	V_{\mathcal{G}}  = - \frac{3 f^2 \Lambda_{HC}^2}{16 \pi^2} \sum_X c_X \Tr \left[ \GG_{X}^L U \GG_{X}^R U^\dagger  \right] = \frac{3 \Lambda_{HC}^2}{16 \pi^2} \sum_{i,\alpha} c_i g_i^2 C_2^i (\phi^\alpha) ~ (\phi^\alpha)^2 + \OO(\phi^3)~,
	\label{eq:SMgaugeMass}
\ee
where the sum is over the three SM gauge groups, $i = s, w, Y$, $c_i$ are non-perturbative $\OO(1)$ coefficients, and $C_2^i(\pi^\alpha)$ is the Casimir of the pNGB $\phi^\alpha$ under the SM gauge group $i$.\footnote{$C_2({\bf F}) =\frac{N^2 - 1}{2N}$ for the fundamental and $C_2({\bf Adj}) = N$ for the adjoint of $SU(N)$, while it corresponds to $Y^2$ under $U(1)_Y$.}
The coefficients in front of the operator are estimated from Eq.~\eqref{eq:NDA} with $L = 1$ and $\mu = 1$, since it arises from one loop and requires insertions of symmetry-breaking spurions.
Since the coefficients $c_{l}$ are expected to be positive \cite{Witten:1983ut}, these terms give positive contributions to the pNGBs mass squared.
In the case of the Higgses one has
\be\begin{split}
	V_{\mathcal{G}}  \supset& \frac{3 \Lambda_{HC}^2}{8 \pi^2} \left( \frac{3}{4} c_w g_w^2 + \frac{1}{4} c_Y g_Y^2 \right) \left( |H_1|^2 + |H_2|^2 \right) + \ldots 
	\label{eq:LHiggsmassGauge}
\end{split}\ee
For all the pNGB irreps, the masses originating from Eq.~\eqref{eq:SMgaugeMass} correspond numerically, up to non-perturbative $\mathcal{O}(1)$ factors, to:
\be\begin{split}
&	\Delta m^2_{\omega} \approx (0.05 \Lambda_{HC})^2~, \quad
	\Delta m^2_{H_{1,2}} \approx (0.08 \Lambda_{HC})^2~, \quad
	\Delta m^2_{\Pi_{L,Q}} \approx (0.13 \Lambda_{HC})^2~, \\
&	\Delta m^2_{S_1} \approx (0.17 \Lambda_{HC})^2~, \quad
	\Delta m^2_{S_3} \approx (0.21 \Lambda_{HC})^2~. \quad
	\Delta m^2_{\tilde R_2, T_2} \approx (0.19 \Lambda_{HC})^2~. \\
&	\Delta m^2_{\tilde \pi_1} \approx (0.26 \Lambda_{HC})^2~, \quad
	\Delta m^2_{\tilde \pi_3} \approx (0.28 \Lambda_{HC})^2~, 	
	\label{eq:gauge_mass_contr}
\end{split}\ee
For $\Lambda_{HC} \approx 10\TeV$ it is immediate to read the numerical value of these contributions, ranging from $\approx 500 \GeV$ for the $\omega^\pm$ state to $\approx 2.8 \TeV$ for the $\tilde \pi_3$.

\subsection{Potential from the four-fermion operators}

The last explicit symmetry-breaking terms to be discussed are due to the four-fermion operators of Eqs.~(\ref{eq:4FYuk},\ref{eq:4FLQ}), responsible for the SM Yukawa and leptoquark couplings. Since their effect on pNGB masses is proportional to the coupling itself, the leading contribution is due to the top quark and the LQ coupling to 3rd generation fermions.

The effects on the pNGB potential from these breaking terms can be traced with the spurions introduced in Eqs.~(\ref{eq:4FermiH12Spurions},\ref{eq:4FermiS13Spurions}). The leading chiral operator generated from the top Yukawa, with its NDA estimate, is
\be\begin{split}
	V_{t}  &=  - \frac{y_{t}^2 N_c f^2 \Lambda_{HC}^2}{16 \pi^2} c_t \sum_i \left| \frac{1}{2\sqrt{2}} \Tr \left[ (\Delta_{H_1}^{i} - \Delta_{H_2}^{i}) (U-U^\dagger) \right]  \right|^2 \\
	 &\supset - \frac{c_t y_{t}^2 N_c \Lambda_{HC}^2}{16 \pi^2} | H_1 - H_2 |^2 + \mathcal{O}(\phi^3)
	\label{eq:4FermiMass}
\end{split}\ee
where $c_t$ is an $\mathcal{O}(1)$ non-perturbative coefficient and $y_{t}$ is the top Yukawa coupling. The $1/2\sqrt{2}$ factor depends on the spurion's normalisation. Although in this case the sign is not fixed, a simple one-loop computation suggests that it could be negative. This is also required to successfully obtain EWSB.
Similar terms arise also from the $S_1$ and $S_3$ leptoquarks couplings to SM fermions:
\bea
	\!\!\!\!\!\! V_{\rm LQ}  \!\!\!&=& \!\!\! - \frac{(c_{1} g_{1}^2 + c_{1}^u g_{1}^{u 2}) f^2 \Lambda_{HC}^2}{16 \pi^2}  \left| \frac{1}{2\sqrt{2}} \Tr \left[ \Delta_{S_1}^{a} (U-U^\dagger) \right]  \right|^2 
	-  \frac{c_{3} g_{3}^2 f^2 \Lambda_{HC}^2}{16 \pi^2} \left| \frac{1}{2\sqrt{2}} \Tr \left[ \Delta_{S_3}^{A,a} (U-U^\dagger) \right]  \right|^2 \nonumber\\
	 \!\!\!&\supset&\!\!\! - \frac{(c_{1} g_{1}^2 + c_{1}^u g_{1}^{u 2}) \Lambda_{HC}^2}{8 \pi^2} |S_1|^2 
	 - \frac{c_{3} g_{3}^2 \Lambda_{HC}^2}{8 \pi^2} |S_3|^2 + \mathcal{O}(\phi^3)~,
	\label{eq:4FermiLQMass}
\eea
where also $c_{1,3}^{(u)} \sim \mathcal{O}(1)$.
Since the (positive) SM gauging contribution to the square pNGB masses is smaller for the Higgs than for the leptoquarks, it is reasonable to expect that these potentially negative terms due to SM fermion loops would be more important for the Higgs than for the LQ, providing a good EWSB.

\subsection{Electroweak Symmetry Breaking and Higgs mass}
\label{sec:EWSB}

For what concerns the dynamics of EWSB, this model reduces to the $\SU(4)_L \times \SU(4)_R \to \SU(4)_D$ case studied in Ref.~\cite{Ma:2015gra}. In fact, neither the LQ nor the other pNGB with valence $\Psi_Q$ HC-fermion enter in any aspect of EWSB. For this reason I can refer to \cite{Ma:2015gra} for most of this discussion, of which I  summarise here only the main aspects.

In the notation used until here, the two Higgs doublets, $H_{1,2} = (H_{1,2}^+, H_{1,2}^0)^T$, are related directly to the valence HC fermions and embedded in the pNGB matrix $U \equiv \exp(i \Pi)$ as (see App.~\ref{app:SU10gen} for this definition)
\be
	\Pi_{4\times 4}(H) =
	\frac{\sqrt{2}}{f} \left( \begin{array}{cccc}
	0 & 0 & H_1^{0*} & H_2^+ \\
	0 & 0 & -H_1^- & H_2^0 \\
	H_1^0 & -H_1^+ & 0 & 0\\
	H_2^- & H_2^{0*} & 0 & 0
	\end{array}\right) ~,
	\label{eq:HiggsPNGBmatr}
\ee
where I focussed only on the lower $4\times 4$ block and set to zero the other fields. A more convenient basis in the two Higgs doublets for studying EWSB is the one adopted in Ref.~\cite{Ma:2015gra}:
\be
	H_1 = \frac{i \tilde H_1 + \tilde H_2}{\sqrt{2}}~, \qquad
	H_2 = \frac{-i \tilde H_1 + \tilde H_2}{\sqrt{2}}~.
\ee
Under $P_H$ one has $\tilde H_1 \to \tilde H_1$ and $\tilde H_2 \to - \tilde H_2$.
In this notation the field which takes the vev is $\langle \tilde H_1 \rangle = (0, v_h / \sqrt{2})^T$, corresponding to $\theta = v_h / \sqrt{2} f$ in Eq.~\eqref{eq:Uvev}. Indeed, since the negative top quark loop contribution to the Higgs potential, Eq.~\eqref{eq:4FermiMass}, is exactly along the direction $|H_1 - H_2|^2 = 2 |\tilde H_1|^2$, this is the field which takes a vev.
The physical fields from the two Higgs doublets are
\be
	\tilde H_1 = \left( G^+, \frac{v_h + h + i G^0}{\sqrt{2}} \right)^T~, \qquad
	\tilde H_2 = \left( H^+, \frac{h_2 + i A_0}{\sqrt{2}} \right)^T~, \qquad
\ee
where $G^{\pm,0}$ are those eaten by the SM $W^\pm$ and $Z$ bosons, $h$ is the physical SM-like $125\GeV$ Higgs as well as the only one which couples linearly to the EW gauge bosons. All the heavy Higgses are embedded in $\tilde H_2$: the two neutral states $h_2$ and $A_0$, and the charged $H^\pm$ one.

In order to minimise the potential and study the Higgs mass I set to zero all the fields except the physical Higgs $h$, in which case the pNGB matrix is given by Eq.~\eqref{eq:Uvev} with $\theta \to (v_h + h) / \sqrt{2} f$. The Higgs potential, from Eqs.~(\ref{eq:LHCmassHiggs},\ref{eq:LHiggsmassGauge},\ref{eq:4FermiMass}) becomes
\be
	V(\theta) =  - C_m f^4 \cos \theta - C_g f^4 \cos 2\theta - 2 C_t f^4 \sin^2 \theta~,
\ee
where
\be
	C_m = \frac{2 B_0}{f^2} (m_E + m_L)~, \quad
	C_g = \frac{3 \Lambda_{HC}^2}{16 \pi^2 f^2} \left( \frac{3}{4} c_w g_w^2 + \frac{1}{4} c_Y g_Y^2\right)~, \quad
	C_t = \frac{N_c y_t^2 c_t \Lambda_{HC}^2}{16 \pi^2 f^2}~,
\ee
and I am assuming $C_{m,g,t} > 0$.
Minimising the potential in $\theta$ gives the EWSB condition
\be
	\frac{v^2}{f^2} \equiv \xi = 2 \sin^2 \theta_{\rm min} = 2 - \frac{C_m^2}{8 \left(C_t - C_g \right)^2}~.
\ee
This condition should be tuned in order to obtain the desired $\xi$. Specifically, one could tune the mass parameters $(m_E + m_L)$ inside $C_m$ to achieve
\be
	C_m = 4 (C_t - C_g) \sqrt{1 - \frac{\xi}{2}}~.
	\label{eq:TuningCond}
\ee
The light Higgs, which in this setup does not mix with the other pNGBs, has a mass
\be
	m_h^2 = (C_t -  C_g) f^2 \xi 
	\sim  N_c c_t m_t^2 - 3  c_w m_W^2~,
	\label{eq:HiggsMass}
\ee
where in the estimate I used $\Lambda_{HC} \sim 4 \pi f$.
It is clear that some degree of cancellation is necessary in order to bring it down to the physical value of $m_h \approx 125 \GeV$.
From the first equality in Eq.~\eqref{eq:HiggsMass}, the tuning condition in Eq.~\eqref{eq:TuningCond}, and the definition of $C_m$ one also obtains
\be
	B_0 (m_E + m_L) = \frac{2 m_h^2}{\xi} \sqrt{1 - \xi/2}~,
	\label{eq:mEmLmh}
\ee
which relates the Higgs mass and the value of $\xi$ to the mass of the singlets $\eta_{1,2}$, Eq.~\eqref{eq:singlmassMatr}.
From the potential one can also derive the triple Higgs coupling:
\be
	\kappa_\lambda \equiv \frac{\lambda_{h^3}}{\lambda_{h^3}^\SM} =  \sqrt{1 - \frac{\xi}{2}}~.
\ee
Up to subleading EWSB corrections, the mass of the heavy Higgs doublet is
\be
	m^2_{\tilde H_2} = f^2 \left( \frac{1}{2} C_m + 2 C_g \right)
	\approx 2 f^2 C_t \sim \frac{2 N_c m_t^2}{\xi}~,
	\label{eq:H2mass}
\ee
where in the last step I used Eq.~\eqref{eq:TuningCond} and the definition of $C_t$.

\subsection{Higgs couplings and electroweak precision tests}
\label{sec:HiggsPheno}

The couplings of the SM-like Higgs boson to the EW gauge bosons and SM fermions are obtained from the pNGB kinetic term, Eq.~\eqref{eq:LOchiralpNGB}, and the SM Yukawa terms, Eq.~\eqref{eq:pNGBYuk}, by substituting in the pNGB matrix in Eq.~\eqref{eq:Uvev} the angle $\theta \to \theta + h / \sqrt{2} f$.
From the pNGB kinetic term one gets, in the unitary gauge,
\be\begin{split}
	\cL^{eff}_{kin} &\supset \left( m_W^2 W^\dagger_\mu W^\mu + \frac{m_Z^2}{2} Z_\mu Z^\mu \right) \frac{2}{\xi} \sin^2 \left( \theta + \frac{h}{\sqrt{2} f} \right)= \\
		&= \left( m_W^2 W^\dagger_\mu W^\mu + \frac{m_Z^2}{2} Z_\mu Z^\mu \right) \left( 1 + 2 \sqrt{1 - \frac{\xi}{2}} \frac{h}{v} + (1 - \xi) \frac{h^2}{v^2} + \ldots \right)~.
	\label{eq:hWWcouplings}
\end{split}\ee
Analogously, from the Yukawa term one has
\be
	 \cL^{Yuk} = - f y_\psi \bar \psi_{SM} \psi_{SM} \sin  \left( \theta + \frac{h}{\sqrt{2} f} \right) 
	 = - m_\psi \bar \psi_{SM} \psi_{SM} \left( 1 + \sqrt{1 - \frac{\xi}{2}} \frac{h}{v} - \frac{h^2 \xi}{4 v^2} + \ldots \right)~.
	 \label{eq:hffcouplings}
\ee
The ratios of the Higgs couplings to the SM prediction can be summarised as
\be
	\kappa_V = \kappa_f = \sqrt{1 - \frac{\xi}{2}}~.
\ee

The contributions of the model to flavour-universal electroweak precision tests is analogous to the one of all composite Higgs models \cite{Giudice:2007fh}. It can be separated in an infrared contribution due to the deviation of the Higgs coupling to electroweak gauge bosons as shown above, a contribution from the other pNGBs, and finally an ultraviolet contribution from the dynamics at the scale $\Lambda_{HC}$. Taken all together, they impose an upper limit on $\xi$ of about $\xi \lesssim 0.08$~\cite{Ma:2015gra}, which also makes the deviations in the Higgs couplings smaller than the present experimental sensitivity.

Other possibly dangerous effects could arise from the term depending on the pNGB matrix in Eq.~\eqref{eq:rhoffCoupl}. In fact, they generate at low energy deviations in the coupling of $\psi$ to SM EW gauge bosons. Particularly dangerous are deviations in the $Z b_L\bar{b}_L$ coupling. The NDA estimate for the relative deviation is $\delta g^{\rm NDA}_{Zb_L} \sim  \xi/4\pi \sim \text{(few)} \times 10^{-3}$, while the experimental limit is at the per-mille level. It might be possible to further suppress the deviation by assigning suitable quantum numbers to $q_L$ under the custodial symmetry group $\SU(2)_L \times \SU(2)_R$ and assume invariance under the parity $P_{LR}$ \cite{Agashe:2006at}. It is also possible that the flavour dynamics at the scale $\Lambda_t$ is such that the operator in Eq.~\eqref{eq:rhoffCoupl} is suppressed. Since it has a different chirality structure than those responsible for the Yukawa and LQ couplings this is not so implausible.
The effects due to the vector resonance $\rho$ are instead further suppressed by its large mass, $m_\rho \approx \Lambda_{HC} \gg m_{\rm pNGB}$.

\section{Flavour phenomenology}
\label{sec:flavour}

The leading effects in flavour observables are mediated by the pNGB leptoquarks $S_1$ and $S_3$. Other possible effects from heavier resonances are further suppressed by the small ratio $m_{LQ}^2 / \Lambda_{HC}^2 \ll 1$.
While this model can reproduce completely the flavour phenomenology described in Ref.~\cite{Buttazzo:2017ixm}, the presence of the $S_1$ coupling to right-handed currents makes the present setup possibly richer. The SMEFT dimension-6 operators obtained by integrating out the leptoquarks at the tree-level are described in Section~\ref{sec:S13LQyuk}.
In this Section I discuss the main aspects of the flavour phenomenology of the model.

\subsection{Muon magnetic moment and $\tau \to \mu \gamma$}
\label{sec:taumugamma}

The presence of $S_1$ couplings to both right- and left-handed top quarks allows the generation of $m_t$-enhanced contributions to both $\tau \to \mu \gamma$ and to the muon anomalous magnetic moment.
The relevant terms from Eq.~\eqref{eq:S13lagr} are
\be
	\LL_{S_{1}} \supset \bar{t}^c \left[ g_1 \beta_{1,b\alpha} P_L  + g_1^u \beta^u_{1,t\alpha} P_R \right] \ell^\alpha S_1 + h.c. ~,
\ee
where $\ell^\alpha = (e, \mu, \tau)$ and I recall that, by definition, $\beta_{1,b\tau} = \beta_{1,t\tau}^u = 1$.
The chirally-enhanced contribution from $S_1$ to $\tau \to \mu \gamma$ is given by (see e.g. Refs.~\cite{Dorsner:2016wpm,Crivellin:2017zlb} and references therein)
\bea
	\!\!\! \mathcal{B}(\tau \to \mu \gamma) &\!\!\! =& \!\!\! \frac{1}{\Gamma_\tau}
	\frac{\alpha N_c^2 m_t^2 m_\tau^3}{64 \pi^4 v^4} \left(1 - \frac{m_\mu^2}{m_\tau^2} \right)
		\left| Q_{S_1} g_S(x_t) - g_F(x_t) \right|^2 |\epsilon_1|^2 |\epsilon_1^u|^2 \left( |\beta_{1,b\mu}|^2 + |\beta^u_{1,t\mu}|^2 \right) = \nonumber\\
	&\!\!\! \approx& \!\!\!(7.0 \times 10^{-2}) \frac{|\epsilon_1|^2}{0.01} |\epsilon_1^u|^2 \left( \frac{|\beta_{1,b\mu}|^2}{0.1^2} + \frac{|\beta^u_{1,t\mu}|^2}{0.1^2} \right) < 4.4 \times 10^{-8}~,
\eea
where $( Q_{S_1} g_S(x_t) - g_F(x_t) ) \stackrel{x_t \ll 1}{\approx} 7/6 + 2/3 \log m_t^2 / m_{S_1}^2$ and I used $m_{S_1} = 1.5 \TeV$. Since the values $|\epsilon_1|^2 \approx 0.01$ and $|\beta_{1,b\mu}| \approx 0.1$ are required to fit the $B$ anomalies \cite{Buttazzo:2017ixm}, this observable puts a bound
\be
	|\epsilon_1^u|^2 \lesssim 10^{-6}~,
	\label{eq:eps1ubound}
\ee
corresponding to $g_1^u \lesssim 10^{-2} g_1$.
From the point of view of the $\SU(2)^5$ flavour symmetry $g_1^u$ and $g_{1,3}$ are expected to be of the same order.
It is interesting to note that by adding the approximate $\U(1)_e$ symmetry, under which all the right-handed leptons transform, in order to suppress the $\tau$ Yukawa coupling \cite{Barbieri:2012uh}, the $g_1^u$ suppression would be automatic since one could predict: $g_1^u / g_1 \sim y_\tau / y_t \sim 10^{-2}$.

The leading contribution to the muon anomalous magnetic moment from $S_1$ is \cite{Dorsner:2016wpm}
\be\begin{split}
	\delta a_\mu &=  - \frac{N_c m_\mu m_t}{12 \pi^2 v^2} \epsilon_1^u \epsilon_1 \beta_{1,b\mu} \beta^u_{1,t\mu} \left( 7 + 4 \log \frac{m_t^2}{m_{S_1}^2} \right) =\\
	&\approx  (7.9 \times 10^{-11}) \times  \frac{\epsilon_1^u}{10^{-3}} \frac{\epsilon_1}{0.1} \frac{\beta_{1,b\mu}}{0.1} \frac{\beta^u_{1,t\mu}}{0.1}~,
\end{split}\ee
while the observed anomaly is $(\delta a_\mu)_{exp} = (2.8 \pm 0.9) \times 10^{-9}$ \cite{Olive:2016xmw}.
One can see that due to the limit in Eq.~\eqref{eq:eps1ubound} the $\gtrsim 3 \sigma$ deviation from the SM observed in $\delta a_\mu$ cannot be explained. The same conclusion was reached in Ref.~\cite{Crivellin:2017zlb}.

\subsection{Charged-current processes}

The observed deviations in charged-current $b \to c \tau \nu$ transitions require the largest new physics contribution.
The effective operators at the $B$-meson mass scale relevant for this model are
\be
	\LL_{\rm eff}^{b\to c \bar \tau\bar\nu_\tau} \supset - \frac{2}{v^2} V_{cb} \left[ (1 + c_{V_L}^\tau) \OO_{V_L}^\tau
	 - c_{S_T}^\tau \OO_{T}^\tau
	 - c_{S_L}^\tau \OO_{S_L}^\tau \right] + h.c. ~,
	 \label{eq:effLagrCC}
\ee
where
\be
	\OO_{V_L}^\tau = (\bar c_L \gamma_\mu b_L) (\bar \tau_L \gamma^\mu \nu_\tau)~, \quad
	\OO_{T}^\tau = (\bar c_R \sigma_{\mu\nu} b_L) (\bar \tau_R \sigma^{\mu\nu} \nu_\tau)~,  \quad
	\OO_{S_L}^\tau = (\bar c_R b_L) (\bar \tau_R \nu_\tau)~, 
\ee
Matching at the tree-level with the SMEFT operators generated by integrating out the $S_1$ and $S_3$ fields, Eq.~(\ref{eq:EFTS13match}), one has:
\be\begin{split}
	c_{V_L}^\tau &= (c_{l q}^{(3)})_{\tau\tau33} + (c_{l q}^{(3)})_{\tau\tau32} \frac{V_{cs}}{V_{cb}} 
	= (|\epsilon_1|^2 - |\epsilon_3|^2) - (|\epsilon_1|^2 \beta_{1, s\tau} - |\epsilon_3|^2 \beta_{3, s\tau}) \frac{V_{tb}^*}{V_{ts}^*}~,\\
	c_{T}^\tau &= (c_{l e q u}^{(3)})_{\tau\tau32} =  \frac{\epsilon_1^u \epsilon_1^*}{4} \, \frac{\beta^u_{1,c\tau}}{V_{cb}}  \\
	c_{S_L}^\tau &= (c_{l e q u}^{(1)})_{\tau\tau32} = - 4 c_T = - \epsilon_1^u \epsilon_1^* \,  \frac{\beta^u_{1,c\tau}}{V_{cb}} ~.
\end{split}\ee
Due to the bound in Eq.~\eqref{eq:eps1ubound}, one can safely neglect the contributions to the tensor and scalar operators proportional to $\epsilon_1^u$ and keep only the vector operator.
The new physics dependence of $R_{D^{(*)}}$ is then simply given by: 
\be
	R_D/R_D^{\SM} = R_{D^*}/R_{D^*}^{\SM} \approx 1 + 2 c_{V_L}^\tau = 1.237 \pm 0.053~.
\ee
The $B_c^- \to \tau \bar\nu_\tau$ branching ratio is very sensitive to the scalar operator $\OO_{S_L}$ and the $B_c$ lifetime can be used to put an upper limit on such terms \cite{Alonso:2016oyd}. However, in this setup the constraint from $\tau \to \mu \gamma$ makes $c_{S_L}^\tau$ completely negligible.

The analogous effects in the muon mode are suppressed by the small coupling to muons, which follows from the $\SU(2)_l$ structure. Deviations from lepton flavour universality in $b \to c \mu (e) \nu$ transitions are constrained at the $\sim\OO(1)\%$ level \cite{Jung:2018lfu}. In this model they are given by \cite{Buttazzo:2017ixm}
\be
	R_{b\to c}^{\mu e} \approx 1 + 2 (|\epsilon_1|^2 - |\epsilon_3|^2) \beta_{b\mu}^2 \left( 1 + \frac{\beta_{s\mu}}{\beta_{b\mu}} \frac{V_{cs}}{V_{cb}} \right)~,
\ee
where I neglected the scalar and tensor contributions. In the natural region of parameter space of the model, this is well within the experimental limit.

\subsection{Neutral-current processes}

\subsubsection*{- $B \to K^{(*)} \mu^+ \mu^-$}

The relevant coefficients of the effective Hamiltonian at the $B$ meson scale and their tree-level matching to the model are (see also Refs.~\cite{Becirevic:2015asa,Becirevic:2016oho,Dorsner:2016wpm,Dorsner:2017ufx}):
\be\begin{split}
	\Delta C_9^\mu &= - \Delta C_{10}^\mu 
		= - \frac{\pi}{\alpha V_{tb} V_{ts}^*} \left( (c_{l q}^{(1)})_{\mu\mu23} + (c_{l q}^{(3)})_{\mu\mu23} \right) 
		= \frac{4 \pi}{\alpha V_{tb} V_{ts}^*} \; |\epsilon_3|^2 \, \beta_{3,b\mu} \beta_{3,s\mu} =\\
	&\approx - 0.69 \frac{|\epsilon_3|^2}{0.01} \frac{\beta_{3,b\mu}}{0.1} \frac{\beta_{3,s\mu}}{0.4|V_{ts}|}
		= - 0.61 \pm 0.12
\end{split}\ee
Given the structure of the $\SU(2)_l$ spurion $V_l$ in Eq.~\eqref{eq:SU2ellspurions}, no contribution to the electron mode is instead generated, implying an effect in the lepton flavour universality ratios $R(K)$ and $R(K^{(*)})$.

\subsubsection*{- $B \to K^{(*)} \nu \bar \nu$}

The relevant effective Lagrangian for this process is \cite{Altmannshofer:2009ma,Buras:2014fpa}
\be
	\LL_{\rm eff}^{b\to s \bar \nu\nu} = \frac{\alpha}{\pi v^2} V_{tb} V_{ts}^* \left( \bar s \gamma_\mu [( C_L^{\SM} \delta^{\alpha\beta} + \Delta C_L^{\alpha\beta}) P_L + \Delta C_R^{\alpha\beta} P_R] b \right) \left( \bar\nu_L^\alpha \gamma^\mu \nu_L^\beta\right)~,
\ee
where $C_L^{\SM} = - 6.38 \pm 0.06$ \cite{Altmannshofer:2009ma,Buras:2014fpa}. The contribution from the leptoquarks is (see also Refs.~\cite{Dorsner:2016wpm,Dorsner:2017ufx})
\be
	\Delta C_L^{\alpha\beta} = - \frac{\pi}{\alpha V_{tb} V_{ts}^*} ( (c_{l q}^{(1)})_{\alpha\beta23} - (c_{l q}^{(3)})_{\alpha\beta23} ) = \frac{2 \pi}{\alpha V_{tb} V_{ts}^*} \left( |\epsilon_1|^2 \; \beta_{1,s\alpha}\beta_{1,b\beta} + |\epsilon_3|^2 \; \beta_{3,s\alpha}\beta_{3,b\beta} \right) ~.
\ee
The relevant observables depend on the EFT coefficient as \cite{Altmannshofer:2009ma,Buras:2014fpa}:
\be
	R_{\nu\nu} = \frac{\mathcal{B}(B \to K^{(*)} \bar \nu \nu)}{\mathcal{B}(B \to K^{(*)} \bar \nu \nu)_{\SM}} 
	\approx \frac{1}{3} \left(2 + \left|1 + \delta c_L^{\tau\tau} \right|^2 \right) <  2.7 ~,
\ee
where
\be
	\delta c_{L}^{\tau\tau} \equiv \frac{\Delta C_{L}^{\tau\tau}}{ C_L^{\SM}} \approx 1.3 \left( \frac{|\epsilon_1|^2 \beta_{1,s\tau} + |\epsilon_3|^2 \beta_{3,s\tau}}{ 0.01 |V_{ts}|} \right)~,
	\label{eq:Rknunu}
\ee
and for simplicity I included only the leading correction due to the tau neutrinos.
The $90\%$ CL limit is taken from Ref.~\cite{Dorsner:2017ufx}.

\subsubsection*{- $B - \bar B$ mixing}

New physics contributions to $B^0 - \overline{B}^0$ mixing via an effective LL operator can be parametrised as
\be
	\Delta \LL_{\Delta B = 2} = - ( C_0^{\SM} + C_0^{\rm NP} ) \frac{(V_{tb} V_{ti}^*)^2}{32 \pi^2 v^2} (\bar b_L \gamma_\mu d_L^i)^2~,
\ee
where $i = d,s$ and $C_0^{\SM} = 4\pi \alpha S_0(x_t) \approx 1.0$.
A loop of the $S_1$ and $S_3$ leptoquarks contributes as (see also Refs.~\cite{Dorsner:2016wpm,Bobeth:2017ecx} for the individual contributions)
\be
	C_0^{S_1 + S_3} = g_1^2 \epsilon_1^2 \left( \frac{\beta_{1,s\tau}}{V_{tb} V_{ts}^*} \right)^2 + 5 g_3^2 \epsilon_3^2 \left( \frac{\beta_{3,s\tau}}{V_{tb} V_{ts}^*} \right)^2 + 2 g_1 g_3 \epsilon_1\epsilon_3 \frac{\beta_{1,s\tau}\beta_{3,s\tau}}{(V_{tb} V_{ts}^*)^2} f\!\left(\frac{m_{S_3}}{m_{s_1}} \right)~,
	\label{eq:BsmixFormula}
\ee
where I neglected SM fermion masses, $f(x) = \frac{x}{x^2 - 1} \log x^2$ (note that $f(x) \in [0,1]$ and $f(1) = 1$), and took into account that $\beta_{1(3),d\tau} / V_{td}^* = \beta_{1(3),s\tau} / V_{ts}^*$ according to the $\U(2)_q$ symmetry structure, implying that the same relative effect is expected in $B_s$ and $B_d$ mixing. The new physics contributions should not exceed $\sim 10\%$ of the SM one, in order to be safe from experimental limits,\footnote{A recent update of lattice calculations is responsible for a shift in the SM prediction which results in a slight tension with the measurement, $(\Delta M_{B_s})^{\rm exp} / (\Delta M_{B_s})^{\SM} = - 0.11 \pm 0.06$. Even though with purely imaginary couplings, $\text{Arg}(g_{1,3}) = \pm \pi/2$, it can be possible to fit this tension, I will not pursue it here since this is an issue still to be settled. See Ref.~\cite{DiLuzio:2017fdq} for a recent detailed discussion.}
\be
	\frac{(\Delta M_{B_{s/d}})^{S_1 + S_3}}{(\Delta M_{B_{s/d}})^{\SM}} = \frac{\eta^{LL}(m_{S_3}) C_0^{S_1 + S_3}}{C_0^\SM} \approx \eta^{LL}(m_{S_3}) \; C_0^{S_1 + S_3} \lesssim 10\%~.
	\label{eq:DMBsLQ}
\ee
where $\eta^{LL}(m_{S_3}) \approx 0.79$ encodes the renormalisation group effects down to $m_b$.

In the limit $g_1 = g_3$, $m_{S_1} = m_{S_3}$, and $\beta_{1,s\tau} = - \beta_{3,s\tau} \gtrsim |V_{ts}|$ one can approximately relate the deviation in $B_s$ mixing to the one in $R_{D^{(*)}}$:
\be
	\frac{(\Delta M_{B_{s/d}})^{S_1 + S_3}}{(\Delta M_{B_{s/d}})^{\SM}} \approx 0.74 \left(\frac{m_{S_{1,3}}}{1 \TeV}\right)^2 \left( \frac{R_{D^{(*)}} / R_{D^{(*)}}^\SM - 1}{0.23}\right)^2~,
	\label{eq:BsMixConstr}
\ee
where in the numerical expression I normalised $R_{D^{(*)}}$ to its best-fit value.
Since the LQ masses cannot be below 1 TeV due to present limits from direct searches (see Section~\ref{sec:LQDirSearchBounds}), the $B_s$ mixing constraint allows only to partially reproduce the charged-current anomalies when taken at face value.
In order to improve the fit, some mild cancellation with other contributions to $B_s$ mixing is required.
As can be seen from the expression above, the required tuning would be of one part in $\sim 10$ or less, for LQ masses not much above 1 TeV. One possibility could be to give complex phases to the LQ couplings in Eq.~\eqref{eq:BsmixFormula} and tune the various terms against each other, or to cancel the LQ contributions with extra ones from the UV theory.

Further contributions to these $\Delta B = 2$ operators can arise via tree-level exchange of heavy resonances at the scale $\Lambda_{HC}$, coupled to SM fermions via UV four-fermion operators such as the one in Eq.~\eqref{eq:rhoffCoupl}. The flavour symmetry protects these effects, giving an MFV-like suppression. The estimate is
\be
	C_0^{\rm UV} \sim g_{\rho \psi}^2 \frac{16 \pi^2 v^2}{\Lambda_{HC}^2} \sim g_{\rho \psi}^2 \xi.
\ee
For $g_{\rho \psi} \sim \OO(1/4\pi)$ these effects are well below the experimental limits. For larger values of the coupling it could be possible to use these extra contributions to partially cancel the one arising at one-loop from the leptoquarks. Also from Eq.~\eqref{eq:rhoffCoupl}, another contribution to the same operator can arise via the flavour-violating  $Zb_Ls_L$ coupling. The coupling can be estimated by NDA to be $\sim g_w V_{ts} \xi / 4\pi$, plus a further suppression should be added due to the $Zbb$ constraint. This gives a contribution to $B_s$-mixing: $C_0^{\rm Z} \sim \xi^2$. Due to the present limits on $\xi$, this is well below the flavour limit.
A stronger constraint can be obtained from lepton-universal contributions to $b s \ell^+ \ell^-$ operators, where the deviation due to this coupling scales like $\Delta C_9^\ell \sim \xi / \alpha$. As the $Zb_L b_L$ constraint, also this shows that the vector operators in Eq.~\eqref{eq:rhoffCoupl} must be suppressed.

\subsection{Radiative corrections to EWPT and $\tau$ decays}

Another relevant set of constraints arise due to renormalization group evolution from $m_{LQ}$ down to the electroweak scale of the semileptonic operators in Eq.~\eqref{eq:semilepOperators} to operators which modify the $Z$ and $W$ couplings to fermions \cite{Feruglio:2016gvd,Feruglio:2017rjo}. In particular, the leading effects are those affecting the $\tau$ and $\nu_\tau$ leptons proportionally to the top Yukawa.
Using the RGE equations from \cite{Jenkins:2013wua} and the results from Ref.~\cite{Feruglio:2017rjo} one gets
\bea
	\delta g_{\tau_L} \!\!\!&\approx&\!\!\! \frac{N_c y_t^2}{16 \pi^2} \log \frac{m_{LQ}}{m_t} \left( (C_{l q}^{(3)})_{\tau\tau 3 3} - (C_{l q}^{(1)})_{\tau\tau 3 3} \right) \approx 0.08 (|\epsilon_1|^2 + |\epsilon_3|^2) = (0.16 \pm 0.58) \times 10^{-3} ~, \nonumber\\
	\delta g_{\tau_R} \!\!\!&\approx&\!\!\! \frac{N_c y_t^2}{16 \pi^2} \log \frac{m_{LQ}}{m_t} (C_{e u})_{\tau\tau 3 3} \approx - 0.08  |\epsilon_1^u|^2 = (0.39 \pm 0.62) \times 10^{-3}~, \\
	\delta g_{\tau}^W \!\!\!&\approx&\!\!\! - \frac{2 N_c y_t^2}{16 \pi^2} \log \frac{m_{LQ}}{m_t} (C_{l q}^{(3)})_{\tau\tau 3 3} \approx - 0.08 (|\epsilon_1|^2 - |\epsilon_3|^2) = (0.97 \pm 0.98) \times 10^{-3}~, \nonumber
\eea
 where in the numerical evaluation I set $m_{LQ} = 1.5 \TeV$, neglected the subleading electroweak contributions, and used the limits from the global fit of Ref.~\cite{Efrati:2015eaa} for $Z\tau\tau$ and from Ref.~\cite{Pich:2013lsa} for the LFU constraints in $\tau$-decays (see the Appendix of Ref.~\cite{Buttazzo:2017ixm} for more details). The deviation in the $Z\nu\nu$ coupling is related by gauge invariance to $\delta g_{\nu_\tau} = \delta g_{\tau_L} + \delta g_{\tau}^W$.

An analogous radiative contribution is generated to lepton-flavour violating (LFV) $Z\tau\mu$ couplings, which can then mediate LFV $\tau$ decays \cite{Buttazzo:2017ixm}:
\be
	\mathcal{B}(\tau \to 3\mu) \approx 5 \times 10^{-4} (|\epsilon_1|^2 + |\epsilon_3|^2)^2 \beta_{b\mu}^2 < 1.2 \times 10^{-8}~.
\ee

\subsection{Fitting the $B$-meson anomalies}
\label{sec:BanomaliesFit}

The $\SU(2)_q \times \SU(2)_l$ flavour structure of the left-handed couplings is well suited to fit the $B$-physics anomalies, as described in Ref.~\cite{Buttazzo:2017ixm}.
Since, as shown above, the relevant effects are very similar to those studied in Ref.~\cite{Buttazzo:2017ixm}, I do not repeat a full numerical global fit here.
Instead, the preferred region in parameter space can be easily understood as follows:
\begin{itemize}
	\item The electroweak constraints put an upper limit $|\epsilon_1|^2 \approx |\epsilon_3|^2 \lesssim 10^{-2}$. 
	\item Fitting the $R(D^{(*)})$ excess while begin at the same time consistent with $R_{\nu\nu}$ then requires $\beta_{1,s\tau} \approx - \beta_{3,s\tau} \approx {\rm(few)} \times |V_{ts}| > 0$.
	If one limits the size of these off-diagonal terms to ${\rm(few)} \times |V_{ts}|$, the EWPT contraints do not allow to completely recover the anomaly \cite{Buttazzo:2017ixm}. Furthermore, the constraint from $B_s$ mixing \eqref{eq:BsMixConstr} makes this tension even stronger if it is not addressed by tuning with some other contribution.
	
	\item The suppression in $B \to K^{(*)} \nu\bar\nu$ required by the previous point corresponds to an enhancement in $B \to K^{(*)} \tau^+ \tau^-$. As shown for example in Ref.~\cite{Buttazzo:2017ixm}, the expected signal could be hundreds of times the SM prediction, bringing it possibly within the expected reach of Belle-II.
	
	\item The fit to the neutral-current $b \to s \mu\mu$ anomalies fixes the remaining parameters: $\beta_{3, b\mu} \approx 0.1$ and $\beta_{3, s\mu} \approx \beta_{3, b\mu} \beta_{1,s\tau}$, consistently with the flavour structure of Eq.~\eqref{eq:LQflavStructure}. The analogous couplings of $S_1$ are expected to be of the same order since the two have the same flavour structure. This value of $\beta_{b\mu}$ and the size of $|\epsilon_{1,3}|^2$ make the contribution to LFV $\tau$ decays much smaller than the present sensitivity.
	
	\item The experimental limit from $\tau \to \mu \gamma$ imposes the constraint $|\epsilon_1^u|^2 \lesssim 10^{-6}$. In terms of couplings this corresponds to $g_1^u \lesssim 10^{-2} g_{1,3}$. This can be naturally linked to the hierarchy $y_\tau / y_t$ by charging the right-handed leptons with an additional approximate $\U(1)_e$ symmetry.
	
\end{itemize}

\section{Collider phenomenology}
\label{sec:collider}

In this section I present the phenomenological aspects of the model more relevant for LHC new physics searches.

\subsection{Possible spectrum}
\label{sec:spectrum}

While the non-perturbative character of the dynamics underlying the model does not allow to make precise predictions for the spectrum of the theory, one can use the pNGB potential and NDA estimates detailed in Section~\ref{sec:potential} to sketch what a typical pNGB spectrum might be like.

\begin{figure}[t]
\centering
\includegraphics[width=1.0\textwidth]{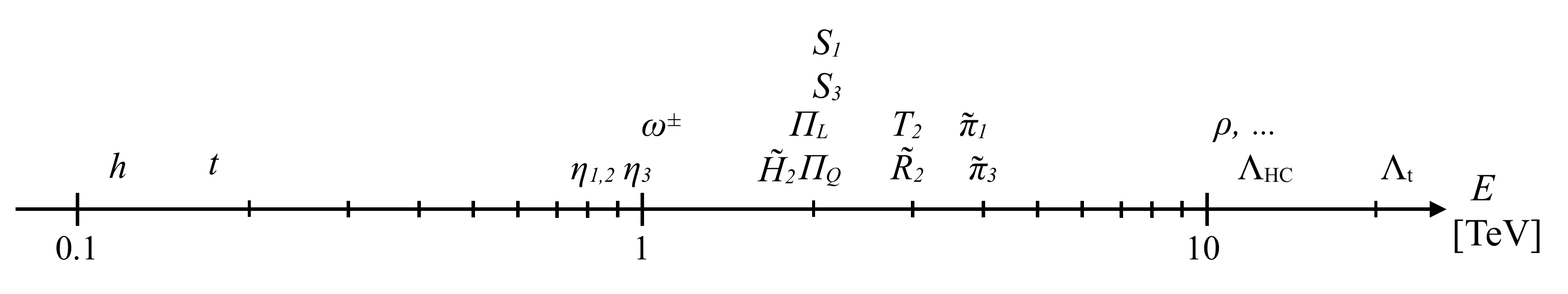}\hfill
\caption{ \small Example of a possible spectrum of the theory.\label{fig:spectrum}}
\end{figure}

For definitiveness in the following I fix
\be
	\xi = 0.05~  \quad (f=1.1\TeV)~,
\ee
corresponding to $\Lambda_{HC} \sim 13\TeV$.
In the simplifying limit $m_E = m_L$, Eq.~\eqref{eq:mEmLmh} relates the Higgs mass and $\xi$ to the mass of the first two singlets $m_{\eta_{1,2}} = \sqrt{2 B_0 m_L}$ = 790 \GeV. Using the QCD value $B_0 \approx 20 f$, one gets $m_L \approx 14\GeV$.
The third singlet mass is $m_{\eta_3} = m_{\eta_{1,2}} \sqrt{\frac{3 + 2 m_Q/m_L}{5}}$, which can be larger than the other two for $m_Q > m_L$, reaching $1\TeV$ for $m_Q \approx 2.5 m_L$.
The mass of the heavy Higgses before EWSB is given by Eq.~\eqref{eq:H2mass}, $m_{\tilde H_2} \sim 1.9 \TeV$.
For the other pNGBs I combine the contributions from the HC-fermion masses, Eq.~\eqref{eq:pNGBmasses}, and from the SM gauging, Eq.~\eqref{eq:gauge_mass_contr}. In the case of the $S_{1,3}$ leptoquarks I also take into account the contribution from the four-fermion operators, Eq.~\eqref{eq:4FermiLQMass}, assumed to be negative.
All the other composite resonances (composite vectors, scalars, HC-baryons, etc.) are expected to be near the $\Lambda_{HC}$ scale, i.e. above 10 TeV. Finally, the sector responsible for generating the four-fermion operators is expected to be not too far above that scale, unless the theory enters a conformal window above $\Lambda_{HC}$.
The resulting spectrum is sketched in Fig.~\ref{fig:spectrum}. The reader should keep in mind that this must be taken with a grain of salt, since $\OO(1)$ deviations from NDA are expected.

In the limit of unbroken EW symmetry, $\theta \to 0$, the only pNGB which mix with each other are the two singlets $\eta_2$ and $\eta_3$, Eq.~\eqref{eq:singlmassMatr}, where the mixing is proportional to the HC fermion mass difference $m_E - m_L$.
For $\theta > 0$, also a small mixing between the $\Pi_L^0$ and the $\eta_1$ singlet arises, proportional to $\propto (c_w g_{w}^2 - c_Y g_Y^2) \sin^2 \theta$, as well as between $S_{1,\frac{1}{3}}$ and  $S_{3,\frac{1}{3}}$ (proportionally to $\propto c_Y g_Y^2 (1 - \cos \theta)$) and between $\tilde R_{2,\frac{1}{3}}$ and  $T_{2,\frac{1}{3}}$ (proportionally to $\propto c_w g_w^2 (1 - \cos \theta)$).
With the specific choice of keeping only the pseudo-scalar combination in the HC bilinears in the four-fermion operators, no other mixing terms is present. In the more general case other mixing terms arise for non-zero $\theta$. A more detailed discussion of this point can be found in \cite{Ma:2015gra}.

\subsection{pNGB anomalous couplings}

Some pNGBs can have a non-zero coupling to two SM gauge bosons via the axial anomaly.
These interactions are fully described at the chiral Lagrangian level by the Wess-Zumino-Witten term \cite{Wess:1971yu,Witten:1983tw}. From that one can extract the relevant coupling of one pNGB to two gauge bosons, which in the class of theories considered here is given by
\be
	\LL_{\rm WZW} \supset - \frac{g_\beta g_\gamma}{16 \pi^2} \frac{\phi^\alpha}{f} 2 N_{HC} A^{\phi^\alpha}_{\beta\gamma} F_{\mu\nu}^\beta \widetilde F^{\gamma\mu\nu}~, \qquad
	A^{\phi^\alpha}_{\beta\gamma} = \text{Tr}\left[ T^\alpha T_{SM}^\beta T_{SM}^\gamma \right]~,
\ee
where $\widetilde F^{\gamma\mu\nu} = \frac{1}{2} \epsilon^{\mu\nu\rho\sigma} F^\gamma_{\rho\sigma}$, $T^\alpha$ is the generator corresponding to the pNGB $\phi^\alpha$ while $g_{\beta}$, $T_{SM}^{\beta}$, and $F_{\mu\nu}^{\beta}$ are the couplings, generators, and field strenght, respectively, of the $A^{\beta}_\mu$ gauge field (as defined in Eq.~\ref{eq:SMgauging}).
The complete list of anomalous couplings for the pNGBs in the theory is the following:
\be
\begin{array}{c | c c c c c c}
	 A^{\phi^\alpha}_{\beta\gamma} & g_1^2 & g_2^2 & g_3^2 & g_1 g_2 & g_1 g_3 & g_2 g_3 \\\hline
	\eta_1 & Y_L & 0 & 0 & 0 & 0 & 0 \\
	\eta_2 & -\frac{1}{4\sqrt{2}} & \frac{1}{4\sqrt{2}} & 0 & 0 & 0 & 0 \\
	\eta_3 & \frac{1 + 48 Y_L}{12\sqrt{30}} & -\frac{\sqrt{3}}{4\sqrt{10}} & -\frac{1}{\sqrt{30}} & 0 & 0 & 0 \\
	\tilde\pi_1 & 0 & 0 & d^{\alpha\beta\gamma}/(2\sqrt{2}) & 0 & \frac{1}{\sqrt{2}} \left( Y_L  - \frac{1}{3} \right) & 0 \\
	\tilde\pi_3 & 0 & 0 & 0 & 0 & 0 & \frac{1}{2\sqrt{2}} \\
	\Pi_L & 0 & 0 & 0 & \frac{Y_L}{2} & 0 & 0 \\
	\Pi_Q & 0 & 0 & 0 & \frac{\sqrt{3}}{2}\left( Y_L  - \frac{1}{3} \right) & 0 & 0 
\end{array},
\label{eq:anomCouplTab}
\ee
where $d^{\alpha\beta\gamma}$ are the $\SU(3)_c$ symmetric structure constants.
Measuring a process involving these coupling would provide information on $N_{HC} / f$. An independent measurement of $f$ (i.e. of $\xi$) could instead be obtained, for example, via Higgs couplings measurements or pNGB scattering.

\subsection{Collider signatures of the pNGBs}
\label{sec:searches}

Here I discuss some of the main aspects of the collider phenomenology of the various pNGBs, listed in Eq.~\eqref{eq:pNGB}, in particular their possible production channels and decay modes. I also present the present bounds and future prospects for the most interesting cases.

\subsubsection{$S_1$ and $S_3$ Leptoquarks}
\label{sec:LQDirSearchBounds}

Due to their linear couplings to SM fermions, the $S_1$ and $S_3$ leptoquarks have a rich phenomenology. The various states are classified under the electromagnetic $\U(1)_{em}$ as:
	\be
		s_{1,-\frac{1}{3}}~, \quad
		s_{3,-\frac{4}{3}}~, \quad
		s_{3,-\frac{1}{3}}~, \quad
		s_{3,\frac{2}{3}}~, 
	\ee
where I defined all in the ${\bf 3}$ of color and the 1 (3) suffix represents the electroweak multiplet they belong to. The NDA estimate puts their mass in the $\sim1.5 - 2.5 \TeV$ range, with $s_3$ being possibly slightly heavier than $s_1$ due to a larger electroweak correction to its mass. The splitting within the electroweak multiplets is subleading. The $B_s$ mixing constraint \eqref{eq:BsMixConstr} favours light leptoquarks.
In presence of EWSB, the $s_{1,-\frac{1}{3}}$ and $s_{3,-\frac{1}{3}}$ states have a small mass mixing. Expanding the $\SU(2)_w$ structure of the interaction Lagrangian of Eq.~\eqref{eq:S13lagr} one gets
\be\begin{split}
	\LL_{LQ} &= g_1 s_{1,-\frac{1}{3}}^\dagger \left( \bar{t}^c_L \tau_L - \bar{b}^c_L \nu_\tau \right) 
	+ g_3 s_{3,-\frac{1}{3}}^\dagger \left( - \bar{t}^c_L \tau_L - \bar{b}^c_L \nu_\tau \right) + h.c. \\
	&+ \sqrt{2} g_3 \left(  s_{3,\frac{2}{3}}^\dagger \bar{t}^c_L \nu_\tau - s_{3,-\frac{4}{3}}^\dagger \bar{b}^c_L \tau_L \right) + h.c.~,
	\label{eq:LQinteractions}
\end{split}\ee
where I neglected flavour-suppressed couplings to light generation fermions as well as those of $S_1$ to right handed fields, due to the $\tau \to \mu \gamma$ constraint discussed in Section~\ref{sec:taumugamma}. The phenomenology of $S_3$ with coupling to muons has been studied in Ref.~\cite{Hiller:2018wbv}.

The following discussion of the collider bounds on these leptoquarks can be also applied to weakly coupled models where these leptoquarks are elementary, since it is only based on the Lagrangian in Eq.~\eqref{eq:LQinteractions}.
Neglecting SM fermion masses, the total decay widths are $\Gamma_{S_{1,3}} = \frac{|g_{1,3}|^2}{8 \pi} m_{S_{1,3}}$ \cite{Dorsner:2016wpm}. The two leptoquarks with charge $\frac{1}{3}$ have equal branching ratio of 1/2 into the two channels $t\tau$ and $b\nu_\tau$, while $s_{3,\frac{2}{3}}$ and $s_{3,-\frac{4}{3}}$ decay to $t\nu$ and $b\tau$, respectively, with unity branching ratio. The deviations from these branching ratios due to multi-body decays, such as those discussed later on and shown in Fig.~\ref{fig:2PhiCoupl}, are suppressed both by the phase space and by the small $\xi$ parameter and can thus be safely neglected.

%
\begin{figure}[t]
\centering
\includegraphics[width=0.65\textwidth]{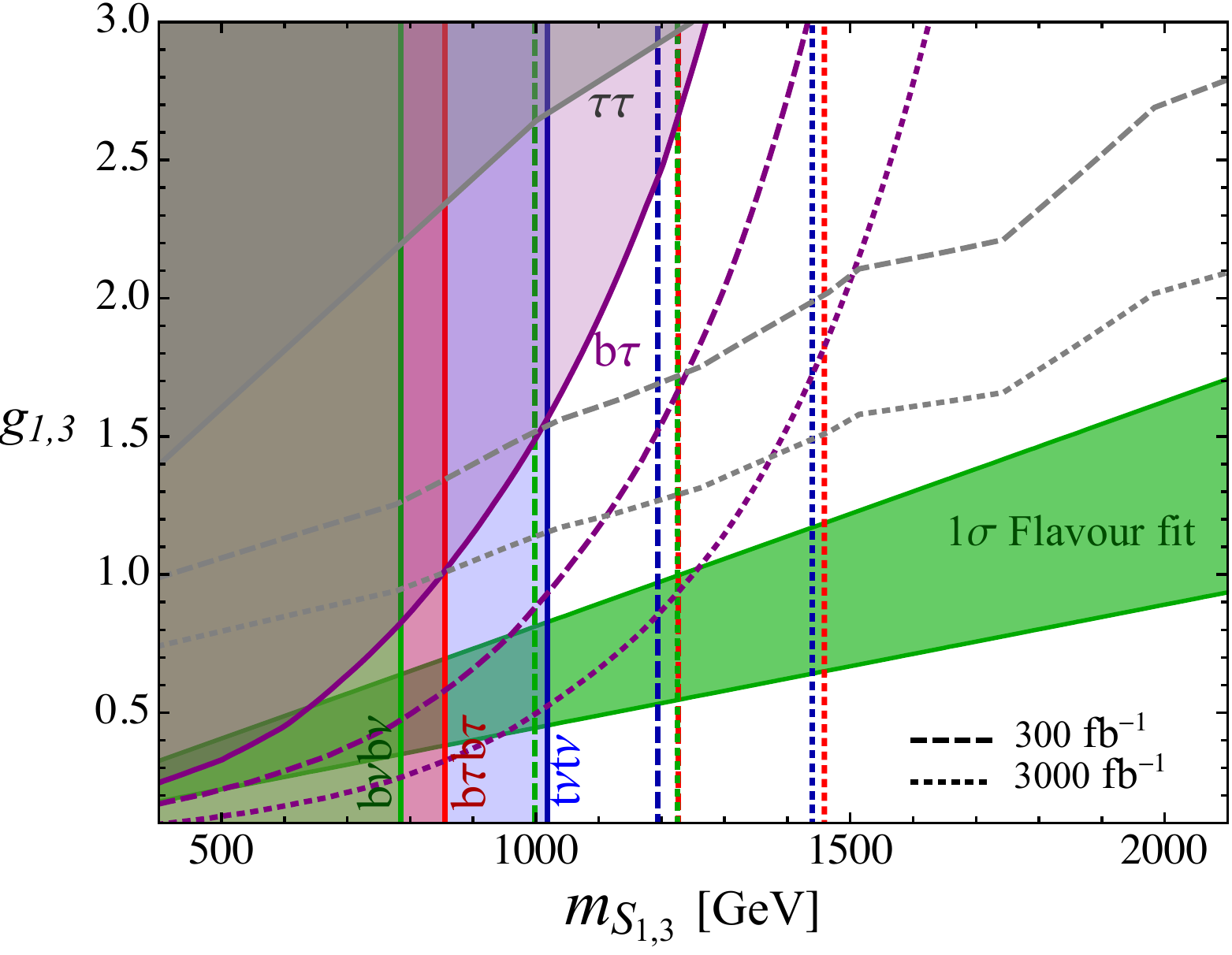}
\caption{ \small
Present and future expected exclusion limits at 95\% CL on the $S_1$ and $S_3$ LQ. Vertical bounds are from various pair-production modes, purple is from single production in the $b\nu$ channel while gray is from the off-shell $\tau\tau$ tail. Dashed and dotted lines are 13 TeV LHC expected limits for 300 and 3000 fb$^{-1}$ of integrated luminosity, respectively. The diagonal green region is the $1\sigma$-favoured one from the flavour fit.
\label{fig:LQbounds}}
\end{figure}

The main production modes at the LHC are pair production via QCD interactions, or single production via the coupling to the $b$ quark. While the former is model-independent, the latter depends on the couplings $g_{1,3}$. For $g_{1,3} = 1$, the single production cross section, via the bottom coupling, becomes larger than pair production for masses $m_{S_{1,3}} \gtrsim 1.4\TeV$ at 13\TeV, in which case $\sigma(p p \to s s^\dagger) \approx \sigma(p p \to s^\dagger \ell + s \bar{\ell} ) \approx 0.37$ fb \cite{Dorsner:2018ynv}. The present experimental limit from CMS \cite{Sirunyan:2017yrk} with 12.9fb$^{-1}$ of integrated luminosity on pair-produced scalar leptoquarks in the final state $b\bar{b} \tau^+ \tau^-$ is $m_{s_{3,-\frac{4}{3}}} > 855 \GeV $ at 95\% CL. This is shown as a solid red line in Fig.~\ref{fig:LQbounds}.
Very recently, the CMS collaboration updated also the searches in the $\tau\tau t t$ \cite{Sirunyan:2018nkj}, $\nu\nu t t$, and $\nu\nu b b$ final states \cite{Sirunyan:2018kzh} with 35.9~fb$^{-1}$ of luminosity. Taking into account the branching ratios described above, the resulting 95\% CL limits in this model are: $m_{1(3),-\frac{1}{3}} > 564 \GeV$ from $t\tau$, $m_{1(3),-\frac{1}{3}} > 795 \GeV$ from $b \nu$ (green vertical line in Fig.~\ref{fig:LQbounds}), and $m_{3,\frac{2}{3}} > 1018 \GeV$ from $t\nu$ (blue line).

The present limit from single-production, in the $b\tau$ channel \cite{Sirunyan:2018jdk}, is shown as a solid purple line in Fig.~\ref{fig:LQbounds}. At present it becomes the most important one for couplings $1.5 \lesssim g_{1,3} \lesssim 3$.

Another relevant search channel for $s_{3,-\frac{4}{3}}$ is in the $\tau^+\tau^-$ final state, where the leptoquark can be exchanged in the $t$-channel \cite{Faroughy:2016osc}. The corresponding 95\% CL limit is shown with a solid gray line in Fig.~\ref{fig:LQbounds} and dominates for large couplings $g_{1,3} \gtrsim 3$. The analogous effect in the $\mu^+ \mu^-$ tail is further suppressed by the small coupling to second generation leptons \cite{Greljo:2017vvb}.

All these limits are collected in Fig.~\ref{fig:LQbounds}, where I also show estimates for the prospects for 300 fb$^{-1}$ (dashed lines) and 3000 fb$^{-1}$ (dotted lines) of luminosity, obtained by rescaling the expected cross section limits with the square root of the luminosity ratios. The green region is the $1\sigma$ preferred one from the flavour fit \cite{Buttazzo:2017ixm}, which assumes that the LQ contribution to $B_s$-mixing is cancelled by some extra terms. Some conclusions can be drawn:
\begin{itemize}
	\item The region relevant for the $B$-physics anomalies and in the mass range $1.5 - 2 \TeV$ will not be tested by the LHC, even with high luminosity. The $28\TeV$ HE-LHC or FCC-hh would be needed. 
	\item For lighter LQ masses and in the region preferred by the flavour fit, the most relevant bound will always come from pair production. The most promising channels are $t\nu t\nu$ and $b\tau b\tau$, since the charge-$2/3$ and charge-$4/3$ LQ decay in these channels with unity branching ratio.
\end{itemize}
%

\subsubsection{Singlets}

\begin{figure}[p]
\centering
\includegraphics[width=0.47\textwidth]{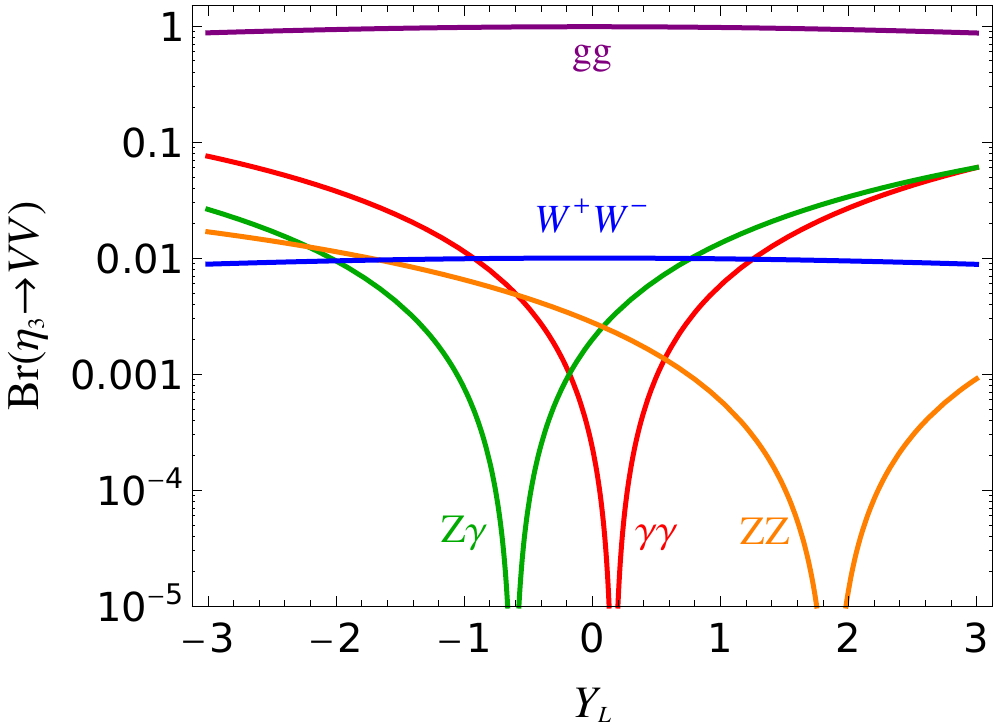}~~
\includegraphics[width=0.47\textwidth]{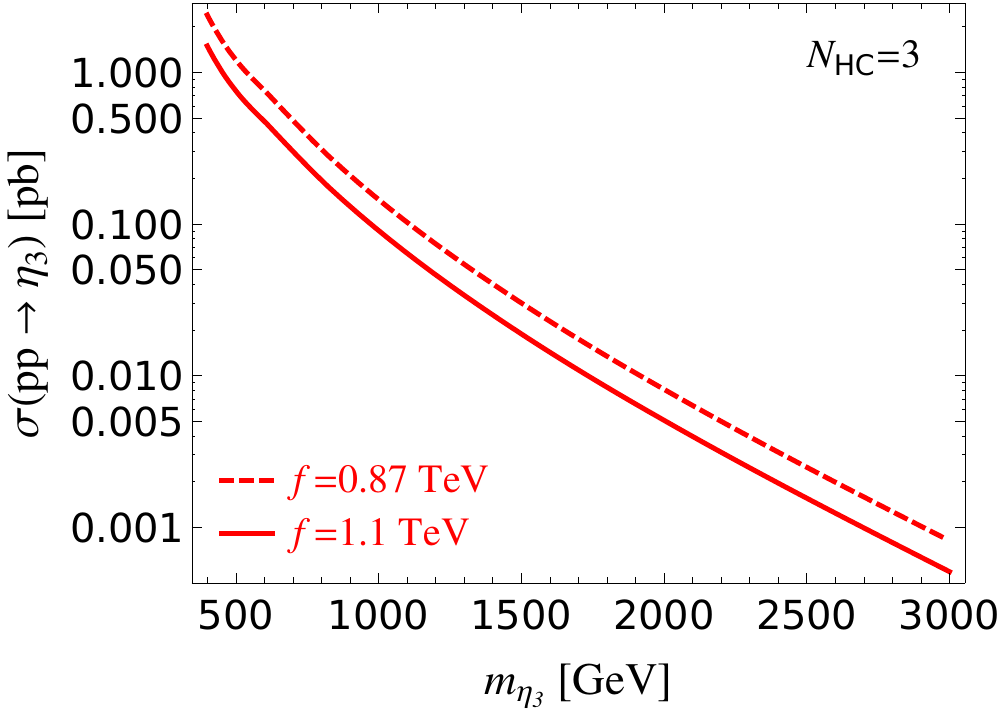}\\\vspace{0.5cm}
\includegraphics[width=0.47\textwidth]{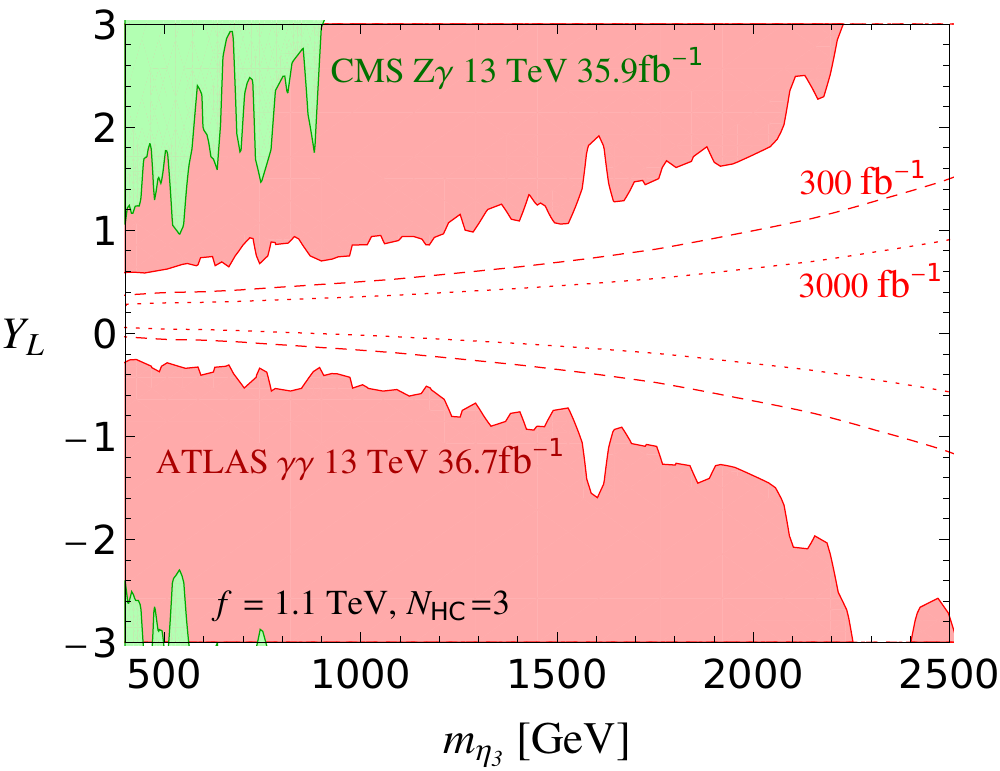}~~
\includegraphics[width=0.47\textwidth]{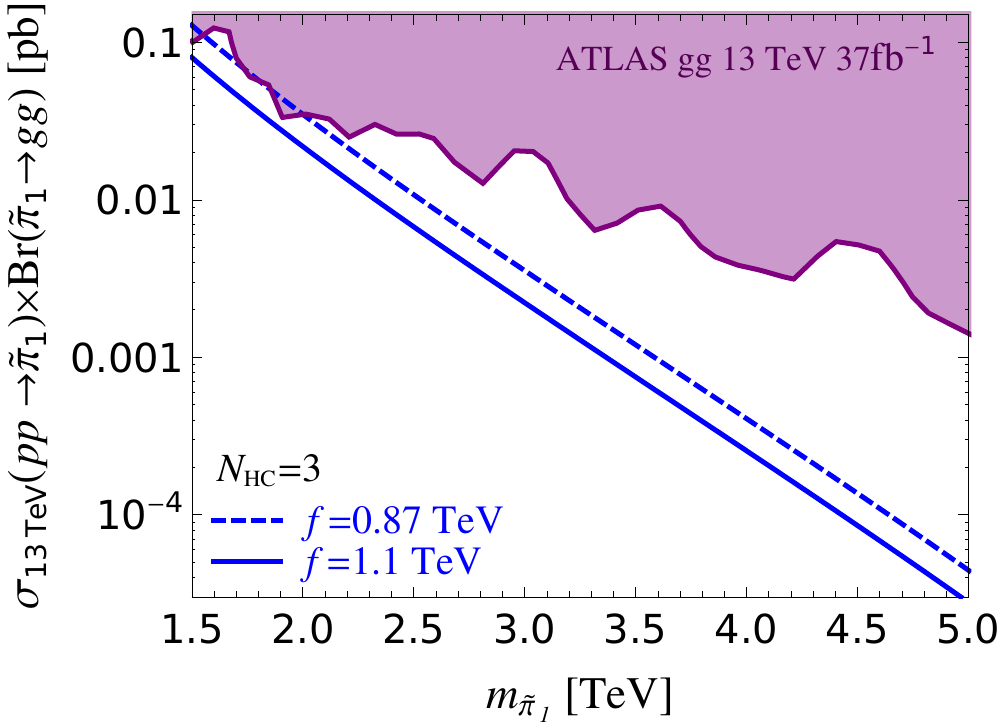}
\caption{ \small
(Top-left) Branching ratios of $\eta_3$ to gauge boson pairs via anomalous couplings, as a function of $Y_L$.\\
(Top-right) Production cross section at 13 TeV LHC via gluon fusion for the singlet $\eta_3$, $N_{HC}=3$, and two values of $f$.\\
(Bottom-left) Excluded region at 95\% CL in the $m_{\eta_3}-Y_L$ plane from the ATLAS $\gamma\gamma$ search \cite{Aaboud:2017yyg}, in red, and from the CMS $Z\gamma$ search \cite{Sirunyan:2017hsb}, in green. The dashed and dotted lines are future LHC prospects for 300 and 3000 fb$^{-1}$ of luminosity.\\
(Bottom-right) Signal cross section for the color octet $\tilde \pi_1$ in dijet ($gg$) as function of its mass, for $f=1.1\; (0.87) \TeV$ in solid (dashed) blue. The purple region is excluded by the ATLAS dijet search \cite{Aaboud:2017yvp}.
\label{fig:eta3_xsec}}
\end{figure}

The two SM singlets $\eta_{1,2}$ are expected to have a mass close to $800 \GeV$, Eqs.~(\ref{eq:singletmasses},\ref{eq:mEmLmh}), while $\eta_3$ can be heavier since its mass depends on $m_Q$. The anomalous couplings in Eq.~\eqref{eq:anomCouplTab} mediate decays of the singlets to pairs of SM gauge bosons.
Assuming these are the leading decay widths, the branching ratios for $\eta_1$ and $\eta_2$ are
\be
\begin{array}{c | c c c c c}
	 \text{Br}	 & gg & \gamma\gamma & Z\gamma & ZZ & W^+ W^- \\\hline
	\eta_1 & 0 & 0.58 & 0.36 & 0.06 & 0 \\
	\eta_2 & 0 & 0 & 0.21 & 0.15 & 0.64
\end{array}\label{eq:singlBr}~,
\ee
while	 for $\eta_3$ it is shown in Fig.~\ref{fig:eta3_xsec} (top-left) as a function of $Y_L$.
The total decay width of $\eta_3$ is $\Gamma_{\eta_3} \approx 50 \MeV$ for a mass of $1\TeV$, $f=1.1\TeV$, and $N_{HC} = 3$. In the numerical results I use the SM gauge couplings evaluated at the scale $m_\eta / 2$ via the one-loop RG equations.
The only singlet to have a production cross section possibly relevant for the LHC is $\eta_3$, via the gluon-gluon coupling. The production cross section for $f = 1.1 \; (0.87) \TeV$ (i.e $\xi=0.05 \; (0.08)$) and $N_{HC}=3$ is shown in Fig.~\ref{fig:eta3_xsec} (top-right). 
To the initial LO result, obtained with MadGraph5\_aMC@NLO \cite{Alwall:2014hca} with the NN23LO pdf set, I applied a constant NNLO (plus partial N$^3$LO) $K$-factor of 2.45 \cite{Ahmed:2015qda}. In the bottom-left panel I show the excluded region in the $m_{\eta_3}-Y_L$ plane from the ATLAS diphoton search \cite{Aaboud:2017yyg}\footnote{I take the efficiency of the fiducial region as ranging from $64\%$ at 200 GeV to $75\%$ at 2700 GeV.} (in red) and from the CMS $Z\gamma$ search \cite{Sirunyan:2017hsb} (in green). The dashed (dotted) red line represents the approximate expected future sensitivity in the $\gamma\gamma$ channel for 300 (3000)~fb$^{-1}$ of integrated luminosity. The $ZZ$ and $WW$ searches are not sensitive enough in this scenario. The limit already excludes values of $|Y_L| \gtrsim 1$ for a wide range of $\eta_3$ masses, implying that an observation of such a singlet could be expected in future searches.

The mass mixing $\eta_2 - \eta_3$, proportional to $m_E - m_L$, could also induce a gluon coupling for the second singlet once the mass matrix is diagonalised.
The only other mixing term involving the singlets is the small one between $\eta_1$ and $\Pi_L^0$, as discussed in Section.~\ref{sec:spectrum}.

\subsubsection{Color-octets}

The pNGB spectrum contains two neutral color-octets $\tilde \pi_{1,3}^0$ and one charged color-octet $\tilde \pi_3^\pm$. Their large QCD charge implies that they are likely the heaviest pNGBs. The NDA estimate puts their mass in the $\sim 3 - 4 \TeV$ range. They couple to pairs of SM gauge bosons via the anomalous interactions in Eq.~\eqref{eq:anomCouplTab}. 
	The main decay mode of $\tilde \pi_1$ is in two gluons, with almost unity branching ratio and $\Gamma(\tilde \pi_1 \to gg) \approx 1.4\GeV$ for $N_{HC} = 3$, $f = 1.1\TeV$, and $m_{\tilde \pi_1} = 4\TeV$. Subleading decay modes are into a gluon and an electroweak gauge boson, including a photon. The color-octet-$\SU(2)_w$-triplet $\tilde \pi_3$, instead, only couples to one gluon and one EW gauge boson.
Therefore, while both can be pair produced by QCD interaction, $\tilde \pi_1$ can also be singly produced via the anomalous couplings to gluons. 
This is the most promising search channel for large masses and the cross section for two different values of $f$ is shown in the bottom-right panel of Fig.~\ref{fig:eta3_xsec} as a function of the mass. 
This is obtained by rescaling the singlet production cross section as $\sigma(p p \to \tilde \pi_1)_{LO} = \frac{8 \Gamma(\tilde \pi_1 \to gg)}{\Gamma(\tilde \eta_3 \to gg)} \sigma(p p \to \tilde \eta_3)_{LO} $, see e.g. Ref.~\cite{Redi:2016kip}.
In the same plot I also show the present experimental limit from the ATLAS dijet resonance search \cite{Aaboud:2017yvp} (I fixed an approximate acceptance $A \approx 0.5$). For a mass of $4\TeV$ the signal is well below the present (as well as future) sensitivity. The present limits for pair-produced scalar color-octets are only in the $\sim 800\GeV$ range \cite{Aaboud:2017nmi}.  Also the limits from $j\gamma$ are still not sensitive, being close to $\sim 1$ fb for a mass of $4\TeV$ \cite{Aaboud:2017nak}. 
See Refs.~\cite{Redi:2016kip,Bai:2016czm} for a more detailed study of the LHC phenomenology of these states. 
	
\subsubsection{Triplets}

The two $\SU(2)_w$ triplets $\Pi_{L,Q}^a$ are expected to have masses around $2 \TeV$. They decay via the anomalous couplings \eqref{eq:anomCouplTab} with branching ratios
\be\begin{split}
	&\BR( \Pi^0_{L,Q} \to \gamma\gamma) = \BR( \Pi^0_{L,Q} \to ZZ) \approx 0.27~,\quad
	\BR( \Pi^0_{L,Q} \to Z\gamma) \approx 0.46~, \\
	&\BR( \Pi^\pm_{L,Q} \to W^\pm \gamma) \approx 0.78~,\BR( \Pi^\pm_{L,Q} \to W^\pm Z) \approx 0.22~.
\end{split}\ee
They can be either singly produced in association with a gauge boson or in vector boson fusion via the same couplings, or pair produced via electroweak gauge interactions.
Due to their large mass and electroweak production modes, they can't be directly detected at the LHC so I do not discuss them further. \\
 
\subsubsection{Other pNGBs}

\begin{figure}[t]
\centering
\includegraphics[width=0.3\textwidth]{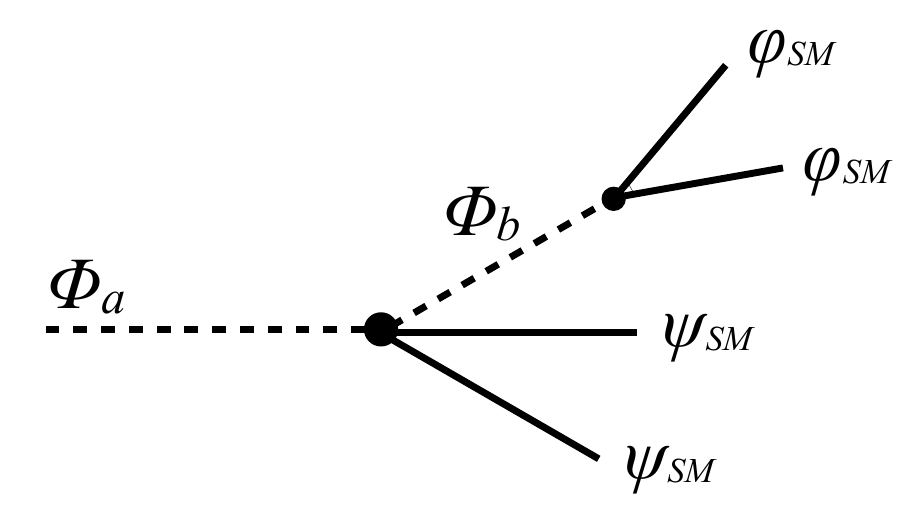} \quad
\includegraphics[width=0.3\textwidth]{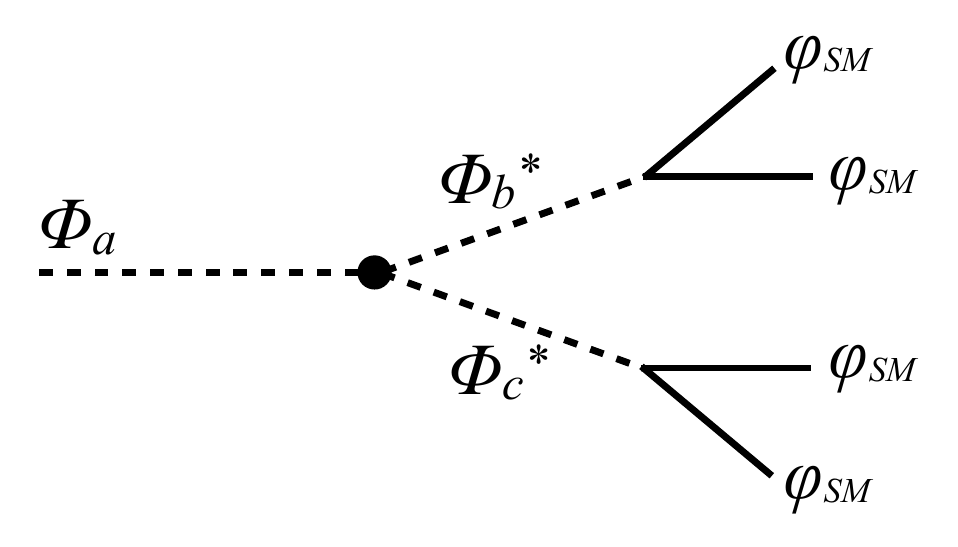}
\caption{ \small Three-body pNGB decay via the LQ coupling Lagrangian (left) and multi-body decays via trilinear interactions in the pNGB potential (right). \label{fig:2PhiCoupl}}
\end{figure}

The other pNGBs do not have any linear coupling to SM states, therefore no allowed decay $\Phi_a \to \varphi_\SM \varphi_\SM$. However, by expanding the Yukawa and LQ-coupling operators of Eqs.~(\ref{eq:pNGBYuk},\ref{eq:S13lagr}) one gets couplings of two pNGB to SM fermions such as $\LL \supset g_x/f  \, \Phi_a \Phi_b \psi_\SM \psi_\SM$. A heavier pNGB can thus have a three-body decay into SM fermions and a lighter pNGB, which in turn could decay to SM states (fermions or gauge bosons) via the processes described above, as shown schematically in Fig.~\ref{fig:2PhiCoupl} (left). Compared to direct two-body decays to fermions, these three-body decays are suppressed by the phase space and by the EWSB parameter $\xi = v^2 / f^2$, since they are absent for $\xi = 0$. For these reasons I do not expect them to modify in an important way the LQ branching ratios described above, contrary to what was recently claimed in Ref.~\cite{Monteux:2018ufc}.
Another possible decay mode is via trilinear pNGB couplings arising from the potential for $\xi>0$ (since there are none in the EW preserving vacuum). This would allow multi-body decays via (possibly off-shell) intermediate pNGBs as shown schematically in Fig.~\ref{fig:2PhiCoupl} (right). I expect these to be further suppressed with respect to the three-body ones by an even smaller phase space and by the fact that the pNGB potential is loop-generated.
Finally, transitions within the same representation of the SM gauge groups are always mediated by couplings to SM gauge bosons.

%

\vspace{1em}
\noindent{\bf $\tilde R_2$ and $T_2$}\\
These states have a mass close to $3 \TeV$. The charges of the individual states are (where all are ${\bf 3}$ of color):
	\be
		r_{\frac{2}{3}}~, \quad
		r_{-\frac{1}{3}}~, \quad
		t_{-\frac{1}{3}}~, \quad
		t_{-\frac{4}{3}}~.
	\ee
	Some of the interactions mediating three-body decays as in Fig.~\ref{fig:2PhiCoupl} (left) are
	\be\begin{split}
		\LL^{\rm eff}_{\rm LQ} &\supset 
		\frac{i g_3}{\sqrt{2} f} \sin \frac{\theta}{2} \; \bar t_L^c \nu_\tau  r_{\frac{2}{3}}^\dagger \tilde \eta_R 
		- \frac{i (g_1 - g_3)}{2 \sqrt{2} f} \sin \frac{\theta}{2} \; \bar b_L^c \nu_\tau  r_{-\frac{1}{3}}^\dagger \tilde \eta_R + h.c. \\
		&+  \frac{i g_3}{\sqrt{2} f} \sin \frac{\theta}{2} \; \bar b_L^c \tau_L  t_{-\frac{4}{3}}^\dagger \tilde \eta_T 
		+ \frac{i (g_1 + g_3)}{2 \sqrt{2} f} \sin \frac{\theta}{2} \; \bar b_L^c \nu_\tau  t_{-\frac{1}{3}}^\dagger \tilde \eta_T + h.c.~. 
		\label{eq:R_eta_psi_psi}
	\end{split}\ee
	where $\tilde \eta_{R,T} = \eta_1 \pm \frac{1}{\sqrt{2}} \eta_2 \mp \frac{1}{\sqrt{30}} \eta_3$ are combinations of the three singlets, which in turn decay to SM gauge bosons via their anomalous couplings. A complete list of the decay modes is beyond the purpose of this paper.
Their main production mode at hadron collider is in pair-production via QCD interactions and the phenomenology is quite similar to the one studied in Ref.~\cite{Monteux:2018ufc}, with bounds in the $\sim 800\GeV$ range. The expected mass in this model is too large to make them observable at the LHC.

\vspace{1em}
\noindent{\bf Heavy Higgs}\\
The states in the heavy Higgs doublet $\tilde H_2$ (i.e. $h_2$, $A_0$, and $H^\pm$) have masses $m_{\tilde H_2} \lesssim 2\TeV$ and do not couple linearly neither to SM fermions nor to gauge bosons. However, they have three-body decays as described above via the Lagrangian
	\be
		\LL^{\rm eff}_{\rm LQ} \supset - \frac{i}{2\sqrt{2} f} \sin \frac{\theta}{2} \; h_2 \left( g_1 S_1 \bar q_L^c \beta_1 \epsilon l_L + g_3 S_3^a \bar q_L^c \beta_3 \epsilon \sigma^a l_L \right) + h.c.~,
	\ee
and similar terms for $H^\pm$ and $A_0$. These terms clearly break the $P_H$ symmetry, thus allowing for $\tilde H_2$ decays. Depending on the masses, the leptoquarks might also be off-shell, thus further suppressing the decay.

\vspace{1em}
\noindent{\bf Charged singlet $\omega^\pm$}\\
The only $\Phi^2 \psi_\SM^2$ coupling linear in $\omega$ involves the $R_2$ scalar, which is expected to be much heavier and itself has three-body decays as shown above. The interaction Lagrangian is
	\be
		\LL^{\rm eff}_{\rm LQ} \supset \frac{i \sin \theta / 2}{2f}  \omega^+ \left( 2 g_3 \bar b_L^c \tau_L  r_{-\frac{1}{3}}^\dagger + (g_1 + g_3) b_L^c \nu_\tau r_{\frac{2}{3}}^\dagger + (g_3 - g_1) t_L^c \tau_L r_{\frac{2}{3}}^\dagger \right) + h.c.~.
	\ee
	Possible other decay modes could arise from next terms in the expansion, \mbox{$\cL^{eff}_{LQ} \supset \Phi^3 \psi_\SM^2$}, or from trilinear couplings in the potential. Some trilinear terms in the potential are (schematically):
	\be
	V_{\rm pNGB} \supset \omega^+ (W^-_L \eta_1,~ h \Pi_L^- ,~ Z_L \Pi_L^-,~ W^-_L \Pi_L^0 ,~ \eta_{2,3} H^-, s_{1(3),-\frac{1}{3}} r_{\frac{2}{3}}^\dagger,~ \ldots)~,
	\ee
	where $W_L$ and $Z_L$ here represent the \emph{eaten} NGBs. Since $\eta_1$ might be the lightest pNGB (other than the Higgs), the decay mode into $W \eta_1$ might be the leading one as both states could be on-shell.
	Similarly to the charged heavy Higgs, also this is pair produced via electroweak interactions and its mass, close to $\sim 1 \TeV$, is too heavy for the LHC. For this reason I do not pursue a more detailed study of its collider phenomenology.

\vspace{1em}
\noindent{\bf Dark matter}\\
From Eq.~\eqref{eq:anomCouplTab} one can note that, in absence of the colored HC fermion and for $Y_L = 0$, the anomalous coupling of $\eta_1$ and $\Pi_L$ vanish. This is a consequence of a symmetry arising in that limit, as discussed in detail in Ref.~\cite{Ma:2015gra}. This would potentially allow the lightest neutral pNGB to be stable and therefore a possible dark matter candidate. In this model, instead, the presence of $\Psi_Q$ and the LQ couplings to fermions, break explicitly this symmetry and allow decays of all pNGBs also for $Y_L = 0$. For example, the terms in Eq.~\eqref{eq:R_eta_psi_psi} mediate the decay of $\eta_1$ via an off-shell $\tilde R_2$ or $T_2$ pNGB: $\eta_1 \to \psi_\SM \psi_\SM \tilde R_2^*$ ($\tilde R_2^* \to \eta_{2,3} \psi_\SM  \psi_\SM$).

Depending on $N_{HC}$, for example for $N_{HC} = 3$, the model can have heavy HC-baryons, $(\Psi_a \Psi_b\Psi_c)$. The lightest neutral one could be stable and a possible dark matter candidate.
A more careful analysis of this possibility, while being beyond the purpose of the present paper, could be an interesting extension of this work.

\section{Summary and conclusions}
\label{sec:conclusions}

The naturalness problem of the electroweak scale is one of the most important unresolved theoretical questions in our understanding of Nature at very small distances. At the same time, we also lack an understanding of the observed pattern of SM fermions masses and mixings. The recent observation of deviations from the SM predictions in some $B$-meson decays could be the first hints of a new sector, which might provide an answer to these questions.

With this underlying motivation, in this paper I presented a composite Higgs model constructed from a fundamental fermionic UV description, based on a $\SU(N_{HC})$ gauge group and vectorlike fermions in its fundamental representation.
The approximate global symmetry of the strongly coupled theory is spontaneously broken by the fermion condensate, inducing a symmetry-breaking pattern $\SU(N_F)_L \times \SU(N_F)_R \to \SU(N_F)_V$, from which the Higgs arises as one of the pNGBs.
The Higgs sector of the theory corresponds to the well-studied case of $N_F = 4$ and the Higgs and electroweak phenomenology is analogous to the one of most composite Higgs models.
By adding an extra HC fermion, charged also under both color and the electroweak group, the number of flavours and the global symmetry groups are extended to $N_F = 10$. This allows also the scalar leptoquarks $S_1$ and $S_3$ to be part of the pNGB spectrum and be possible mediators for the observed $B$-physics anomalies.

The model is however not UV complete, the coupling of the composite sector to SM fermions arises from four-fermion operators with structure $\bar\psi_\SM\psi_\SM \bar\Psi_{HC}\Psi_{HC}$, assumed to be generated by some unspecified dynamics at a scale not too far above the confinement scale $\Lambda_{HC}$. The study of such a UV dynamics will be the focus of future work.
An approximate $\SU(2)^5$ flavour symmetry is introduced to protect the theory from unwanted effects in flavour physics and the conservation of a combination of baryon and lepton number is imposed to avoid proton decay. Below the HC confinement scale these operators generate the SM Higgs Yukawa couplings as well as the LQ couplings to fermions. Interestingly enough, the required symmetries forbid a linear coupling to SM fermions for any other pNGB.

These couplings allow the two scalar LQ to contribute to flavour observables. The natural hierarchy $m_{LQ} \ll \Lambda_{HC}$ and the approximate flavour symmetry protect the model from unwanted effects (both in flavour and high-$p_T$ physics) from the heavy resonances.
An important constraint on the LQ couplings is obtained from $\tau \to \mu \gamma$, which imposes a strong limit on the LQ coupling to right-handed fermions. Other relevant bounds are those from $Z$ and $W$ couplings to $\tau$ and $\nu_\tau$, from $B \to K^* \nu\nu$, and $B_s-\bar{B}_s$ mixing. Improvements in any of these will be crucial to test the coupling-structure of the model.
As described in Section~\ref{sec:BanomaliesFit}, while the neutral-current anomalies can be completely addressed, the charged-current ones can be only partially recovered due to the $B_s$ mixing constraint.
Assuming this is solved by a mild tuning with some extra contribution, a residual $\sim 1.5 \sigma$ discrepancy remains in $R(D^{(*)})$ due to tension with the EWPT, see also Ref.~\cite{Buttazzo:2017ixm}.

This difficulty in reproducing to the full extent the $b\to c \tau \bar\nu$ excess is common in many new physics interpretations of the anomalies, suggesting that those in charged-current, if confirmed, might decrease in size when more data will be analysed.

Spurion analysis and NDA estimates allow to study the pNGB potential, generated by the explicit global-symmetry breaking terms. With this, the conditions for a successful electroweak symmetry breaking are identified and the pNGB spectrum can be estimated, see Fig.~\ref{fig:spectrum}.
The leading source of pNGB masses is due to SM gauge contributions. These make the colored states the heaviest ($m_{\SU(3)}\sim 2 - 4 \TeV$), followed by the electroweak ones ($m_{\SU(2)\times \U(1)} \sim 1 -2 \TeV$), and finally the singlets ($m_{\eta} \sim 0.8-1 \TeV$), which receive a mass only from the fundamental HC fermion masses.

In general, such a heavy spectrum might prove challenging to test at the LHC.
The NDA estimate for the masses of the two scalar LQ mediating the flavour anomalies is $m_{S_{1,3}}\sim 1.5 \TeV$.
At present, the most sensitive search channel at the LHC is via pair-production, with decays into third-generation fermions. The most recent limits crossed the $1 \TeV$ threshold but from Fig.~\ref{fig:LQbounds} it is clear that the region in masses and couplings relevant to this model might not be probed even at HL-LHC.

The pNGB with most promising prospects for a LHC observation is the singlet $\eta_3$, which couples to pairs of SM gauge bosons via the anomaly, so that it can be produced in gluon-fusion and decay in two photons. Present diphoton searches already put relevant constraints on the model parameters, as can be seen in Fig.~\ref{fig:eta3_xsec}.
All the other states, including the heavy resonances at the $\Lambda_{HC}$ scale, might require the next generation of colliders in order to be observed.

In the next few years the LHCb and Belle II experiments will provide a conclusive answer as to the nature of the present $B$-physics anomalies. If they will be confirmed as genuine new physics effects, our understanding of Nature at the TeV scale will be revolutionised. In order to uncover the UV dynamics underlying these flavour effects, a multi-pronged approach must be adopted, combining flavour physics, Higgs and electroweak precision measurements, and high-$p_T$ direct searches. This can only be done in concrete UV models, and the composite scenario described here represents an interesting case where all three provide crucial pieces of information.

\subsection*{Acknowledgements}

It is a great pleasure to acknowledge all the discussions and collaborations with D. Buttazzo, A. Greljo, and G. Isidori, which sparkled the main ideas of this work. I am also deeply grateful to A. Azatov, D. Buttazzo, A. Greljo, G. Isidori, and M. Nardecchia for the encouraging support and for precious feedback on the work and the manuscript.

\appendix

\section{Requirements for a UV description}
\label{app:Requirements}

The choices of viable UV theories for a composite Higgs model with a fundamental fermionic strongly interacting theory are constrained by a set of theoretical and phenomenological requirements.

I begin by introducing a new gauge interaction, called \emph{hypercolor} (HC), with a simple group $\GG_{HC}$, and a set of new Weyl fermions $\psi_i$ charged under this symmetry as well as under the SM gauge group:
\be
	\psi_i \sim ({\bf r}^i_{HC}, {\bf r}^i_{c}, {\bf r}^i_{w})_{Y_i}~.
\ee
In the absence of SM gauging and other explicit symmetry-breaking terms, the theory enjoys a global symmetry $G = \SU(n_1) \times \SU(n_2) \times \ldots \times \SU(n_p) \times \U(1)^{p-1}$, where $p$ is the number of different types of representations ${\bf r}_{HC}$ under $\GG_{HC}$ of the HC fermions, each with $n_i$ fermions \cite{Ferretti:2013kya,Cacciapaglia:2014uja}. The number $n_i$ corresponds to the dimension of the SM representation of $\psi_i$. This also fixes the embedding of the SM gauge group in $G$.
I assume hypercolor confines at a scale $\Lambda_{HC}$ and that the theory forms a condensate, which breaks spontaneously the global symmetry $G$ to a subgroup $H$.
The type of spontaneous breaking of the global symmetry depends on the type of representation ${\bf r}^i_{HC}$ of the fermions under $\GG_{HC}$ \cite{Ferretti:2013kya,Cacciapaglia:2014uja}:
\begin{enumerate}
	\item[a] \emph{Complex and vectorlike.} In this case one has $n$ fermions $\psi_i$ in a complex representation and $n$ fermions $\psi^c_i$ in the conjugate one.  The condensate $\langle \psi^c_i \psi_j \rangle = - B_0 f^2 \delta_{ij}$ breaks $\SU(n)_1 \times \SU(n)_2 \rightarrow \SU(n)_D$.
	\item[b] \emph{Real.} In this case $\langle \psi_i \psi_j \rangle = - B_0 f^2 \phi_{ij}$, where $\phi_{ij} = \phi_{ji}$, breaks $\SU(n) \rightarrow \SO(n)$.
	\item[c] \emph{Pseudo-real.} In this case $\langle \psi_i \psi_j \rangle = - B_0 f^2 \phi_{ij}$, with $\phi_{ij} = - \phi_{ji}$, breaks $\SU(n) \rightarrow \Sp(n)$.
\end{enumerate}
For each of these, the minimal scenarios which include the Higgs doublet as pNGB as well as custodial symmetry are listed in Refs.~\cite{Ferretti:2013kya,Cacciapaglia:2014uja,Ferretti:2016upr} (see also the examples in \cite{Schmaltz:2010ac,Ferretti:2014qta,Vecchi:2015fma,Ma:2015gra})\footnote{It should be noted that Refs.~\cite{Ferretti:2013kya,Ferretti:2014qta,Ferretti:2016upr,Vecchi:2015fma} also require partial compositeness for SM fermions via a linear mixing with HC-baryons. This imposes some important restrictions on the possible models.}: a) $\SU(4)_1 \times \SU(4)_2 \rightarrow \SU(4)_D$, b) $\SU(5) \rightarrow \SO(5)$, and c) $\SU(4) \rightarrow \Sp(4)$, which is isomorphic to $\SO(6) \rightarrow \SO(5)$.

The goal is to find the viable hypercolor groups and set of fermions $\psi_i$ which satisfy a number of requirements:
\begin{enumerate}
\item $H \supset \SU(3)_c \times \SU(2)_L \times \SU(2)_R \times \U(1)_X$. This includes the requirement of custodial symmetry, unbroken color, and correct hypercharge assignment.
\item $G/H$ should include at least one Higgs doublet $H \sim ({\bf 1}, {\bf 2})_{1/2}$ and the scalar leptoquarks $S_1 \sim ({\bf \bar{3}}, {\bf 1})_{1/3}$ and $S_3 \sim ({\bf \bar{3}}, {\bf 3})_{1/3}$.
\item Absence of gauge anomalies.

\end{enumerate}

Requiring the presence of scalar leptoquarks among the pNGB, some of the HC fermions have to be colored. In fact, these states arise as $\GG_{HC}$-invariant bilinears of HC fermions: $(\psi_i^c \psi_j)$ in the complex case or $(\psi_i \psi_j)_{s(a)}$ (where $s$ ($a$) stands for the (a)symmetric part) in the (pseudo-)real case.
At the same time, $\SU(3)_c$ should be contained in the unbroken group $H$, i.e. the condensate should be a singlet of color. I aim to find a field content which satisfies these conditions with the smallest number of flavours.
Let us focus first on the HC-fermions which carry color. For minimality I consider only fields in the fundamental of $\SU(3)_c$.

In the case of real or pseudo-real representations of $\GG_{HC}$ it has been recognised \cite{Ferretti:2013kya,Ferretti:2016upr} that a solution, free from gauge anomalies, can be found by introducing two HC-fermions in the fundamental and anti-fundamental of color: $\chi_1 \sim ({\bf 3}, {\bf r}^\chi_w)$, $\chi_2 \sim ({\bf \bar 3}, {\bf \bar r}^\chi_w)$. The global group, $\SU(6 \times \text{dim}({\bf r}^\chi_w))$, is then broken by the $\langle\chi_i \chi_j\rangle$ condensate to $\SO(6 \times \text{dim}({\bf r}_w))$, if $\chi_i$ are in the real representation, or to $\Sp(6 \times \text{dim}({\bf r}_w))$ if they are in the pseudo-real. Of course, the unbroken group should be aligned in the $\SU(3)_c$-invariant direction.
Since $S_{3}$ is charged under $\SU(2)_w$, a way to obtain such a colored triplet must be also found, either by having a non-trivial ${\bf r}^\chi_w$, or by introducing extra triplets.
By inspection it is easy to show that, in either case, the number of flavours quickly grows quite large, being always above at least $14$ once also the Higgs and custodial symmetry requirements are added. One should then also check explicitly that the $S_{1,3}$ representations are indeed present among the broken generators. I do not consider further possible solutions in this direction.

Let us assume instead that the HC fermions carrying color sit in a complex representation of $\GG_{HC}$, coming in vectorlike pairs to guarantee absence of gauge anomalies: $\psi_Q = ({\bf r}_{HC}^Q, {\bf 3}, {\bf r}^Q_w)_{Y_Q}$ and $\psi_Q^c = ({\bf \bar{r}}_{HC}^Q, {\bf \bar{3}}, {\bf \bar{r}}^Q_w)_{-Y_Q}$. The models in Refs.~\cite{Ferretti:2014qta,Vecchi:2015fma} fall in this category.
In order to have the $S_1$ and $S_3$ LQ then requires other uncolored HC fermions in the same complex $\GG_{HC}$ representation: $S_{1,3} \sim (\psi^c_Q \eta)$. 
Particularly appealing is then the case where ${\bf r}_w^Q \sim {\bf 2}$, so that $\eta = \psi_L \sim ({\bf r}_{HC}^Q, {\bf 1}, {\bf 2})$ and it can be identified as the same HC fermion also responsible for the Higgs sector. In this case the minimal number of flavours compatible with points 1) and 2) is $N_F = 10$, giving a symmetry-breaking pattern $\SU(10)_1 \times \SU(10)_2 \to \SU(10)_D$.\footnote{Note that another solution, with same number of flavours, could be obtained by substituting $\psi_Q$ with two fields: $\psi_U = ({\bf r}_{HC}^Q, {\bf 3}, {\bf 1})$ and $\psi_T = ({\bf r}_{HC}^Q, {\bf 1}, {\bf 3})$, in which case the LQs are given by $S_{3} \sim (\psi^c_U \psi_T)$, $S_{1} \sim (\psi^c_U \psi_{E,N})$. The case described in the main text is more minimal in the sense of requiring less irreducible SM representations and gives a closer relation between $S_1$ and $S_3$ since they have the same valence HC-fermions.} The vectorlike character of the theory also ensures automatically that all gauge anomalies vanish.
 This is the solution considered in the main text.

\section{RG evolution of the gauge couplings}
\label{app:RGevolution}

For $n$ Weyl spinors in the fundamental representation of $\GG = \SU(N)$, the 1-loop $\beta$-function of the gauge coupling, $\beta(g) = \mu \frac{d g}{d\mu}$, is given by
\be
	\beta(g) \approx \frac{g^3}{16 \pi^2} b_0 ~, \qquad
	b_0 = - \left( \frac{11}{3} N - \frac{2}{3} \frac{1}{2} n \right)~.
\ee
For the four gauge couplings in this model, with $\GG_{HC} = \SU(N_{HC})$, one has:
\be\begin{split}
	b_0^{HC} &= - \frac{11}{3} N_{HC} + \frac{20}{3} ~,\\
	b_0^{c} &= - 11 + \frac{12 + 4 N_{HC}}{3}~,\\
	b_0^{w} &= - \frac{22}{3} + \frac{12 + 8 N_{HC}}{3}~,\\
	b_0^{Y} &= \frac{20}{3} + \frac{2N_{HC}}{9} (7 - 24 Y_L + 60 Y_L^2)~.\\
\end{split}\ee
While HC is always asymptotically free (for any $N_{HC} \geq 2$), requiring color to be asymptotically free as well fixes the upper bound $N_{HC} \leq 5$. Instead, both $\SU(2)_w$ and $\U(1)_Y$ always grow in the UV.
For $N_{HC} = 3$ and for $Y_L = 0$ or $\frac{1}{2}$ the theory remains perturbative up to the Plank scale. For $Y_L = -\frac{1}{2}$, instead, the hypercharge reaches the Landau pole close to $\sim 10^{12} \GeV$.

\section{$\SU(10)$ generators}
\label{app:SU10gen}

\subsection{List of generators}
\label{app:SU10generators}

The embedding of $\GG_{SM}$ in $\SU(10)_{L,R}$ is diagonal in the two groups and given by the following expression of the HC fermion
\be
	\Psi = \left( \Psi_{Q, 1}^1, \Psi_{Q, 2}^1, \Psi_{Q, 3}^1, \Psi_{Q, 1}^2, \Psi_{Q, 2}^2, \Psi_{Q, 3}^2, \Psi_{L}^1, \Psi_{L}^2, \Psi_{N}, \Psi_{E} \right)~,
\ee
where the index shown explicitely are $\Psi^{\SU(2)_w}_{\SU(3)_c}$~.
All the $\SU(10)$ generators are normalised as $\text{Tr}[T^\alpha T^\beta] = \frac{1}{2} \delta^{\alpha \beta}$. 

In this notation $i = 1, 2, 3$ indicate $\SU(2)_w$ indices while $A = 1, \ldots, 8$ indicate $\SU(3)_c$ indices, both in the adjoint representations.
It is useful to classify the generators in the $\SU(6) \times \SU(4)$ subgroups of $\SU(10)$.
The generators of the $\SU(4)$ subgroup are:
\bea
	T^{i} = \frac{1}{2} \left( \begin{array}{ccc}
		0_{6\times 6} &  0_{6\times 2} & 0_{6\times 2} \\
		0_{2\times 6} & \sigma^i & 0_{2\times 2} \\
		0_{2\times 6} & 0_{2\times 2} & 0_{2\times 2}
	\end{array}\right),
	& &
	T^{3 + i}  = \frac{1}{2} \left( \begin{array}{ccc}
		0_{6\times 6} &  0_{6\times 2} & 0_{6\times 2} \\
		0_{2\times 6} & 0_{2\times 2} & 0_{2\times 2} \\
		0_{2\times 6} & 0_{2\times 2} & \sigma^i
	\end{array}\right), \nonumber\\
	T^{5 + 2a} = \frac{1}{2} \left( \begin{array}{ccc}
		0_{6\times 6} &  0_{6\times 2} & 0_{6\times 2} \\
		0_{2\times 6} & 0_{2\times 2} & M_{2\times 2}^a  \\
		0_{2\times 6} & (M_{2\times 2}^a)^T  & 0_{2\times 2}
	\end{array}\right),
	& &
	T^{5 + 2a + 1} = \frac{1}{2} \left( \begin{array}{ccc}
		0_{6\times 6} &  0_{6\times 2} & 0_{6\times 2} \\
		0_{2\times 6} & 0_{2\times 2} & i M_{2\times 2}^a  \\
		0_{2\times 6} & -i (M_{2\times 2}^a)^T  & 0_{2\times 2}
	\end{array}\right), \nonumber\\
	T^{15} = \frac{1}{2\sqrt{2}} \left( \begin{array}{ccc}
		0_{6\times 6} &  0_{6\times 2} & 0_{6\times 2} \\
		0_{2\times 6} & 1_{2\times 2} & 0_{2\times 2}  \\
		0_{2\times 6} & 0_{2\times 2}  & - 1_{2\times 2}
	\end{array}\right), &&
\eea
where $a = 1, 2, 3, 4$ and $M_{2\times 2}^a$ are defined as
\be
M^1= \left( \begin{array}{cc}
		0 & 0 \\
		1 & 0
	\end{array}\right), \quad
M^2= \left( \begin{array}{cc}
		1 & 0 \\
		0 & 0
	\end{array}\right), \quad
M^3= \left( \begin{array}{cc}
		0 & 1 \\
		0 & 0
	\end{array}\right), \quad
M^4= \left( \begin{array}{cc}
		0 & 0 \\
		0 & 1
	\end{array}\right)~.
\ee
The $\SU(6)$ generators are (I take the Gell-Mann matrices normalised as $\text{Tr}[\lambda^A \lambda^B] = \delta^{AB}/2$):
\bea
	T^{15 + A} &\!\!\! =& \!\!\! \frac{1}{\sqrt{2}} \left( \begin{array}{ccc}
		\lambda^A & 0_{3\times 3} & 0_{3\times 4}\\
		0_{3\times 3} & \lambda^A & 0_{3\times 4}\\
		0_{4\times 3} & 0_{4\times 3} & 0_{4\times 4}
	\end{array}\right), \quad
	T^{15 + i \times 8 + A} = \frac{1}{\sqrt{2}} \left( \begin{array}{ccc}
		\sigma^i_{11}\lambda^A & \sigma^i_{12}\lambda^A & 0_{3\times 4}\\
		\sigma^i_{21} (\lambda^A)^\dagger & \sigma^i_{22}\lambda^A & 0_{3\times 4}\\
		0_{4\times 3} & 0_{4\times 3} & 0_{4\times 4}
	\end{array}\right), \nonumber\\
	T^{15 + 4 \times 8 + i} &\!\!\! =& \!\!\! \frac{1}{2\sqrt{3}} \left( \begin{array}{ccc}
		\sigma^i_{11} 1_{3\times 3} & \sigma^i_{12} 1_{3\times 3} & 0_{3\times 4}\\
		\sigma^i_{21} 1_{3\times 3} & \sigma^i_{22} 1_{3\times 3} & 0_{3\times 4}\\
		0_{4\times 3} & 0_{4\times 3} & 0_{4\times 4}
	\end{array}\right). \quad 
\eea
The generators associated with the leptoquarks $S_{1,3}$ can be written as
\be\begin{split}
	t_{S_{1,3}^{i,A}} &= \frac{1}{2\sqrt{2}} \left( \begin{array}{ccccc}
		0_{3\times 3} & 0_{3\times 3} & \sigma^i_{11} M_{31}^\alpha & \sigma^i_{12} M_{31}^\alpha & 0_{3\times 2} \\
		0_{3\times 3} & 0_{3\times 3} & \sigma^i_{12} M_{31}^\alpha & \sigma^i_{22} M_{31}^\alpha & 0_{3\times 2} \\
		\sigma^{i,*}_{11} M_{31}^{T,\alpha} & \sigma^{i,*}_{21} M_{31}^{T,\alpha} & 0 & 0 & 0_{1\times 2}\\
		\sigma^{i,*}_{12} M_{31}^{T,\alpha} & \sigma^{i,*}_{22} M_{31}^{T,\alpha} & 0 & 0 & 0_{1\times 2}\\
	\end{array}\right), \\
	t_{(S_{1,3}^{i,A})*} &= \frac{1}{2\sqrt{2}} \left( \begin{array}{ccccc}
		0_{3\times 3} & 0_{3\times 3} & i \sigma^i_{11} M_{31}^\alpha & i \sigma^i_{12} M_{31}^\alpha & 0_{3\times 2} \\
		0_{3\times 3} & 0_{3\times 3} & i \sigma^i_{12} M_{31}^\alpha & i \sigma^i_{22} M_{31}^\alpha & 0_{3\times 2} \\
		-i \sigma^{i,*}_{11} M_{31}^{T,\alpha} & -i \sigma^{i,*}_{21} M_{31}^{T,\alpha} & 0 & 0 & 0_{1\times 2}\\
		-i \sigma^{i,*}_{12} M_{31}^{T,\alpha} & -i \sigma^{i,*}_{22} M_{31}^{T,\alpha} & 0 & 0 & 0_{1\times 2}\\
	\end{array}\right), \\
\end{split}\ee
where $\alpha = 1,2,3$, $(M_{31}^\alpha)_i = \delta^{\alpha,i}$ and for $i = 1,2,3$ this describes the generators associated with $S_3$ while with $i = 4$ (i.e. $\sigma^4 = {\bf 1}$) this describes the generator of $S_1$.
They are assigned as:
\be
	T^{50 + 6 (i - 1) + (2\alpha - 1)} = t_{S_{1,3}^{i,A}} ~, \qquad
	T^{50 + 6 (i - 1) + (2\alpha)} = t_{(S_{1,3}^{i,A})*} ~, \qquad
\ee
The generators associated with the states $T_2$ and $\tilde R_2$ are
\be\begin{split}
	T^{74 + 4 X + (2\alpha - 1)} &= \frac{1}{2} \left( \begin{array}{cccc}
		0_{3\times 3} & 0_{3\times 3} & 0_{3\times 2} & M_{32}^{X + \alpha} \\
		0_{3\times 3} & 0_{3\times 3} & 0_{3\times 2} & 0_{3\times 2} \\
		0_{2\times 3} & 0_{2\times 3} & 0_{2\times 2} & 0_{2\times 2} \\
		M_{32}^{X + \alpha,T} & 0_{2\times 3} & 0_{2\times 2} & 0_{2\times 2} \\
	\end{array}\right), \\
	T^{74 + 4 X + (2\alpha)} &= \frac{1}{2} \left( \begin{array}{cccc}
		0_{3\times 3} & 0_{3\times 3} & 0_{3\times 2} & i M_{32}^{X + \alpha} \\
		0_{3\times 3} & 0_{3\times 3} & 0_{3\times 2} & 0_{3\times 2} \\
		0_{2\times 3} & 0_{2\times 3} & 0_{2\times 2} & 0_{2\times 2} \\
		-i M_{32}^{X + \alpha,T} & 0_{2\times 3} & 0_{2\times 2} & 0_{2\times 2} \\
	\end{array}\right), \\
	T^{74 + 4 X + 6 + (2\alpha - 1)} &= \frac{1}{2} \left( \begin{array}{cccc}
		0_{3\times 3} & 0_{3\times 3} & 0_{3\times 2} & 0_{3\times 2} \\
		0_{3\times 3} & 0_{3\times 3} & 0_{3\times 2} & M_{32}^{X + \alpha} \\
		0_{2\times 3} & 0_{2\times 3} & 0_{2\times 2} & 0_{2\times 2} \\
		0_{2\times 3} & M_{32}^{X + \alpha,T} & 0_{2\times 2} & 0_{2\times 2} \\
	\end{array}\right), \\
	T^{74 + 4 X + 6 + (2\alpha)} &= \frac{1}{2} \left( \begin{array}{cccc}
		0_{3\times 3} & 0_{3\times 3} & 0_{3\times 2} & 0_{3\times 2} \\
		0_{3\times 3} & 0_{3\times 3} & 0_{3\times 2} & i M_{32}^{X + \alpha} \\
		0_{2\times 3} & 0_{2\times 3} & 0_{2\times 2} & 0_{2\times 2} \\
		0_{2\times 3}  & -i M_{32}^{X + \alpha,T} & 0_{2\times 2} & 0_{2\times 2} \\
	\end{array}\right), \\
\end{split}\ee
where $X = 0~(3)$ for $T_2$ $\tilde R_2$, $\alpha = 1,2,3$, and $(M_{32}^b)_{xy} = \delta^{b, 3(y-1) + x}$, with $x=1,2,3$, $y = 1,2$, and $b = 1,\ldots, 6$.

The last generator (which is a singlet under the $\SU(6) \times \SU(4)$ subgroups) is:
\be
	T^{99} = \sqrt{\frac{3}{40}} \left( \begin{array}{cc}
		- \frac{2}{3} 1_{6\times 6} & 0_{6\times 4}\\
		0_{4\times 6} & 1_{4\times 4}
	\end{array}\right)~.
\ee

In this basis, the 3 Cartan generators of $\SU(10)_D$ singlets under $\GG_{SM}$ are (in a block-diagonal notation)
\be\begin{split}
	T^{\eta_1} &= T^{6~} = \left( 0_{6\times 6} \right) \times \left( 0_{2\times 2} \right) \times \frac{1}{2}  \left( \sigma^3 \right) ~, \\
	T^{\eta_2} &= T^{15} = \left( 0_{6\times 6} \right) \times \frac{1}{\sqrt{8}} \left( 1_{2\times 2} \right) \times  \frac{-1}{\sqrt{8}} \left(1_{2\times 2} \right)~, \\
	T^{\eta_3} &=  T^{99} = \frac{-1}{\sqrt{30}} \left( 1_{6\times 6} \right) \times \sqrt{\frac{3}{40}} \left( 1_{4\times 4}\right) ~.
	\label{eq:singlGen}
\end{split}\ee

\subsubsection*{Standard Model gauging}
\label{app:SMgauging}

The SM gauge generators are embedded in the unbroken global symmetry group, $\GG_{SM} \subset \SU(10)_D \times U(1)_X$, and are defined from how they act on the HC fermions $\Psi$, for example $T_Y \Psi_i = Y_i \Psi_i$ and analogously for the $\SU(2)_w$ and $\SU(3)_c$ generators. This implies that, while being defined from $\SU(10)$ generators, they are normalised as $\text{Tr}[T^\alpha_{G_\SM} T^\beta_{G_\SM}] = \delta^{\alpha\beta} n_r/2$, where $n_r$ is the number of irreducible representations of the specific gauge group $G_\SM$ inside $\Psi$. In particular, they are given by:
\be\begin{split}
	T^A_{\SU(3)_c} &= \sqrt{2} \, T^{15 + A}~, \\
	T^i_{\SU(2)_w} &= T^i + \sqrt{3} \, T^{15 + 4 \times 8 + i}~, \\
	T_Y &= T^6 + \frac{2\sqrt{30}}{15} \, T^{99} + \left( Y_L - \frac{1}{5} \right) {\bf 1}_{10\times 10} ~.
	\label{eq:SMgauging}
\end{split}\ee

\subsection{Defining the pNGB}
\label{app:pNGBdef}

The 99 real pNGB are labelled from their association with the respective $\SU(10)$ generator: $\phi^\alpha T^\alpha$. From this, I define those with specific SM quantum numbers, as listed in Eq.~\eqref{eq:pNGB}.

\be
	\eta_1 = \phi^{6}~, \qquad
	\eta_2 = \phi^{15}~, \qquad
	\eta_3 = \phi^{99}~,
\ee
\be
	\tilde\pi_1^A  = \phi^{15 + A}~, \quad
	\tilde\pi_3^{A,i}  = \phi^{15 + i \times 8 + A}~, \quad (A=1,\ldots,8, ~~ i = 1,2,3)\\
\ee
\be
	\Pi_L^i  = \phi^{i}~, \quad
	\Pi_Q^i  = \phi^{15 + 4 \times 8 + i}~,\quad (i = 1,2,3)~, \quad
	\omega^{\pm} = (\phi^4 \mp i \phi^5) /\sqrt{2}~,
\ee
\be
	h_1^{1,4} = -\phi^{7,10}~,\quad
	h_1^{2,3} = \phi^{8,9}~,\quad
	h_2^a = \phi^{10 + a}~,\quad(a=1,\ldots,4)~,
\ee
\be
	s_{3}^{A,\alpha} = \phi^{50 + 6 (A - 1) + \alpha}~,\quad
	s_{1}^{\alpha} = \phi^{50 + 6\times3 + \alpha}~, \quad (A=1,2,3, ~~\alpha = 1, \ldots, 6)
\ee
\be
	t_2^\beta = \phi^{74 + \beta}~, \quad
	\tilde r_2^\beta = \phi^{74 + 12 + \beta}~, \quad (\beta = 1, \ldots, 12)~.
\ee
From these real (except $\omega^\pm$) fields one can get the complex ones as:
\be
	H_{1,2} = \frac{1}{\sqrt{2}}\left(\begin{array}{c} h_{1,2}^1 + i h_{1,2}^2 \\ h_{1,2}^3 + i h_{1,2}^4 \end{array} \right)~, \quad
	\Pi_{L,Q}^+ = \frac{1}{\sqrt{2}} ( \Pi_{L,Q}^1 - i \Pi_{L,Q}^2)~, \quad
	\Pi_{L,Q}^0 = \Pi_{L,Q}^3~,
\ee
\be
	T_2^a = \frac{1}{\sqrt{2}}\left(\begin{array}{c} t_{2}^{5 + 2a} - i t_{2}^{6 + 2a}  \\
				t_{2}^{2a - 1} - i t_{2}^{2a} \end{array} \right)~, \quad
	\tilde R_2^a = \frac{1}{\sqrt{2}}\left(\begin{array}{c} \tilde r_{2}^{2a - 1} + i \tilde r_{2}^{2a}  \\
				\tilde r_{2}^{5 + 2a} + i \tilde r_{2}^{6 + 2a}\end{array} \right)~,
\ee
\be\begin{split}
&	S_1^a = \frac{1}{\sqrt{2}}\left( s_{1}^{2a - 1} - i s_{1}^{2a}  \right)~, \quad
	S_3^{a,+\frac{1}{3}} = \frac{1}{\sqrt{2}}\left( s_{3}^{3, 2a - 1} - i s_{3}^{3, 2a}  \right)~, \\
&	S_3^{a,+\frac{4}{3}} = \frac{1}{2} \left( s_{3}^{1, 2a - 1} - i s_{3}^{2, 2a-1} - i s_{3}^{1, 2a} - s_{3}^{2, 2a}  \right)~, \\
&	S_3^{a,-\frac{2}{3}} = \frac{1}{2} \left( s_{3}^{1, 2a - 1} + i s_{3}^{2, 2a-1} - i s_{3}^{1, 2a} + s_{3}^{2, 2a}  \right)~,
\end{split}\ee
where $a = 1,2,3$ is a $\SU(3)_c$ index.

\subsection{Higgs and Leptoquark spurions}
\label{app:spurions}

The spurions corresponding to the two Higgs doublets are defined, in a block-diagonal matrix notation, as: 
\be\begin{split}
	\Delta_{H_1}^1 = {\bf 1}_{6\times 6} \otimes
	\left( \begin{array}{cccc}
	0 & 0 & 0 & 0 \\
	0 & 0 & -1 & 0 \\
	0 & 0 & 0 & 0 \\
	0 & 0 & 0 & 0
	\end{array}\right)~,& \qquad
	\Delta_{H_1}^2 = {\bf 1}_{6\times 6} \otimes
	\left( \begin{array}{cccc}
	0 & 0 & 1 & 0 \\
	0 & 0 & 0 & 0 \\
	0 & 0 & 0 & 0 \\
	0 & 0 & 0 & 0
	\end{array}\right)~, \\
	\Delta_{H_2}^1 = {\bf 1}_{6\times 6} \otimes
	\left( \begin{array}{cccc}
	0 & 0 & 0 & 0 \\
	0 & 0 & 0 & 0 \\
	0 & 0 & 0 & 0 \\
	1 & 0 & 0 & 0
	\end{array}\right)~,& \qquad
	\Delta_{H_2}^2 = {\bf 1}_{6\times 6} \otimes
	\left( \begin{array}{cccc}
	0 & 0 & 0 & 0 \\
	0 & 0 & 0 & 0 \\
	0 & 0 & 0 & 0 \\
	0 & 1 & 0 & 0
	\end{array}\right)~.
\end{split}\ee
Those of the two scalar leptoquarks, instead, have non-vanishing elements only in the $6\times 2$ block in positions \text{B = [1-6,7-8]} of the $10\times 10$ matrix:
\be\begin{split}
	[\Delta_{S_1}^{a}]_{B}: & 
	\left(
\begin{array}{cc}
 1 & 0 \\
 0 & 0 \\
 0 & 0 \\
 0 & 1 \\
 0 & 0 \\
 0 & 0 \\
\end{array}
\right), ~
	\left(
\begin{array}{cc}
 0 & 0 \\
 1 & 0 \\
 0 & 0 \\
 0 & 0 \\
 0 & 1 \\
 0 & 0 \\
\end{array}
\right), ~
	\left(
\begin{array}{cc}
 0 & 0 \\
 0 & 0 \\
 1 & 0 \\
 0 & 0 \\
 0 & 0 \\
 0 & 1 \\
\end{array}
\right),
\end{split}\ee
\be\begin{split}
	[\Delta_{S_3}^{i \, a}]_{B}: &
	\left(
\begin{array}{cc}
 0 & 1 \\
 0 & 0 \\
 0 & 0 \\
 1 & 0 \\
 0 & 0 \\
 0 & 0 \\
\end{array}
\right), ~ 
	\left(
\begin{array}{cc}
 0 & 0 \\
 0 & 1 \\
 0 & 0 \\
 0 & 0 \\
 1 & 0 \\
 0 & 0 \\
\end{array}
\right), ~ 
	\left(
\begin{array}{cc}
 0 & 0 \\
 0 & 0 \\
 0 & 1 \\
 0 & 0 \\
 0 & 0 \\
 1 & 0 \\
\end{array}
\right), \\
	&
	\left(
\begin{array}{cc}
 0 & -i \\
 0 & 0 \\
 0 & 0 \\
 i & 0 \\
 0 & 0 \\
 0 & 0 \\
\end{array}
\right), ~ 
	\left(
\begin{array}{cc}
 0 & 0 \\
 0 & -i \\
 0 & 0 \\
 0 & 0 \\
 i & 0 \\
 0 & 0 \\
\end{array}
\right), ~ 
	\left(
\begin{array}{cc}
 0 & 0 \\
 0 & 0 \\
 0 & -i \\
 0 & 0 \\
 0 & 0 \\
 i & 0 \\
\end{array}
\right), \\
	&
	\left(
\begin{array}{cc}
 1 & 0 \\
 0 & 0 \\
 0 & 0 \\
 0 & -1 \\
 0 & 0 \\
 0 & 0 \\
\end{array}
\right), ~ 
	\left(
\begin{array}{cc}
 0 & 0 \\
 1 & 0 \\
 0 & 0 \\
 0 & 0 \\
 0 & -1 \\
 0 & 0 \\
\end{array}
\right), ~ 
	\left(
\begin{array}{cc}
 0 & 0 \\
 0 & 0 \\
 1 & 0 \\
 0 & 0 \\
 0 & 0 \\
 0 & -1 \\
\end{array}
\right),
\end{split}\ee
where $a = 1,2,3$ (columns) is a color index while $i = 1,2,3$ (rows) is a $\SU(2)_w$ index in the adjoint.

\bibliographystyle{JHEP}

{\footnotesize
\bibliography{biblio}}

\providecommand{\href}[2]{#2}\begingroup\raggedright\begin{thebibliography}{100}

\bibitem{Lees:2012xj}
{\bf BaBar} Collaboration, J.~P. Lees et~al. {\em Phys. Rev. Lett.} {\bf 109}
  (2012) 101802, [\href{http://arxiv.org/abs/1205.5442}{{\tt
  arXiv:1205.5442}}].

\bibitem{Lees:2013uzd}
{\bf BaBar} Collaboration, J.~P. Lees et~al. {\em Phys. Rev.} {\bf D88} (2013),
  no.~7 072012, [\href{http://arxiv.org/abs/1303.0571}{{\tt arXiv:1303.0571}}].

\bibitem{Aaij:2015yra}
{\bf LHCb} Collaboration, R.~Aaij et~al. {\em Phys. Rev. Lett.} {\bf 115}
  (2015), no.~11 111803, [\href{http://arxiv.org/abs/1506.08614}{{\tt
  arXiv:1506.08614}}]. [Addendum: Phys. Rev. Lett.115,no.15,159901(2015)].

\bibitem{Huschle:2015rga}
{\bf Belle} Collaboration, M.~Huschle et~al. {\em Phys. Rev.} {\bf D92} (2015),
  no.~7 072014, [\href{http://arxiv.org/abs/1507.03233}{{\tt
  arXiv:1507.03233}}].

\bibitem{Sato:2016svk}
{\bf Belle} Collaboration, Y.~Sato et~al. {\em Phys. Rev.} {\bf D94} (2016),
  no.~7 072007, [\href{http://arxiv.org/abs/1607.07923}{{\tt
  arXiv:1607.07923}}].

\bibitem{Hirose:2016wfn}
{\bf Belle} Collaboration, S.~Hirose et~al. {\em Phys. Rev. Lett.} {\bf 118}
  (2017), no.~21 211801, [\href{http://arxiv.org/abs/1612.00529}{{\tt
  arXiv:1612.00529}}].

\bibitem{Amhis:2016xyh}
{\bf HFLAV} Collaboration, Y.~Amhis et~al. {\em Eur. Phys. J.} {\bf C77}
  (2017), no.~12 895, [\href{http://arxiv.org/abs/1612.07233}{{\tt
  arXiv:1612.07233}}].

\bibitem{Aaij:2013qta}
{\bf LHCb} Collaboration, R.~Aaij et~al. {\em Phys. Rev. Lett.} {\bf 111}
  (2013) 191801, [\href{http://arxiv.org/abs/1308.1707}{{\tt
  arXiv:1308.1707}}].

\bibitem{Aaij:2015oid}
{\bf LHCb} Collaboration, R.~Aaij et~al. {\em JHEP} {\bf 02} (2016) 104,
  [\href{http://arxiv.org/abs/1512.04442}{{\tt arXiv:1512.04442}}].

\bibitem{Ciuchini:2015qxb}
M.~Ciuchini, M.~Fedele, E.~Franco, S.~Mishima, A.~Paul, L.~Silvestrini, and
  M.~Valli {\em JHEP} {\bf 06} (2016) 116,
  [\href{http://arxiv.org/abs/1512.07157}{{\tt arXiv:1512.07157}}].

\bibitem{Aaij:2014ora}
{\bf LHCb} Collaboration, R.~Aaij et~al. {\em Phys. Rev. Lett.} {\bf 113}
  (2014) 151601, [\href{http://arxiv.org/abs/1406.6482}{{\tt
  arXiv:1406.6482}}].

\bibitem{Aaij:2017vbb}
{\bf LHCb} Collaboration, R.~Aaij et~al. {\em JHEP} {\bf 08} (2017) 055,
  [\href{http://arxiv.org/abs/1705.05802}{{\tt arXiv:1705.05802}}].

\bibitem{Altmannshofer:2015sma}
W.~Altmannshofer and D.~M. Straub in {\em {Proceedings, 50th Rencontres de
  Moriond Electroweak Interactions and Unified Theories: La Thuile, Italy,
  March 14-21, 2015}}, pp.~333--338, 2015.
\newblock \href{http://arxiv.org/abs/1503.06199}{{\tt arXiv:1503.06199}}.

\bibitem{Descotes-Genon:2015uva}
S.~Descotes-Genon, L.~Hofer, J.~Matias, and J.~Virto {\em JHEP} {\bf 06} (2016)
  092, [\href{http://arxiv.org/abs/1510.04239}{{\tt arXiv:1510.04239}}].

\bibitem{Altmannshofer:2017yso}
W.~Altmannshofer, P.~Stangl, and D.~M. Straub
  \href{http://arxiv.org/abs/1704.05435}{{\tt arXiv:1704.05435}}.

\bibitem{DAmico:2017mtc}
G.~D'Amico, M.~Nardecchia, P.~Panci, F.~Sannino, A.~Strumia, R.~Torre, and
  A.~Urbano \href{http://arxiv.org/abs/1704.05438}{{\tt arXiv:1704.05438}}.

\bibitem{Capdevila:2017bsm}
B.~Capdevila, A.~Crivellin, S.~Descotes-Genon, J.~Matias, and J.~Virto
  \href{http://arxiv.org/abs/1704.05340}{{\tt arXiv:1704.05340}}.

\bibitem{Ciuchini:2017mik}
M.~Ciuchini, A.~M. Coutinho, M.~Fedele, E.~Franco, A.~Paul, L.~Silvestrini, and
  M.~Valli {\em Eur. Phys. J.} {\bf C77} (2017), no.~10 688,
  [\href{http://arxiv.org/abs/1704.05447}{{\tt arXiv:1704.05447}}].

\bibitem{Hiller:2017bzc}
G.~Hiller and I.~Nisandzic {\em Phys. Rev.} {\bf D96} (2017), no.~3 035003,
  [\href{http://arxiv.org/abs/1704.05444}{{\tt arXiv:1704.05444}}].

\bibitem{Bhattacharya:2014wla}
B.~Bhattacharya, A.~Datta, D.~London, and S.~Shivashankara {\em Phys. Lett.}
  {\bf B742} (2015) 370--374, [\href{http://arxiv.org/abs/1412.7164}{{\tt
  arXiv:1412.7164}}].

\bibitem{Alonso:2015sja}
R.~Alonso, B.~Grinstein, and J.~Martin~Camalich {\em JHEP} {\bf 10} (2015) 184,
  [\href{http://arxiv.org/abs/1505.05164}{{\tt arXiv:1505.05164}}].

\bibitem{Greljo:2015mma}
A.~Greljo, G.~Isidori, and D.~Marzocca {\em JHEP} {\bf 07} (2015) 142,
  [\href{http://arxiv.org/abs/1506.01705}{{\tt arXiv:1506.01705}}].

\bibitem{Calibbi:2015kma}
L.~Calibbi, A.~Crivellin, and T.~Ota {\em Phys. Rev. Lett.} {\bf 115} (2015)
  181801, [\href{http://arxiv.org/abs/1506.02661}{{\tt arXiv:1506.02661}}].

\bibitem{Bauer:2015knc}
M.~Bauer and M.~Neubert {\em Phys. Rev. Lett.} {\bf 116} (2016), no.~14 141802,
  [\href{http://arxiv.org/abs/1511.01900}{{\tt arXiv:1511.01900}}].

\bibitem{Fajfer:2015ycq}
S.~Fajfer and N.~Kosnik {\em Phys. Lett.} {\bf B755} (2016) 270--274,
  [\href{http://arxiv.org/abs/1511.06024}{{\tt arXiv:1511.06024}}].

\bibitem{Barbieri:2015yvd}
R.~Barbieri, G.~Isidori, A.~Pattori, and F.~Senia {\em Eur. Phys. J.} {\bf C76}
  (2016), no.~2 67, [\href{http://arxiv.org/abs/1512.01560}{{\tt
  arXiv:1512.01560}}].

\bibitem{Buttazzo:2016kid}
D.~Buttazzo, A.~Greljo, G.~Isidori, and D.~Marzocca {\em JHEP} {\bf 08} (2016)
  035, [\href{http://arxiv.org/abs/1604.03940}{{\tt arXiv:1604.03940}}].

\bibitem{Das:2016vkr}
D.~Das, C.~Hati, G.~Kumar, and N.~Mahajan {\em Phys. Rev.} {\bf D94} (2016)
  055034, [\href{http://arxiv.org/abs/1605.06313}{{\tt arXiv:1605.06313}}].

\bibitem{Boucenna:2016qad}
S.~M. Boucenna, A.~Celis, J.~Fuentes-Martin, A.~Vicente, and J.~Virto {\em
  JHEP} {\bf 12} (2016) 059, [\href{http://arxiv.org/abs/1608.01349}{{\tt
  arXiv:1608.01349}}].

\bibitem{Becirevic:2016yqi}
D.~Becirevic, S.~Fajfer, N.~Kosnik, and O.~Sumensari {\em Phys. Rev.} {\bf D94}
  (2016), no.~11 115021, [\href{http://arxiv.org/abs/1608.08501}{{\tt
  arXiv:1608.08501}}].

\bibitem{Hiller:2016kry}
G.~Hiller, D.~Loose, and K.~Schoenwald {\em JHEP} {\bf 12} (2016) 027,
  [\href{http://arxiv.org/abs/1609.08895}{{\tt arXiv:1609.08895}}].

\bibitem{Bhattacharya:2016mcc}
B.~Bhattacharya, A.~Datta, J.-P. Gu\'evin, D.~London, and R.~Watanabe {\em
  JHEP} {\bf 01} (2017) 015, [\href{http://arxiv.org/abs/1609.09078}{{\tt
  arXiv:1609.09078}}].

\bibitem{Barbieri:2016las}
R.~Barbieri, C.~W. Murphy, and F.~Senia {\em Eur. Phys. J.} {\bf C77} (2017),
  no.~1 8, [\href{http://arxiv.org/abs/1611.04930}{{\tt arXiv:1611.04930}}].

\bibitem{Becirevic:2016oho}
D.~Becirevic, N.~Kosnik, O.~Sumensari, and R.~Zukanovich~Funchal {\em JHEP}
  {\bf 11} (2016) 035, [\href{http://arxiv.org/abs/1608.07583}{{\tt
  arXiv:1608.07583}}].

\bibitem{Bordone:2017anc}
M.~Bordone, G.~Isidori, and S.~Trifinopoulos {\em Phys. Rev.} {\bf D96} (2017),
  no.~1 015038, [\href{http://arxiv.org/abs/1702.07238}{{\tt
  arXiv:1702.07238}}].

\bibitem{Megias:2017ove}
E.~Megias, M.~Quiros, and L.~Salas {\em JHEP} {\bf 07} (2017) 102,
  [\href{http://arxiv.org/abs/1703.06019}{{\tt arXiv:1703.06019}}].

\bibitem{Crivellin:2017zlb}
A.~Crivellin, D.~M{\"u}ller, and T.~Ota {\em JHEP} {\bf 09} (2017) 040,
  [\href{http://arxiv.org/abs/1703.09226}{{\tt arXiv:1703.09226}}].

\bibitem{Cai:2017wry}
Y.~Cai, J.~Gargalionis, M.~A. Schmidt, and R.~R. Volkas
  \href{http://arxiv.org/abs/1704.05849}{{\tt arXiv:1704.05849}}.

\bibitem{Altmannshofer:2017poe}
W.~Altmannshofer, P.~S. Bhupal~Dev, and A.~Soni {\em Phys. Rev.} {\bf D96}
  (2017), no.~9 095010, [\href{http://arxiv.org/abs/1704.06659}{{\tt
  arXiv:1704.06659}}].

\bibitem{Sannino:2017utc}
F.~Sannino, P.~Stangl, D.~M. Straub, and A.~E. Thomsen
  \href{http://arxiv.org/abs/1712.07646}{{\tt arXiv:1712.07646}}.

\bibitem{Faroughy:2016osc}
D.~A. Faroughy, A.~Greljo, and J.~F. Kamenik {\em Phys. Lett.} {\bf B764}
  (2017) 126--134, [\href{http://arxiv.org/abs/1609.07138}{{\tt
  arXiv:1609.07138}}].

\bibitem{Feruglio:2016gvd}
F.~Feruglio, P.~Paradisi, and A.~Pattori {\em Phys. Rev. Lett.} {\bf 118}
  (2017), no.~1 011801, [\href{http://arxiv.org/abs/1606.00524}{{\tt
  arXiv:1606.00524}}].

\bibitem{Feruglio:2017rjo}
F.~Feruglio, P.~Paradisi, and A.~Pattori {\em JHEP} {\bf 09} (2017) 061,
  [\href{http://arxiv.org/abs/1705.00929}{{\tt arXiv:1705.00929}}].

\bibitem{Cornella:2018tfd}
C.~Cornella, F.~Feruglio, and P.~Paradisi
  \href{http://arxiv.org/abs/1803.00945}{{\tt arXiv:1803.00945}}.

\bibitem{Buttazzo:2017ixm}
D.~Buttazzo, A.~Greljo, G.~Isidori, and D.~Marzocca {\em JHEP} {\bf 11} (2017)
  044, [\href{http://arxiv.org/abs/1706.07808}{{\tt arXiv:1706.07808}}].

\bibitem{Assad:2017iib}
N.~Assad, B.~Fornal, and B.~Grinstein {\em Phys. Lett.} {\bf B777} (2018)
  324--331, [\href{http://arxiv.org/abs/1708.06350}{{\tt arXiv:1708.06350}}].

\bibitem{Calibbi:2017qbu}
L.~Calibbi, A.~Crivellin, and T.~Li \href{http://arxiv.org/abs/1709.00692}{{\tt
  arXiv:1709.00692}}.

\bibitem{DiLuzio:2017vat}
L.~Di~Luzio, A.~Greljo, and M.~Nardecchia {\em Phys. Rev.} {\bf D96} (2017),
  no.~11 115011, [\href{http://arxiv.org/abs/1708.08450}{{\tt
  arXiv:1708.08450}}].

\bibitem{Bordone:2017bld}
M.~Bordone, C.~Cornella, J.~Fuentes-Martin, and G.~Isidori {\em Phys. Lett.}
  {\bf B779} (2018) 317--323, [\href{http://arxiv.org/abs/1712.01368}{{\tt
  arXiv:1712.01368}}].

\bibitem{Greljo:2018tuh}
A.~Greljo and B.~A. Stefanek {\em Phys. Lett.} {\bf B782} (2018) 131--138,
  [\href{http://arxiv.org/abs/1802.04274}{{\tt arXiv:1802.04274}}].

\bibitem{Blanke:2018sro}
M.~Blanke and A.~Crivellin \href{http://arxiv.org/abs/1801.07256}{{\tt
  arXiv:1801.07256}}.

\bibitem{Barbieri:2017tuq}
R.~Barbieri and A.~Tesi {\em Eur. Phys. J.} {\bf C78} (2018), no.~3 193,
  [\href{http://arxiv.org/abs/1712.06844}{{\tt arXiv:1712.06844}}].

\bibitem{Cline:2017aed}
J.~M. Cline {\em Phys. Rev.} {\bf D97} (2018), no.~1 015013,
  [\href{http://arxiv.org/abs/1710.02140}{{\tt arXiv:1710.02140}}].

\bibitem{Gripaios:2009dq}
B.~Gripaios {\em JHEP} {\bf 02} (2010) 045,
  [\href{http://arxiv.org/abs/0910.1789}{{\tt arXiv:0910.1789}}].

\bibitem{Gripaios:2014tna}
B.~Gripaios, M.~Nardecchia, and S.~A. Renner {\em JHEP} {\bf 05} (2015) 006,
  [\href{http://arxiv.org/abs/1412.1791}{{\tt arXiv:1412.1791}}].

\bibitem{Georgi:1984af}
H.~Georgi and D.~B. Kaplan {\em Phys. Lett.} {\bf B145} (1984) 216--220.

\bibitem{Kaplan:1983fs}
D.~B. Kaplan and H.~Georgi {\em Phys. Lett.} {\bf B136} (1984) 183--186.

\bibitem{Sakaki:2013bfa}
Y.~Sakaki, M.~Tanaka, A.~Tayduganov, and R.~Watanabe {\em Phys. Rev.} {\bf D88}
  (2013), no.~9 094012, [\href{http://arxiv.org/abs/1309.0301}{{\tt
  arXiv:1309.0301}}].

\bibitem{Hiller:2014yaa}
G.~Hiller and M.~Schmaltz {\em Phys. Rev.} {\bf D90} (2014) 054014,
  [\href{http://arxiv.org/abs/1408.1627}{{\tt arXiv:1408.1627}}].

\bibitem{Dorsner:2017ufx}
I.~Dor{\v s}ner, S.~Fajfer, D.~A. Faroughy, and N.~Ko{\v s}nik {\em JHEP} {\bf
  10} (2017) 188, [\href{http://arxiv.org/abs/1706.07779}{{\tt
  arXiv:1706.07779}}].

\bibitem{Fajfer:2018bfj}
S.~Fajfer, N.~Ko{\v s}nik, and L.~Vale~Silva {\em Eur. Phys. J.} {\bf C78}
  (2018), no.~4 275, [\href{http://arxiv.org/abs/1802.00786}{{\tt
  arXiv:1802.00786}}].

\bibitem{Galloway:2010bp}
J.~Galloway, J.~A. Evans, M.~A. Luty, and R.~A. Tacchi {\em JHEP} {\bf 10}
  (2010) 086, [\href{http://arxiv.org/abs/1001.1361}{{\tt arXiv:1001.1361}}].

\bibitem{Schmaltz:2010ac}
M.~Schmaltz, D.~Stolarski, and J.~Thaler {\em JHEP} {\bf 09} (2010) 018,
  [\href{http://arxiv.org/abs/1006.1356}{{\tt arXiv:1006.1356}}].

\bibitem{Barnard:2013zea}
J.~Barnard, T.~Gherghetta, and T.~S. Ray {\em JHEP} {\bf 02} (2014) 002,
  [\href{http://arxiv.org/abs/1311.6562}{{\tt arXiv:1311.6562}}].

\bibitem{Ferretti:2013kya}
G.~Ferretti and D.~Karateev {\em JHEP} {\bf 03} (2014) 077,
  [\href{http://arxiv.org/abs/1312.5330}{{\tt arXiv:1312.5330}}].

\bibitem{Cacciapaglia:2014uja}
G.~Cacciapaglia and F.~Sannino {\em JHEP} {\bf 04} (2014) 111,
  [\href{http://arxiv.org/abs/1402.0233}{{\tt arXiv:1402.0233}}].

\bibitem{Ferretti:2014qta}
G.~Ferretti {\em JHEP} {\bf 06} (2014) 142,
  [\href{http://arxiv.org/abs/1404.7137}{{\tt arXiv:1404.7137}}].

\bibitem{Vecchi:2015fma}
L.~Vecchi {\em JHEP} {\bf 02} (2017) 094,
  [\href{http://arxiv.org/abs/1506.00623}{{\tt arXiv:1506.00623}}].

\bibitem{Ma:2015gra}
T.~Ma and G.~Cacciapaglia {\em JHEP} {\bf 03} (2016) 211,
  [\href{http://arxiv.org/abs/1508.07014}{{\tt arXiv:1508.07014}}].

\bibitem{Ferretti:2016upr}
G.~Ferretti {\em JHEP} {\bf 06} (2016) 107,
  [\href{http://arxiv.org/abs/1604.06467}{{\tt arXiv:1604.06467}}].

\bibitem{Dimopoulos:1979qi}
S.~Dimopoulos and L.~Susskind {\em PRINT-79-0196 (COLUMBIA)} (1979).

\bibitem{Luty:2004ye}
M.~A. Luty and T.~Okui {\em JHEP} {\bf 09} (2006) 070,
  [\href{http://arxiv.org/abs/hep-ph/0409274}{{\tt hep-ph/0409274}}].

\bibitem{Luty:2008vs}
M.~A. Luty {\em JHEP} {\bf 04} (2009) 050,
  [\href{http://arxiv.org/abs/0806.1235}{{\tt arXiv:0806.1235}}].

\bibitem{Hasenfratz:2009ea}
A.~Hasenfratz {\em Phys. Rev.} {\bf D80} (2009) 034505,
  [\href{http://arxiv.org/abs/0907.0919}{{\tt arXiv:0907.0919}}].

\bibitem{Fodor:2009wk}
Z.~Fodor, K.~Holland, J.~Kuti, D.~Nogradi, and C.~Schroeder {\em Phys. Lett.}
  {\bf B681} (2009) 353--361, [\href{http://arxiv.org/abs/0907.4562}{{\tt
  arXiv:0907.4562}}].

\bibitem{Aoki:2012eq}
Y.~Aoki, T.~Aoyama, M.~Kurachi, T.~Maskawa, K.-i. Nagai, H.~Ohki, A.~Shibata,
  K.~Yamawaki, and T.~Yamazaki {\em Phys. Rev.} {\bf D86} (2012) 054506,
  [\href{http://arxiv.org/abs/1207.3060}{{\tt arXiv:1207.3060}}].

\bibitem{Aoki:2013xza}
{\bf LatKMI} Collaboration, Y.~Aoki, T.~Aoyama, M.~Kurachi, T.~Maskawa, K.-i.
  Nagai, H.~Ohki, A.~Shibata, K.~Yamawaki, and T.~Yamazaki {\em Phys. Rev.}
  {\bf D87} (2013), no.~9 094511, [\href{http://arxiv.org/abs/1302.6859}{{\tt
  arXiv:1302.6859}}].

\bibitem{Hasenfratz:2014rna}
A.~Hasenfratz, D.~Schaich, and A.~Veernala {\em JHEP} {\bf 06} (2015) 143,
  [\href{http://arxiv.org/abs/1410.5886}{{\tt arXiv:1410.5886}}].

\bibitem{Appelquist:2014zsa}
{\bf LSD} Collaboration, T.~Appelquist et~al. {\em Phys. Rev.} {\bf D90}
  (2014), no.~11 114502, [\href{http://arxiv.org/abs/1405.4752}{{\tt
  arXiv:1405.4752}}].

\bibitem{Aoki:2014oha}
{\bf LatKMI} Collaboration, Y.~Aoki et~al. {\em Phys. Rev.} {\bf D89} (2014)
  111502, [\href{http://arxiv.org/abs/1403.5000}{{\tt arXiv:1403.5000}}].

\bibitem{Fodor:2015baa}
Z.~Fodor, K.~Holland, J.~Kuti, S.~Mondal, D.~Nogradi, and C.~H. Wong {\em JHEP}
  {\bf 06} (2015) 019, [\href{http://arxiv.org/abs/1503.01132}{{\tt
  arXiv:1503.01132}}].

\bibitem{Raby:1979my}
S.~Raby, S.~Dimopoulos, and L.~Susskind {\em Nucl. Phys.} {\bf B169} (1980)
  373--383.

\bibitem{Preskill:1981sr}
J.~Preskill and S.~Weinberg {\em Phys. Rev.} {\bf D24} (1981) 1059.

\bibitem{Vafa:1983tf}
C.~Vafa and E.~Witten {\em Nucl. Phys.} {\bf B234} (1984) 173--188.

\bibitem{Gasser:1983yg}
J.~Gasser and H.~Leutwyler {\em Annals Phys.} {\bf 158} (1984) 142.

\bibitem{Pich:1995bw}
A.~Pich {\em Rept. Prog. Phys.} {\bf 58} (1995) 563--610,
  [\href{http://arxiv.org/abs/hep-ph/9502366}{{\tt hep-ph/9502366}}].

\bibitem{Sannino:2016sfx}
F.~Sannino, A.~Strumia, A.~Tesi, and E.~Vigiani {\em JHEP} {\bf 11} (2016) 029,
  [\href{http://arxiv.org/abs/1607.01659}{{\tt arXiv:1607.01659}}].

\bibitem{Coleman:1969sm}
S.~R. Coleman, J.~Wess, and B.~Zumino {\em Phys. Rev.} {\bf 177} (1969)
  2239--2247.

\bibitem{Callan:1969sn}
C.~G. Callan, Jr., S.~R. Coleman, J.~Wess, and B.~Zumino {\em Phys. Rev.} {\bf
  177} (1969) 2247--2250.

\bibitem{Mrazek:2011iu}
J.~Mrazek, A.~Pomarol, R.~Rattazzi, M.~Redi, J.~Serra, and A.~Wulzer {\em Nucl.
  Phys.} {\bf B853} (2011) 1--48, [\href{http://arxiv.org/abs/1105.5403}{{\tt
  arXiv:1105.5403}}].

\bibitem{Manohar:1983md}
A.~Manohar and H.~Georgi {\em Nucl. Phys.} {\bf B234} (1984) 189--212.

\bibitem{Panico:2011pw}
G.~Panico and A.~Wulzer {\em JHEP} {\bf 09} (2011) 135,
  [\href{http://arxiv.org/abs/1106.2719}{{\tt arXiv:1106.2719}}].

\bibitem{Contino:2011np}
R.~Contino, D.~Marzocca, D.~Pappadopulo, and R.~Rattazzi {\em JHEP} {\bf 10}
  (2011) 081, [\href{http://arxiv.org/abs/1109.1570}{{\tt arXiv:1109.1570}}].

\bibitem{Kaplan:1991dc}
D.~B. Kaplan {\em Nucl. Phys.} {\bf B365} (1991) 259--278.

\bibitem{Matsedonskyi:2012ym}
O.~Matsedonskyi, G.~Panico, and A.~Wulzer {\em JHEP} {\bf 01} (2013) 164,
  [\href{http://arxiv.org/abs/1204.6333}{{\tt arXiv:1204.6333}}].

\bibitem{Redi:2012ha}
M.~Redi and A.~Tesi {\em JHEP} {\bf 10} (2012) 166,
  [\href{http://arxiv.org/abs/1205.0232}{{\tt arXiv:1205.0232}}].

\bibitem{Marzocca:2012zn}
D.~Marzocca, M.~Serone, and J.~Shu {\em JHEP} {\bf 08} (2012) 013,
  [\href{http://arxiv.org/abs/1205.0770}{{\tt arXiv:1205.0770}}].

\bibitem{Caracciolo:2012je}
F.~Caracciolo, A.~Parolini, and M.~Serone {\em JHEP} {\bf 02} (2013) 066,
  [\href{http://arxiv.org/abs/1211.7290}{{\tt arXiv:1211.7290}}].

\bibitem{Marzocca:2013fza}
D.~Marzocca, A.~Parolini, and M.~Serone {\em JHEP} {\bf 03} (2014) 099,
  [\href{http://arxiv.org/abs/1312.5664}{{\tt arXiv:1312.5664}}].

\bibitem{Agashe:2004rs}
K.~Agashe, R.~Contino, and A.~Pomarol {\em Nucl. Phys.} {\bf B719} (2005)
  165--187, [\href{http://arxiv.org/abs/hep-ph/0412089}{{\tt hep-ph/0412089}}].

\bibitem{Dimopoulos:1979es}
S.~Dimopoulos and L.~Susskind {\em Nucl. Phys.} {\bf B155} (1979) 237--252.
  [2,930(1979)].

\bibitem{Eichten:1979ah}
E.~Eichten and K.~D. Lane {\em Phys. Lett.} {\bf 90B} (1980) 125--130.

\bibitem{Hill:2002ap}
C.~T. Hill and E.~H. Simmons {\em Phys. Rept.} {\bf 381} (2003) 235--402,
  [\href{http://arxiv.org/abs/hep-ph/0203079}{{\tt hep-ph/0203079}}]. [Erratum:
  Phys. Rept.390,553(2004)].

\bibitem{Lane:2002wv}
K.~Lane \href{http://arxiv.org/abs/hep-ph/0202255}{{\tt hep-ph/0202255}}.

\bibitem{Barbieri:2011ci}
R.~Barbieri, G.~Isidori, J.~Jones-Perez, P.~Lodone, and D.~M. Straub {\em Eur.
  Phys. J.} {\bf C71} (2011) 1725, [\href{http://arxiv.org/abs/1105.2296}{{\tt
  arXiv:1105.2296}}].

\bibitem{Barbieri:2012uh}
R.~Barbieri, D.~Buttazzo, F.~Sala, and D.~M. Straub {\em JHEP} {\bf 07} (2012)
  181, [\href{http://arxiv.org/abs/1203.4218}{{\tt arXiv:1203.4218}}].

\bibitem{Barbieri:2012tu}
R.~Barbieri, D.~Buttazzo, F.~Sala, D.~M. Straub, and A.~Tesi {\em JHEP} {\bf
  05} (2013) 069, [\href{http://arxiv.org/abs/1211.5085}{{\tt
  arXiv:1211.5085}}].

\bibitem{deBlas:2017xtg}
J.~de~Blas, J.~C. Criado, M.~Perez-Victoria, and J.~Santiago {\em JHEP} {\bf
  03} (2018) 109, [\href{http://arxiv.org/abs/1711.10391}{{\tt
  arXiv:1711.10391}}].

\bibitem{Witten:1983ut}
E.~Witten {\em Phys. Rev. Lett.} {\bf 51} (1983) 2351.

\bibitem{Giudice:2007fh}
G.~F. Giudice, C.~Grojean, A.~Pomarol, and R.~Rattazzi {\em JHEP} {\bf 06}
  (2007) 045, [\href{http://arxiv.org/abs/hep-ph/0703164}{{\tt
  hep-ph/0703164}}].

\bibitem{Agashe:2006at}
K.~Agashe, R.~Contino, L.~Da~Rold, and A.~Pomarol {\em Phys. Lett.} {\bf B641}
  (2006) 62--66, [\href{http://arxiv.org/abs/hep-ph/0605341}{{\tt
  hep-ph/0605341}}].

\bibitem{Dorsner:2016wpm}
I.~Dor{\v s}ner, S.~Fajfer, A.~Greljo, J.~F. Kamenik, and N.~Ko{\v s}nik {\em
  Phys. Rept.} {\bf 641} (2016) 1--68,
  [\href{http://arxiv.org/abs/1603.04993}{{\tt arXiv:1603.04993}}].

\bibitem{Olive:2016xmw}
{\bf Particle Data Group} Collaboration, C.~Patrignani et~al. {\em Chin. Phys.}
  {\bf C40} (2016), no.~10 100001.

\bibitem{Alonso:2016oyd}
R.~Alonso, B.~Grinstein, and J.~Martin~Camalich {\em Phys. Rev. Lett.} {\bf
  118} (2017), no.~8 081802, [\href{http://arxiv.org/abs/1611.06676}{{\tt
  arXiv:1611.06676}}].

\bibitem{Jung:2018lfu}
M.~Jung and D.~M. Straub \href{http://arxiv.org/abs/1801.01112}{{\tt
  arXiv:1801.01112}}.

\bibitem{Becirevic:2015asa}
D.~Be{\v c}irevi{\'c}, S.~Fajfer, and N.~Ko{\v s}nik {\em Phys. Rev.} {\bf D92}
  (2015), no.~1 014016, [\href{http://arxiv.org/abs/1503.09024}{{\tt
  arXiv:1503.09024}}].

\bibitem{Altmannshofer:2009ma}
W.~Altmannshofer, A.~J. Buras, D.~M. Straub, and M.~Wick {\em JHEP} {\bf 04}
  (2009) 022, [\href{http://arxiv.org/abs/0902.0160}{{\tt arXiv:0902.0160}}].

\bibitem{Buras:2014fpa}
A.~J. Buras, J.~Girrbach-Noe, C.~Niehoff, and D.~M. Straub {\em JHEP} {\bf 02}
  (2015) 184, [\href{http://arxiv.org/abs/1409.4557}{{\tt arXiv:1409.4557}}].

\bibitem{Bobeth:2017ecx}
C.~Bobeth and A.~J. Buras {\em JHEP} {\bf 02} (2018) 101,
  [\href{http://arxiv.org/abs/1712.01295}{{\tt arXiv:1712.01295}}].

\bibitem{DiLuzio:2017fdq}
L.~Di~Luzio, M.~Kirk, and A.~Lenz {\em Phys. Rev.} {\bf D97} (2018), no.~9
  095035, [\href{http://arxiv.org/abs/1712.06572}{{\tt arXiv:1712.06572}}].

\bibitem{Jenkins:2013wua}
E.~E. Jenkins, A.~V. Manohar, and M.~Trott {\em JHEP} {\bf 01} (2014) 035,
  [\href{http://arxiv.org/abs/1310.4838}{{\tt arXiv:1310.4838}}].

\bibitem{Efrati:2015eaa}
A.~Efrati, A.~Falkowski, and Y.~Soreq {\em JHEP} {\bf 07} (2015) 018,
  [\href{http://arxiv.org/abs/1503.07872}{{\tt arXiv:1503.07872}}].

\bibitem{Pich:2013lsa}
A.~Pich {\em Prog. Part. Nucl. Phys.} {\bf 75} (2014) 41--85,
  [\href{http://arxiv.org/abs/1310.7922}{{\tt arXiv:1310.7922}}].

\bibitem{Wess:1971yu}
J.~Wess and B.~Zumino {\em Phys. Lett.} {\bf 37B} (1971) 95--97.

\bibitem{Witten:1983tw}
E.~Witten {\em Nucl. Phys.} {\bf B223} (1983) 422--432.

\bibitem{Hiller:2018wbv}
G.~Hiller, D.~Loose, and I.~Ni{\v s}and{\v z}i{\'c} {\em Phys. Rev.} {\bf D97}
  (2018), no.~7 075004, [\href{http://arxiv.org/abs/1801.09399}{{\tt
  arXiv:1801.09399}}].

\bibitem{Dorsner:2018ynv}
I.~Dor{\v s}ner and A.~Greljo {\em JHEP} {\bf 05} (2018) 126,
  [\href{http://arxiv.org/abs/1801.07641}{{\tt arXiv:1801.07641}}].

\bibitem{Sirunyan:2017yrk}
{\bf CMS} Collaboration, A.~M. Sirunyan et~al. {\em JHEP} {\bf 07} (2017) 121,
  [\href{http://arxiv.org/abs/1703.03995}{{\tt arXiv:1703.03995}}].

\bibitem{Sirunyan:2018nkj}
{\bf CMS} Collaboration, A.~M. Sirunyan et~al.
  \href{http://arxiv.org/abs/1803.02864}{{\tt arXiv:1803.02864}}.

\bibitem{Sirunyan:2018kzh}
{\bf CMS} Collaboration, A.~M. Sirunyan et~al.
  \href{http://arxiv.org/abs/1805.10228}{{\tt arXiv:1805.10228}}.

\bibitem{Sirunyan:2018jdk}
{\bf CMS} Collaboration, A.~M. Sirunyan et~al. {\em Submitted to: JHEP} (2018)
  [\href{http://arxiv.org/abs/1806.03472}{{\tt arXiv:1806.03472}}].

\bibitem{Greljo:2017vvb}
A.~Greljo and D.~Marzocca {\em Eur. Phys. J.} {\bf C77} (2017), no.~8 548,
  [\href{http://arxiv.org/abs/1704.09015}{{\tt arXiv:1704.09015}}].

\bibitem{Alwall:2014hca}
J.~Alwall, R.~Frederix, S.~Frixione, V.~Hirschi, F.~Maltoni, O.~Mattelaer,
  H.~S. Shao, T.~Stelzer, P.~Torrielli, and M.~Zaro {\em JHEP} {\bf 07} (2014)
  079, [\href{http://arxiv.org/abs/1405.0301}{{\tt arXiv:1405.0301}}].

\bibitem{Ahmed:2015qda}
T.~Ahmed, M.~C. Kumar, P.~Mathews, N.~Rana, and V.~Ravindran {\em Eur. Phys.
  J.} {\bf C76} (2016), no.~6 355, [\href{http://arxiv.org/abs/1510.02235}{{\tt
  arXiv:1510.02235}}].

\bibitem{Aaboud:2017yyg}
{\bf ATLAS} Collaboration, M.~Aaboud et~al. {\em Phys. Lett.} {\bf B775} (2017)
  105--125, [\href{http://arxiv.org/abs/1707.04147}{{\tt arXiv:1707.04147}}].

\bibitem{Sirunyan:2017hsb}
{\bf CMS} Collaboration, A.~M. Sirunyan et~al.
  \href{http://arxiv.org/abs/1712.03143}{{\tt arXiv:1712.03143}}.

\bibitem{Aaboud:2017yvp}
{\bf ATLAS} Collaboration, M.~Aaboud et~al. {\em Phys. Rev.} {\bf D96} (2017),
  no.~5 052004, [\href{http://arxiv.org/abs/1703.09127}{{\tt
  arXiv:1703.09127}}].

\bibitem{Redi:2016kip}
M.~Redi, A.~Strumia, A.~Tesi, and E.~Vigiani {\em JHEP} {\bf 05} (2016) 078,
  [\href{http://arxiv.org/abs/1602.07297}{{\tt arXiv:1602.07297}}].

\bibitem{Aaboud:2017nmi}
{\bf ATLAS} Collaboration, M.~Aaboud et~al. {\em Eur. Phys. J.} {\bf C78}
  (2018), no.~3 250, [\href{http://arxiv.org/abs/1710.07171}{{\tt
  arXiv:1710.07171}}].

\bibitem{Aaboud:2017nak}
{\bf ATLAS} Collaboration, M.~Aaboud et~al. {\em Eur. Phys. J.} {\bf C78}
  (2018), no.~2 102, [\href{http://arxiv.org/abs/1709.10440}{{\tt
  arXiv:1709.10440}}].

\bibitem{Bai:2016czm}
Y.~Bai, V.~Barger, and J.~Berger {\em Phys. Rev.} {\bf D94} (2016), no.~1
  011701, [\href{http://arxiv.org/abs/1604.07835}{{\tt arXiv:1604.07835}}].

\bibitem{Monteux:2018ufc}
A.~Monteux and A.~Rajaraman \href{http://arxiv.org/abs/1803.05962}{{\tt
  arXiv:1803.05962}}.

\end{thebibliography}\endgroup

\end{document}